\documentclass[a4paper,11pt]{article}

\usepackage{classeJBArxive}

\usepackage{arydshln}
\usepackage{movie15}

\newcommand*\unit[1]{\bigl[\, \mathsf{#1} \,\bigr]}

\newcommand{\DF}{\textsc{Du}$\,$\textsc{Fort}--\textsc{Frankel}}
\newcommand{\Eu}{\textsc{Euler}}
\newcommand{\ADI}{Alternating Direction Implicit}

\newcommand{\cpu}{\mathrm{cpu}}

\newcommand{\Bi}{\mathrm{Bi}}
\newcommand{\Fo}{\mathrm{Fo}}
\newcommand{\hs}{\mathsf{h}}

\newcommand{\tf}{t_{\,\mathrm{f}}}
\newcommand{\tref}{t_{\,0}}

\newcommand{\ii}{\mathrm{i}}

\title{
\vspace{-1.5cm}
An efficient two-dimensional heat transfer model for building envelopes \\	
\vspace{4pt}
}

\author{Julien Berger \textsuperscript{a}$^{\ast}$, Suelen Gasparin\textsuperscript{a}, Walter Mazuroski\textsuperscript{b}, Nathan Mendes\textsuperscript{c}\\
\date{\today\vspace{-0.5cm}}}

\begin{document}

\maketitle

\begin{center}
\small
\textsuperscript{a} Laboratoire des Sciences de l’Ingénieur pour l’Environnement (LaSIE), UMR 7356 CNRS, La Rochelle Université, CNRS, 17000, La Rochelle, France\\
\textsuperscript{b} Univ. Savoie Mont Blanc, LOCIE, 73000 Chambéry, France \\
\textsuperscript{c} Thermal Systems Laboratory, Mechanical Engineering Graduate Program, \\
Pontif\'icia Universidade Cat\'olica do Paran\'a, Rua Imaculada Conceição, 1155, CEP : 80215-901,
Curitiba, Brazil\\
\end{center}


\begin{abstract}

A two-dimensional model is proposed for energy efficiency assessment through the simulation of heat transfer in building envelopes, considering the influence of the surrounding environment. The model is based on the \DF ~approach that provides an explicit scheme with a relaxed stability condition. The model is first validated using an analytical solution and then compared to three other standard schemes. Results show that the proposed model offers a good compromise in terms of high accuracy and reduced computational efforts. Then, a more complex case study is investigated, considering non-uniform shading effects due to the neighboring buildings. In addition, the surface heat transfer coefficient varies with wind velocity and height, which imposes an addition non-uniform boundary condition. After showing the reliability of the model prediction, a comparison over almost $120$ cities in France is carried out between the two- and the one-dimensional approaches of the current building simulation programs. Important discrepancies are observed for regions with high magnitudes of solar radiation and wind velocity. Last, a sensitivity analysis is carried out using a derivative-based approach. It enables to assess the variability of the solution according to the modeling of the two-dimensional boundary conditions. Moreover, the proposed model computes efficiently the solution and its sensitivity to the modeling of the urban environment.

\textbf{Key words:} \DF ~method; Two-dimensional heat transfer; Numerical model; Building energy efficiency; Non-uniform boundary conditions; Surface solar fraction.

\end{abstract}

\section{Introduction}

The building sector is responsible for almost $33\%$ of the world global energy consumption and the current environmental context imposes an improvement of the energy efficiency of building envelopes \cite{IEA_2015}. For this, several tools, called building simulation programs, have been developed over the last 50 years to assess building energy performance. A review of such models has been proposed in \cite{Woloszyn_2008} with a recent update in \cite{Mendes_2017}. Among the most contemporary, one can cite \texttt{Domus} \cite{Mendes_2008} or \texttt{EnergyPlus} \cite{EnergyPlus} as examples that employ modern techniques of shading assessment, for instance, but have building envelope engines limited to one-dimensional heat transfer modeling.

Among all the phenomena involved in building physics, the heat transfer process through the envelope is one of the most important since it represents a major part of the energy consumption. The conduction loads through the envelope require fine and accurate modeling to guarantee the reliability of the building simulation programs. However, several drawbacks can be outlined. 

First, generally, the building simulation programs mentioned in \cite{Woloszyn_2008,Mendes_2017} model the heat transfer process through the building envelope in one-dimension, as mentioned above. Indeed, for simulation at large scales (district or urbanity), the reliability of one-dimensional envelope models is reduced. Furthermore, most common approaches are the resistance-capacitance model or response-factor method to simulate the heat transfer through the building envelope \cite{Frayssinet_2018}. As reported in \cite{Rodler_2017,Martinez_2019}, detailed models based on two- or three-dimensional approaches are required to increase the accuracy of the predictions. Some attempts have been made to include two-dimensional modeling in \cite{Berger_2016} considering both heat and mass transfer. In \cite{Berger_2018,Mazuroski_2018}, a two-dimensional model has been proposed based on an intelligent co-simulation approach. However, those works assume only simple time-varying boundary conditions.

The second drawback arises from the modeling of the outside boundary conditions. Those are given by time-varying climatic data with a time step of one hour, which may increase the inaccuracy particularly for the modeling of the outside incident radiation flux \cite{Rodler_2017}. Moreover, as reported in \cite{Rocha_2017}, many tools use simple trigonometric methods for shading assessment. Some alternative techniques have been proposed to increase the accuracy of the methods. Particularly, in \cite{Rocha_2017}, a pixel counting technique is developed and validated using experimental data. Even if this approach has been integrated into the \texttt{Domus} building simulation program, the simulation still considers one-dimensional transfer through the envelope. Similarly, the heat transfer coefficient at the interface between the wall and the outside environment is modeled using empirical models. An extensive literature review is given in \cite{Mirsadeghi_2013}. One can note in Table~1 of the mentioned reference \cite{Mirsadeghi_2013} that most programs assume constant values. In building simulation programs such as \texttt{EnergyPlus}, the coefficient may vary according to the wind speed velocity and/or the height. Nevertheless, as mentioned in \cite{Lauzet_2019}, the building simulation programs cannot handle spatially variable boundary conditions. 

Even with the drawbacks identified, the development of two- or three-dimensional heat transfer model in building envelope is still a difficult task. Indeed, the physical phenomena in buildings are generally observed over (at least) one year. Besides, building physical domains scale with several meters. Thus, the characteristic time and space lengths may induce significant computational cost. Thus, efficient numerical models are worth of investigation. In this paper, an innovative numerical model based on the \DF ~scheme is studied, which has already demonstrated a promising efficiency in \cite{Gasparin_2017a,Gasparin_2017b,Berger_2019c} for the simulation of one-dimensional heat and mass transfer through porous materials in building envelopes. Here, the model is extended to simulate two-dimensional heat transfer in a building facade over one year. It considers time and space varying convective and radiative boundary conditions at the external surface. A comparison is performed to analyze the influence of two-dimensional modeling to predict building energy efficiency. For this, a derivative-based approach is used to compute efficiently the time-varying sensitivity of the critical outputs. 

To assess this study, the article is organized as follows. The mathematical model for the solution and its sensitivity is described in Section~\ref{sec:mathematical_model}. Then, the \DF ~numerical scheme is presented in Section~\ref{sec:numerical_model}. A validation procedure is carried out using an analytical solution in Section~\ref{sec:validation_case}. Then, a more realistic case study of a building within aurban environment is treated in Section~\ref{sec:real_case_study}. To conclude, final remarks are addressed in Section~\ref{sec:conclusion}.

\section{Description of the mathematical model}
\label{sec:mathematical_model}

\begin{figure}
\centering
\includegraphics[width=.95\textwidth]{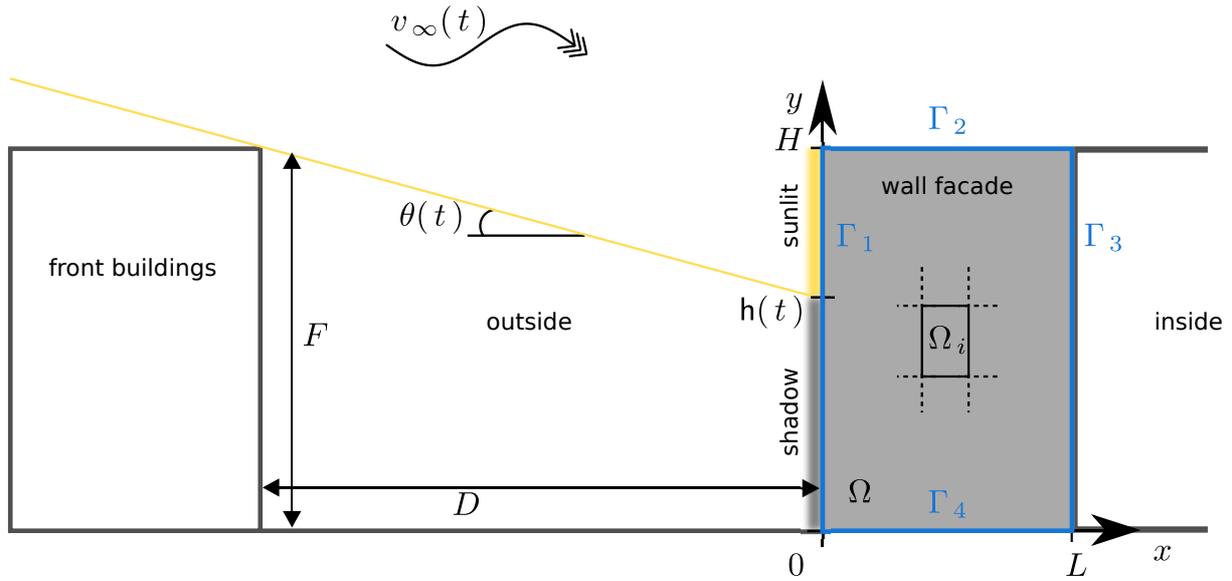}
\caption{Illustration of the problem considered and the physical domain of heat transfer in the building facade.}
\label{fig:domain}
\end{figure}

The problem involves heat transfer through the facade of a building located in an urban area where the studied building faces another one (Figure~\ref{fig:domain}). The front building is located at a distance $D \ \unit{m}$ and it has a height $F \ \unit{m}$ and induces a time-varying shadow on the studied building.

\subsection{Governing equations}

The two-dimensional heat diffusion transfer is considered in a facade composed of a multi-layered wall. The process occurs over the time domain $\Omega_{\,t} \egal \bigl[\, 0 \,,\, \tf \,\bigr]\,$. The space domain is illustrated in Figure~\ref{fig:domain}, where $L\,$, $H$ and $W$ are the length, height and width of the wall. The space coordinates $x$ and $y$ belong to the domains:
\begin{align*}
\Omega_{\,x} \, \eqdef \, \bigl\{\, x \,\big|\, x \,\in \, \bigl[\,0 \,,\, L \,\bigr] \,\bigr\} \qquad \text{and} \qquad
\Omega_{\,y} \, \eqdef \, \bigl\{\, y \,\big|\, y \,\in \, \bigl[\,0 \,,\, H \,\bigr] \,\bigr\}\,.
\end{align*}
Thus, the spatial domain of the wall is $\Omega \, \eqdef \, \Omega_{\,x} \, \cup \, \Omega_{\,y}\,$. The four boundaries of the domains are defined such as:
\begin{align*}
\Gamma_{\,1} &\, \eqdef \, \Bigl\{\, \bigl(\,x\,,\,y\,\bigr) \,\Big|\, 
y \,\in \, \Omega_{\,y} \,,\, x \egal 0  \,\Bigr\} \,,  &&
\Gamma_{\,2} \, \eqdef \, \Bigl\{\, \bigl(\,x\,,\,y\,\bigr) \,\Big|\, 
x \,\in \, \Omega_{\,x} \,,\, y \egal H  \,\Bigr\} \,, \\[4pt]
\Gamma_{\,3} &\, \eqdef \, \Bigl\{\, \bigl(\,x\,,\,y\,\bigr) \,\Big|\, 
y \,\in \, \Omega_{\,y} \,,\, x \egal L  \,\Bigr\} \,,  &&
\Gamma_{\,4} \, \eqdef \, \Bigl\{\, \bigl(\,x\,,\,y\,\bigr) \,\Big|\, 
x \,\in \, \Omega_{\,x} \,,\, y \egal 0  \,\Bigr\} \,.
\end{align*}
Thus, the whole boundary of the spatial domain is $\Gamma \, \eqdef \, \displaystyle \bigcup_{i=1}^{4} \Gamma_{\,i} \,$. The governing equation of heat transfer is:
\begin{align}
\label{eq:heat}
c\,(\,x\,,\,y\,) \cdot \pd{T}{t} \moins \pd{}{x} \biggl(\, k\,(\,x\,,\,y\,) \cdot \pd{T}{x} \,\biggr)
\moins \pd{}{y} \biggl(\, k\,(\,x\,,\,y\,) \cdot \pd{T}{y} \,\biggr) \egal 0 \,,
\end{align}
where $c \ \unit{J\,.\,m^{\,-3}\,.\,K^{\,-1}}$ and $k \ \unit{W\,.\,m^{\,-1}\,.\,K^{\,-1}}$ are the volumetric heat capacity and the thermal conductivity. The wall is composed of $N$ materials and $\Omega \egal \ \displaystyle \bigcup_{i=1}^{N} \Omega_{\,i} \,$, with $\Omega_{\,i}$ being the space domain of the material $i\,$. Thus, both $k$ and $c$ depend on space coordinates:
\begin{align*}
k\,(\,x\,,\,y\,) \egal \sum_{i\egal 1}^{N} k_{\,i} \cdot \phi_{\,i}(\,x\,,\,y\,) \,, \qquad
c\,(\,x\,,\,y\,) \egal \sum_{i\egal 1}^{N} c_{\,i} \cdot \phi_{\,i}(\,x\,,\,y\,)  \,,
\end{align*}
where $k_{\,i}$ and $c_{\,i}$ are the thermal conductivity and the heat capacity of the material $i$ assumed as constant. The function $\phi_{\,i}$ corresponds to a piece wise function basis:
\begin{align*}
\phi_{\,i} \,(\,x\,,\,y\,) \egal 
\begin{cases} 
\ 1 \,, & \quad (\,x\,,\,y\,) \, \in \, \Omega_{\,i} \,, \\
\ 0  \,, & \quad (\,x\,,\,y\,) \, \notin \, \Omega_{\,i} \,.
\end{cases} 
\end{align*}
Initially, the wall is assumed in steady-state condition:
\begin{align*}
T \egal T_{\,0} \,(\,x\,,\,y\,) 
\,,  \quad \forall \, (\,x\,,\,y\,) \, \in \, \Omega \,,\, \quad t \egal 0 \,,
\end{align*}
where $T_{\,0}$ is a given function dependent on space coordinates. The latter requires to be consistent with the boundary conditions. One important output is the heat flux $j \ \unit{W\,.\,m^{\,-2}}$ defined by:
\begin{align*}
j (\,x\,,\,y\,,\,t\,) \egal - \, k(\,x\,,\,y\,) \cdot \grad T \scal \boldsymbol{n}_{\,x\,,\,y} \,,
\end{align*}
with $\boldsymbol{n}_{\,x}$ and $\boldsymbol{n}_{\,y}$ being the unitary normal vector of $x$ and $y$ axis, respectively. The total heat flux $\mathcal{J} \ \unit{W\,.\,m^{\,-2}}$ impacting at the inside of the ambient zone is computed by:
\begin{align}
\label{eq:_total_flux}
\mathcal{J}(\,t\,) \egal \frac{1}{H} \ \int_{\,\Gamma_{\,3}} \ 
- \, k(\,x \, = \, L\,,\,y\,) \cdot
\pd{T}{x}\,\biggr|_{\,x \, = \, L} \ \mathrm{d}y \,.
\end{align}
The last interesting output is the integrated thermal gain (also called thermal or conduction loads):
\begin{align}
\label{thermal_loads}
E(\,\,t\,)  \egal \int_{\,\overline{\Omega_{\,t}}} \ \mathcal{J}(\,t\,) \ \mathrm{d}t \,,
\end{align}
where $\overline{\Omega_{\,t}} \, \subset \, \Omega_{\,t}$ is a time interval generally defined as one month. The thermal loads indicate the amount of thermal energy transferred through the wall.

\subsection{Boundary conditions}

At the interface between two materials, the continuity of the heat flux and temperature field are assumed. At the interface between the wall and the air, the diffusive heat flux entering is equal to the convective and radiative ones. Thus, a \textsc{Robin} type condition are assumed at the boundary $\Gamma_{\,i}\,$: 
\begin{align*}
k_{\,i} \ \pd{T}{n_{\,i}}  \plus h_{\,\infty\,,\,i}\cdot T \egal h_{\,\infty\,,\,i} \cdot T_{\,\infty\,,\,i} \plus q_{\,\infty\,,\,i} \,, 
\quad \forall \, (\,x\,,\,y\,) \, \in \, \Gamma_{\,i} \,, \quad \forall \, i \, \in \, \bigl\{\, 1 \,,\, \ldots \,,\, 4 \,\bigr\}\,,
\end{align*}
where $\displaystyle \pd{T}{n} \, \eqdef \, \pd{T}{x} \cdot \boldsymbol{n} \scal \boldsymbol{n}_{\,x}$ or $ \displaystyle  \pd{T}{n} \, \eqdef \, \pd{T}{y} \cdot \boldsymbol{n} \scal \boldsymbol{n}_{\,y}$ with $ \boldsymbol{n}$ being the outward normal of the considered boundary, $h_{\,\infty\,,\,i} \ \unit{W\,.\,m^{\,-2}\,.\,K^{\,-1}}$ is the surface heat transfer coefficient between the material and the surrounding ambient air and $q_{\,\infty\,,\,i} \ \unit{W\,.\,m^{\,-2}}$ is the incident short-wave radiation flux.
The air temperature $T_{\,\infty\,,\,i}$ depends on time:
\begin{align*}
T_{\,\infty\,,\,i} \,:\, t \longmapsto T_{\,\infty\,,\,i}\,(\,t\,) \,. 
\end{align*}

\begin{figure}
\centering
\includegraphics[width=.45\textwidth]{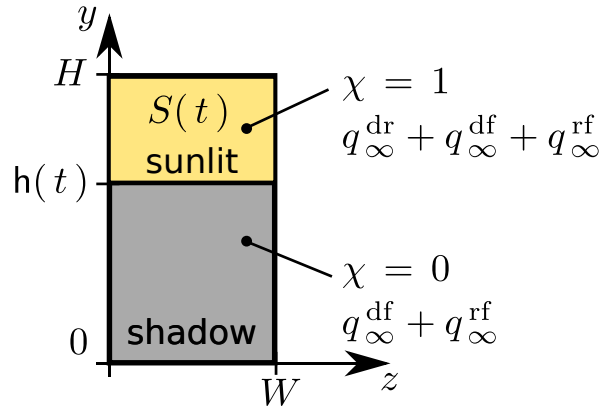}
\caption{Illustration of the indicator function $\chi \,$, using the sunlit area ratio computed by \texttt{Domus}.}
\label{fig:comp_shadow}
\end{figure}

For the external boundary $\Gamma_{\,1}$, the surface heat transfer coefficient and the incident short-wave radiation flux depend on both time and space:
\begin{align*}
h_{\,\infty\,,\,1}\,:\,  (\,y\,,\,t\,) \longmapsto h_{\,\infty\,,\,1}\,(\,y\,,\,t\,) \,, \qquad
q_{\,\infty\,,\,1} \,:\, (\,y\,,\, t\,) \longmapsto q_{\,\infty\,,\,1}\,(\,y\,,\,t\,) \,.
\end{align*}
The coefficient $h_{\,\infty\,,\,1}$ depends on height $y$ and time according to the wind velocity $v_{\,\infty}\ \unit{m\,.\,s^{\,-1}}$ \cite{McAdams_1985}:
\begin{align}
\label{eq:h1}
h_{\,\infty\,,\,1}(\,y\,,\,t\,) \egal 
h_{\,10} \plus h_{\,11} \cdot \frac{v_{\,\infty}}{v_{\,0}} \cdot \biggl(\, \frac{y}{y_{\,0}} \,\biggr)^{\,\beta} \,,
\end{align}
where 
\begin{align*}
v_{\,\infty} \,:\, t \longmapsto v_{\,\infty}\,(\,t\,) \,,
\end{align*}
and $\bigl(\, h_{\,10}\,,\, h_{\,11} \,\bigr)  \ \unit{W\,.\,m^{\,-2}\,.\,K^{\,-1}}\,$ and $\beta \ \unit{-}$ are given coefficients. The reference quantities $v_{\,0}$ and $y_{\,0}$ are set to $v_{\,0} \egal 1 \ \mathsf{m\,.\,s^{\,-1}}$ and $y_{\,0} \egal 1 \ \mathsf{m}\,$. The mean value of the surface heat transfer coefficient is defined by:
\begin{align*}
\overline{h}_{\,\infty\,,\,1} \, \eqdef \, \frac{1}{H \cdot \tf} \, \int_{\,\Omega_{\,t}} \ \int_{\,\Omega_{\,y}} h_{\,\infty\,,\,1}(\, x\,,\, t \,) \ \mathrm{d}x \, \mathrm{d}t \,.
\end{align*}
The incident short-wave radiation flux $q_{\,\infty\,,\,1}$ also depends on space and time according to the variation of the sunlit on the facade. It is constituted with the direct $q_{\,\infty}^{\,\mathrm{dr}}$, diffuse $q_{\,\infty}^{\,\mathrm{df}}$ and reflected $q_{\,\infty}^{\,\mathrm{rf}}$ components. The direct flux $q_{\,\infty}^{\,\mathrm{dr}}(\,t\,)$ depends on the total direct solar radiation $I \ \unit{W\,.\,m^{\,-2}}\,$:
\begin{align*}
q_{\,\infty}^{\,\mathrm{dr}}(\,t\,) \egal I(\,t\,) \cdot \cos \Bigl(\, \theta(\,t\,) \,\Bigr) \,,
\end{align*}
where $\theta \ \unit{-}$ is the angle between the wall normal and the solar beam. The magnitude of the direct heat flux depends on the position of the shadow on the facade. The latter can be the consequence of different shading elements such as screens, trees or other buildings. Thus, the incident radiation flux is decomposed as:
\begin{align}
\label{eq:incident_flux_bound_1}
q_{\,\infty\,,\,1} (\,y\,,\,t\,) \egal \alpha \cdot \Bigl(\, q_{\,\infty}^{\,\mathrm{dr}}(\,t\,) \cdot \chi (\,y\,,\,t\,) 
\plus q_{\,\infty}^{\,\mathrm{df}}(\,t\,) 
\plus q_{\,\infty}^{\,\mathrm{rf}}(\,t\,) \,\Bigr) \,,
\end{align}
where $\alpha \ \unit{-}$ is the wall absorptivity and $\chi (\,y\,,\,t\,)$ is an indicator function which feature is illustrated in Figure~\ref{fig:comp_shadow}. It returns $1$ if $y$ is out of the shadow and $0$ if $y$ is in the shadow. Thus, the indicator function is defined as: 
\begin{align}
\label{eq:chi}
\chi (\,y\,,\,t\,) \egal
\begin{cases}
\ 0 \,, \qquad  y \, \leqslant \, \hs  (\,t\,)  \,, \\[4pt]
\ 1 \,, \qquad  y \, > \, \hs (\,t\,)  \,,
\end{cases}
\end{align}
where $\hs  (\,t\,) \ \unit{m}$ is the height of the shadow on the outside wall facade. It is computed according to:
\begin{align*}
\hs (\,t\,) \egal \displaystyle H \cdot \bigl(\,1 \moins S(\,t\,) \,\bigr) \,,
\end{align*}
where $\displaystyle S(\,t\,) \unit{-}$  is the sunlit area ratio perpendicular to the ground. It corresponds to the ratio between the sunlit area $A_{\,s} \ \unit{m^{\,2}}$ and the total area $A$ of the wall facade. Assuming that the frontier between the sunlit and shadow area as a straight line, the sunlit area ratio is given by:
\begin{align}
\label{eq:S}
S(\,t\,) \egal \frac{\bigl(\, H \moins \hs(\,t\,)\,\bigr)}{H} \,.
\end{align}
It is calculated using the pixel counting technique described in \cite{Jones_2012} and implemented in the \texttt{Domus} building simulation program \cite{Mendes_2008,Rocha_2017}.

\subsection{Sensitivity analysis of the two dimensional aspect of the boundary conditions}

The outside boundary conditions are modeled in two dimensions, \emph{i.e.}, varying according to the time $t$ and the height $y\,$. To evaluate the influence of such modeling on the assessment of the energy efficiency of the facade, a derivative-based sensitivity analysis is carried out \cite{Sobol_2009,Kucherenko_2016,Jumabekova_2019,Berger_2020}. The four essential parameters in modeling the outside boundary conditions are the height of the front building $F\,$, the distance of the front building $D\,$, the first-order coefficient $h_{\,11}$ and the coefficient $\beta$ of the power law described in Eq.~\eqref{eq:h1}. A \textsc{Taylor} development of the temperature is expressed: 
\begin{align}
\label{eq:Taylor_dev}
 T (\,x\,,\,t\,,\,&h_{\,11}\,,\,\beta\,,\,\,F\,,\,D\,) \nonumber
\egal T (\,x\,,\,t\,,\,h_{\,11}^{\,\circ}\,,\,\beta^{\,\circ}\,,\,F^{\,\circ}\,,\,D^{\,\circ}\,)\\[4pt]
& 
\plus \pd{T}{h_{\,11}}\,\biggl|_{\,h_{\,11} \egal h_{\,11}^{\,\circ}} \cdot \, \delta h_{\,11}  
\plus \pd{T}{\beta}\,\biggl|_{\,\beta \egal \beta^{\,\circ}} \cdot \, \delta \beta 
\plus \pd{T}{F}\,\biggl|_{\,F \egal F^{\,\circ}} \cdot \,  \delta F
\plus \pd{T}{D}\,\biggl|_{\,D \egal D^{\,\circ}} \cdot \, \delta D  \nonumber \\[4pt]
& \plus \mathcal{O}\bigl(\, \delta h_{\,11}^{\,2} \,,\, \delta \beta^{\,2} \,,\, \delta F^{\,2} \,,\, \delta D^{\,2} \,\bigr)\,,
\end{align}
where 
\begin{align*}
\delta h_{\,11} \, \eqdef \, h_{\,11} \moins h_{\,11}^{\,\circ} \,,\quad
\delta \beta \, \eqdef \, \beta \moins \beta^{\,\circ}  \,,\quad
\delta F \, \eqdef \, F \moins F^{\,\circ}  \,,\quad
\delta D \, \eqdef \, D \moins D^{\,\circ}  \,.
\end{align*}
This development enables to assess the variability of the temperature for any value of parameters $\bigl(\,\beta\,,\,h_{\,11}\,,\,F\,,\,D\,\bigr)$ around the given ones $\bigl(\,\beta^{\,\circ} \,,\,h_{\,11}^{\,\circ} \,,\,F^{\,\circ} \,,\,D^{\,\circ} \,\bigr)\,$. Note that the \textsc{Taylor} development -Eq.~\eqref{eq:Taylor_dev}- can be written for any other chosen output such as, for instance, the total heat flux $\mathcal{J}$ from Eq.~\eqref{eq:_total_flux} or the thermal loads $E$ from Eq.~\eqref{thermal_loads}. 

\begin{figure}
\centering
\subfigure[\label{fig:ext_F}]{\includegraphics[width=.45\textwidth]{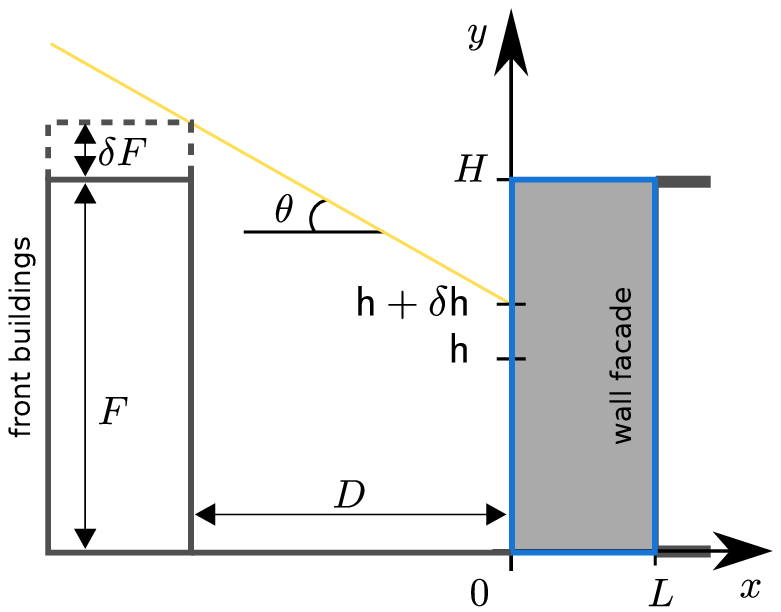}}  \hspace{0.2cm}
\subfigure[\label{fig:ext_D}]{\includegraphics[width=.44\textwidth]{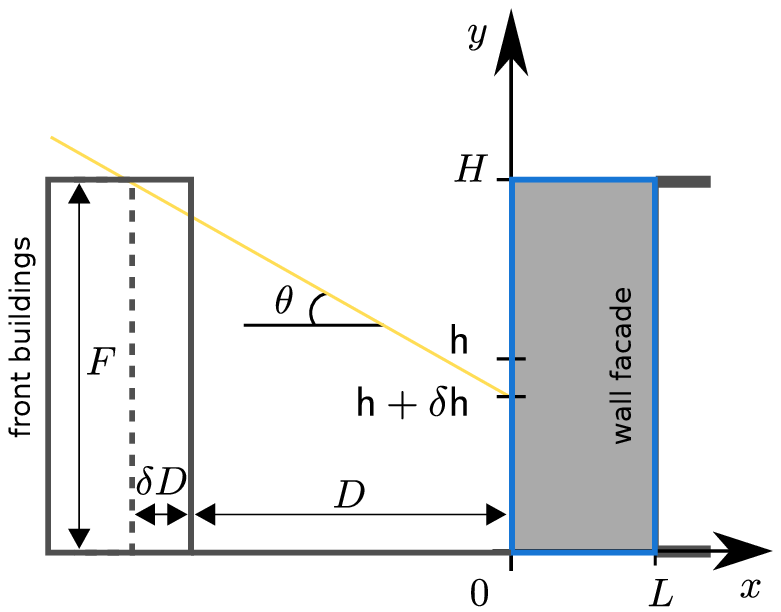}} 
\caption{Variation of the shadow height according to a slight variation of the front building height \emph{(a)} or distance \emph{(b)}.}
\end{figure}

Instead of performing costly discrete sampling to assess the partial derivative relative to each of the four parameters $\bigl(\,\beta\,,\,h_{\,11}\,,\,F\,,\,D\,\bigr)\,$, the governing equation~\eqref{eq:heat} is directly differentiated with respect to the selected parameter. For this, we denote the four sensitivity coefficients by:
\begin{align*}
\Theta_{\,1} \egal \pd{T}{h_{\,11}} \,, \qquad 
\Theta_{\,2} \egal \pd{T}{\beta}\,, \qquad 
\Theta_{\,3} \egal \pd{T}{F}\,, \qquad 
\Theta_{\,4} \egal \pd{T}{D}\,.
\end{align*}
Each of them is the solution to the following partial differential equations:
\begin{align}
\label{eq:sensitivity_coefficient}
c \cdot \pd{\Theta_{\,i}}{t} 
\moins \pd{}{x} \biggl(\, k \cdot \pd{\Theta_{\,i}}{x} \,\biggr)
\moins \pd{}{y} \biggl(\, k \cdot \pd{\Theta_{\,i}}{y} \,\biggr) 
\egal 0 \,, \qquad \forall \, i \,\in \, \bigl\{\,1 \,,\, \ldots \,,\, 4 \,\bigr\} \,.
\end{align}
The initial condition is $\theta_{\,i} \egal 0 \,,  \quad \forall \, (\,x\,,\,y\,) \, \in \, \Omega \,,\, t \egal 0 \,, \forall \, i \,\in \, \bigl\{\,1 \,,\, \ldots \,,\, 4 \,\bigr\} \,$. The differences  in the computation of the sensitivity coefficients $\Theta_{\,i}$ arise in the boundary conditions. For $h_{\,11}$ and $\beta$, the boundary conditions are:
\begin{align*}
k_{\,i} \cdot \pd{\Theta_{\,1}}{n_{\,i}}  \plus h_{\,\infty\,,\,i}\cdot \Theta_{\,1} 
& \egal \pd{h_{\,\infty\,,\,i}}{h_{\,11}} \cdot \bigl(\, T_{\,\infty\,,\,i} \moins T \,\bigr) \,, \\[4pt]
k_{\,i} \cdot \pd{\Theta_{\,2}}{n_{\,i}}  \plus h_{\,\infty\,,\,i}\cdot \Theta_{\,2} 
& \egal \pd{h_{\,\infty\,,\,i}}{\beta} \cdot \bigl(\, T_{\,\infty\,,\,i} \moins T \,\bigr) \,, 
\quad \forall \, (\,x\,,\,y\,) \, \in \, \Gamma_{\,i} \,,
\quad \forall i \, \in \, \bigl\{\, 1 \,,\, \ldots \,,\, 4 \,\bigr\}\,,
\end{align*}
with 
\begin{align*}
\pd{h_{\,\infty\,,\,i}}{h_{\,11}} \,(\,y\,,\,t\,)
\egal
\begin{cases}
& 0 \,, \qquad  \forall \, i \, \in \, \bigl\{\, 2 \,,\, 3 \,,\, 4 \,\bigr\}\,, \\[4pt]
& \displaystyle \frac{v_{\,\infty}}{v_{\,0}} \cdot \biggl(\, \frac{y}{y_{\,0}} \,\biggr)^{\,\beta} \,, \qquad i \egal 1 \,.
\end{cases} 
\end{align*}
and
\begin{align*}
\displaystyle \pd{h_{\,\infty\,,\,i}}{\beta} \,(\,y\,,\,t\,)
\egal
\begin{cases}
& 0 \,, \qquad  \forall \, i \, \in \, \bigl\{\, 2 \,,\, 3 \,,\, 4 \,\bigr\}\,, \\[4pt]
& \displaystyle h_{\,11} \cdot \frac{v_{\,\infty}}{v_{\,0}} \cdot
\ln \biggl(\, \frac{y}{y_{\,0}} \,\biggr)
\cdot \biggl(\, \frac{y}{y_{\,0}} \,\biggr)^{\,\beta} 
\,, \qquad i \egal 1 \,.
\end{cases} 
\end{align*}
For $\Theta_{\,3}$ and $\Theta_{\,4}\,$, the boundary conditions are:
\begin{align*}
k_{\,i} \cdot \pd{\Theta_{\,3}}{n_{\,i}}  \plus h_{\,\infty\,,\,i}\cdot \Theta_{\,3} 
& \egal \pd{q_{\,\infty\,,\,i}}{F}  \,, \\[4pt]
k_{\,i} \cdot \pd{\Theta_{\,4}}{n_{\,i}}  \plus h_{\,\infty\,,\,i}\cdot \Theta_{\,4} 
& \egal \pd{q_{\,\infty\,,\,i}}{D} \,,
\quad \forall \, (\,x\,,\,y\,) \, \in \, \Gamma_{\,i} \,,
\quad \forall i \, \in \, \bigl\{\, 1 \,,\, \ldots \,,\, 4 \,\bigr\}\,.
\end{align*}
For those two sensitivity coefficients, the purpose is to obtain the derivative of $q_{\,\infty\,,\,i}$ according to $F$ or $D\,$. First, it should be noted that the incident flux on the boundaries $\Gamma_{\,2}\,$, $\Gamma_{\,3}$ and $\Gamma_{\,4}$ do not vary with those parameters:
\begin{align*}
\pd{q_{\,\infty\,,\,i}}{F} \egal \pd{q_{\,\infty\,,\,i}}{D} \egal 0 \,, \qquad \forall \, i \, \in \, \bigl\{\, 2 \,,\, 3 \,,\, 4 \,\bigr\}\,.
\end{align*}
Then, from Eq.\eqref{eq:incident_flux_bound_1}, we have:
\begin{align*}
\pd{q_{\,\infty\,,\,1}}{F} \egal \alpha \cdot q_{\,\infty}^{\,\mathrm{dr}}\cdot \pd{\chi}{F} \,,
\end{align*}
and similarly
\begin{align*}
\pd{q_{\,\infty\,,\,1}}{D} \egal \alpha \cdot q_{\,\infty}^{\,\mathrm{dr}}\cdot \pd{\chi}{D} \,.
\end{align*}
Since there is no direct analytical relation between the indicator function $\chi$ and the geometric parameters $F$ and $D\,$, the partial derivatives $\displaystyle \pd{\chi}{F}$ and $\displaystyle \pd{\chi}{D}$ are obtained  using a discrete modeling and geometric considerations. As illustrated in Figures~\ref{fig:ext_F} and \ref{fig:ext_D}, the increase of the building front height and distance is denoted by $\delta F$ and $\delta D\,$, respectively. Thus, we have
\begin{align*}
\pd{\chi}{F} \, \approx \, \frac{\chi(\,F \plus \delta F\,,\,t\,) \moins \chi(\,F\,) }{\delta F} 
\plus \mathcal{O}\bigl(\,\delta F\,\bigr) \,, \qquad
\pd{\chi}{D} \, \approx \, \frac{\chi(\,D \plus \delta D\,,\,t\,) \moins \chi(\,D\,) }{\delta D} 
\plus \mathcal{O}\bigl(\,\delta D\,\bigr) \,.
\end{align*}
Note that discrete derivative of higher-order accuracy can be defined if required. The derivative according to $F$ is first treated. An increase of the front building height $\delta F$ induces an increase $\delta \hs $ of the shadow height. Thus, the new shadow height is given by:
\begin{align*}
\tilde{\hs} \egal \hs \plus \delta \hs \egal 
\min \, \bigl(\, \hs \plus \delta F \,,\, H \,\bigr) \,.
\end{align*}
For an increase of the front building height, the new shadow height verifies $\hs \, < \, \tilde{\hs} \,< \, H\,$.
Thus, the indicator function at $F \plus \delta F$ can be evaluated by:
\begin{align*}
\chi(\,F \plus \delta F\,,\,t\,)
\egal
\begin{cases}
\ 0 \,, \qquad  y \, \leqslant \, \tilde{\hs} (\,t\,) \,, \\[4pt]
\ 1 \,, \qquad  y \, > \, \tilde{\hs} (\,t\,) \,,
\end{cases}
\end{align*}
to obtain the discrete derivative of the indicator function:
\begin{align*}
\pd{\chi}{F} \, \approx \,
\frac{-1}{\delta F} \
\begin{cases}
\ 0 \,, \qquad  y \, \leqslant \, \hs  (\,t\,) \,, \\[4pt]
\ 1 \,, \qquad \hs  (\,t\,) \, < \, y \, \leqslant \, \tilde{\hs} (\,t\,) \,, \\[4pt]
\ 0 \,, \qquad  y \, > \, \tilde{\hs} (\,t\,) \,.
\end{cases}
\end{align*}
Similarly, the derivative according to $D$ is assessed. Using geometrical consideration from Figure~\ref{fig:ext_D}, an increase of the front building distance $\delta D$ implies a decrease of the shadow height:
\begin{align*}
\tilde{\hs}  \egal 
\max \, \bigl(\, \hs \moins \delta D \cdot \tan \theta\,,\, 0 \,\bigr) \,.
\end{align*}
The new shadow height verified $0 \, < \, \tilde{\hs} \, < \, \hs \,$. The discrete derivative of the indicator function according to parameter $D$ is: 
\begin{align*}
\pd{\chi}{D} \, \approx \,
\frac{1}{\delta D}  \
\begin{cases}
\ 0 \,, \qquad  y \, \leqslant \, \tilde{\hs} \,, \\[4pt]
\ 1 \,, \qquad \tilde{\hs} \, < \, y \, \leqslant \, \hs (\,t\,) \,, \\[4pt]
\ 0 \,, \qquad  y \, > \, \hs (\,t\,) \,.
\end{cases}
\end{align*}
A similar development can be done for a decrease in the height or distance of the front building. Using the governing equation combined with the initial and boundary conditions, the four sensitivity coefficients can be computed to perform a \textsc{Taylor} development of the interesting output. The computation is carried out with the governing equation of heat transfer~\eqref{eq:heat}. Note that Eq.~\eqref{eq:sensitivity_coefficient} and \eqref{eq:heat} are equal from a mathematical point of view. The same efficient numerical model can be employed to compute the solution. Within an explicit time scheme, the  total cost to assess the sensitivity of the output scales with $5$ times the cost of the direct problem ($4$ sensitivity coefficient plus the equation of heat transfer). This cost is strongly reduced compared to sampling approaches. 

\subsection{Dimensionless formulation}

To perform efficient numerical computations, it is of major importance to elaborate a dimensionless formulation of the problem. For this, the temperature is transformed into the dimensionless variable $u\,$:
\begin{align}
\label{eq:transformation_T_u}
u \,\eqdef \, \frac{T \moins T_{\,0}}{\delta T} \,,
\end{align}
where $T_{\,0}$ and $\delta T$ are chosen reference temperatures. This transformation is also applied to the initial condition $T_{\,0}$ and to the boundary condition $T_{\,\infty}\,$. The space and time coordinates are also changed:
\begin{align*}
t^{\,\star} & \eqdef \, \frac{t}{t_{\,0}} \,,  
&&x^{\,\star} \eqdef \, \frac{x}{L_{\,x\,,\,0}} \,,
&& y^{\,\star} \eqdef \, \frac{x}{L_{\,y\,,\,0}} \,.
\end{align*}
where $t_{\,0}\,$, $L_{\,x\,,\,0}\egal L$ and $L_{\,y\,,\,0}\egal H$ are reference time and length quantities. A different reference length is chosen for $x$ and $y$ coordinates to re-scale the dimensionless problem on the plate $\bigl[\,0\,,\,1\,\bigr] \times \bigl[\,0\,,\,1\,\bigr]\,$. The material properties are converted to:
\begin{align*}
k^{\,\star} & \eqdef \, \frac{k}{k_{\,0}} \,,
&& c^{\,\star} \eqdef \, \frac{c}{c_{\,0}} \,,
\end{align*}
where $k_{\,0}$ and $c_{\,0}$ are reference thermal conductivity and volumetric heat capacity. The coefficients $c^{\,\star}$ and $k^{\,\star}$ are called distortion ones according to the reference conditions. Through these transformations, dimensionless numbers are enhanced. Namely, the \textsc{Fourier} number characterizes the diffusion process through the $x$ or $y$ directions. The \textsc{Biot} number translates the intensity of the heat penetration at the interface between the air and the material. Both are defined such as:
\begin{align*}
\Fo_{\,x} & \eqdef \, \frac{k_{\,0} \ t_{\,0}}{c_{\,0} \ L_{\,x\,,\,0}^{\,2}} \,,
&& \Fo_{\,y} \eqdef \, \frac{k_{\,0} \ t_{\,0}}{c_{\,0} \ L_{\,y\,,\,0}^{\,2}} \,,
&& \Bi_{\,x} \eqdef \, \frac{h \ L_{\,x\,,\,0}}{k_{\,0}} \,,
&& \Bi_{\,y} \eqdef \, \frac{h \ L_{\,y\,,\,0}}{k_{\,0}} \,.
\end{align*}
At the boundaries, the heat flux is changed such as:
\begin{align*}
q_{\,\infty\,,\,x}^{\,\star} & \eqdef \, \frac{q_{\,\infty} \ L_{\,x\,,\,0}}{k_{\,0} \ \delta T} \,,
&& q_{\,\infty\,,\,y}^{\,\star} \eqdef \, \frac{q_{\,\infty} \ L_{\,y\,,\,0}}{k_{\,0} \ \delta T} \,.
\end{align*}
In the end, the dimensionless problem is defined as: 
\begin{align}
\label{eq:heat_2D_dimensionless}
c^{\,\star} \cdot \pd{u}{t^{\,\star}} \moins 
\Fo_{\,x} \cdot \pd{}{x^{\,\star}} \biggl(\, k^{\,\star} \cdot \pd{u}{x^{\,\star}} \,\biggr)
\moins 
\Fo_{\,y} \cdot \pd{}{y^{\,\star}} \biggl(\, k^{\,\star}\cdot \pd{u}{y^{\,\star}} \,\biggr) \egal 0 \,,
\end{align}
with the boundary conditions:
\begin{align*}
k_{\,i}^{\,\star} \cdot \pd{u}{n_{\,i}}  \plus \Bi_{\,i}\cdot u \egal \Bi_{\,i}\cdot u_{\,\infty\,,\,i} \plus q^{\,\star}_{\,\infty\,,\,i} \,, 
\end{align*}
and the initial condition $u \egal u_{\,0}\,$.

\section{Direct numerical model}
\label{sec:numerical_model}

\subsection{The \DF ~numerical method}

\subsubsection{Numerical scheme}

A uniform discretisation is considered for space and time lines. For the sake of clarity, the super-script $^{\,\star}$ is removed in this section for the description of the numerical method. The discretisation parameters are denoted using $\Delta t$ for the time, $\Delta x$ for the $x$ space and $\Delta y$ for the $y$ one. The discrete values of the function $u \, (\,x\,,\,y\,,\,t\,)$ are written as $u_{\,j\,i}^{\,n} \ \eqdef \ u\,(\,x_{\,j}\,,\,y_{\,i}\,,\,t^{\,n}\,)$ with $i \egal \bigl\{\, 1 \,, \ldots \,, N_{\,y}\,\bigr\}\,$, $j \egal \bigl\{\, 1 \,, \ldots \,, N_{\,x}\,\bigr\}$ and $n \egal \bigl\{\, 1 \,, \ldots \,, N_{\,t} \,\bigr\} \,$. 

\begin{figure}
\centering
\includegraphics[width=.5\textwidth]{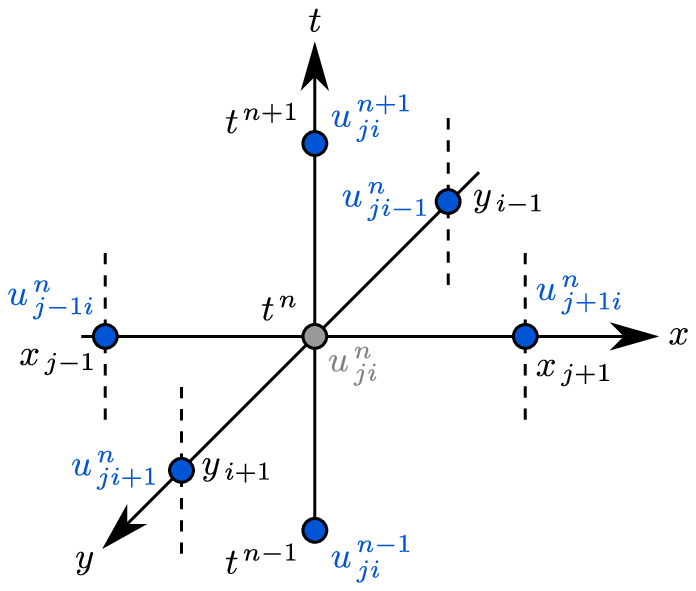}
\caption{\DF ~numerical scheme stencil.}
\label{fig:DF_stencil}
\end{figure}

The \DF ~scheme is employed to build an efficient numerical model for the two-dimensional heat diffusion equation. For the sake of simplicity, to explain the numerical scheme the latter is written as:
\begin{align}
\label{eq:heat_2D}
c \cdot \pd{u}{t} \egal \pd{}{x} \biggl(\, k_{\,x} \cdot \pd{u}{x}\,\biggr) \plus \pd{}{y} \biggl(\, k_{\,y} \cdot \pd{u}{y}\,\biggr) \,.
\end{align}
According to Eq.~\eqref{eq:heat_2D_dimensionless}, we have $k_{\,x} \, \eqdef \, \Fo_{\,x} \cdot k$ and $k_{\,y} \, \eqdef \, \Fo_{\,x} \cdot k\,$ . The coefficients $c\,$, $k_{\,x}$ and $k_{\,y}$ are assumed as constant, independent on time or space. First, Eq.~\eqref{eq:heat_2D} is discretized using finite central differences and forward \textsc{Euler} approach:
\begin{align}
\label{eq:heat_2D_FECD}
\frac{c}{\Delta t} \cdot \Bigl(\, u_{\,j\,i}^{\,n+1} \moins u_{\,j\,i}^{\,n} \,\Bigr) 
\egal &
\frac{k_{\,x}}{\Delta x^{\,2}} \cdot 
\Bigl(\, u_{\,j-1\,i}^{\,n} \moins 2 \, u_{\,j\,i}^{\,n} \plus u_{\,j+1\,i}^{\,n}  \,\Bigr)
\plus \frac{k_{\,y}}{\Delta y^{\,2}} \cdot 
\Bigl(\, u_{\,j\,i-1}^{\,n} \moins 2 \, u_{\,ji}^{\,n} \plus u_{\,j\,i+1}^{\,n}  \,\Bigr)  \,.
\end{align}
Then, to obtain the \DF ~scheme, the term $u_{\,j\,i}^{\,n}$ is replaced by $\displaystyle \frac{1}{2} \Bigl(\,u_{\,j\,i}^{n+1} \plus u_{\,j\,i}^{\,n-1} \,\Bigr)$ in Eq.~\eqref{eq:heat_2D_FECD}. It yields to the following explicit expression:
\begin{align}
\label{eq:DF_num_scheme}
u_{\,j\,i}^{\,n+1} \egal
\Sigma_{\,x} \cdot \Bigl(\, u_{\,j-1\,i}^{\,n} \plus u_{\,j+1\,i}^{\,n} \,\Bigr)
\plus
\Sigma_{\,y} \cdot \Bigl(\, u_{\,j\,i-1}^{\,n} \plus u_{\,j\,i+1}^{\,n} \,\Bigr)
\plus 
\Sigma_{\,xy} \cdot u_{\,j\,i}^{\,n-1} \,,
\end{align}
where the coefficient $\Sigma_{\,x}\,$, $\Sigma_{\,y}$ and $\Sigma_{\,xy}$ are given by:
\begin{align*}
\Sigma_{\,x} & \eqdef \, \frac{\lambda_{\,x}}{1 \plus \lambda_{\,x} \plus \lambda_{\,y}} \,,
&& \Sigma_{\,y} \, \eqdef \, \frac{\lambda_{\,y}}{1 \plus \lambda_{\,x} \plus \lambda_{\,y}} \,,
&& \Sigma_{\,xy} \, \eqdef \, \frac{1 \moins \lambda_{\,x} \moins \lambda_{\,y}}{1 \plus \lambda_{\,x} \plus \lambda_{\,y}} \,, \\[4pt]
 \lambda_{\,x} & \eqdef \, \frac{2 \, \Delta t}{\Delta x^{\,2}} \, \frac{k_{\,x}}{c} \,,
&& \lambda_{\,y} \, \eqdef \, \frac{2 \, \Delta t}{\Delta y^{\,2}} \, \frac{k_{\,y}}{c} \,.
\end{align*}
The stencil of the scheme is illustrated in Figure~\ref{fig:DF_stencil}. The scheme is explicit expressed so no costly inversion of matrix is required, as in implicit approaches. Furthermore, as demonstrated in next section, it has an extended stability region, so the so-called \textsc{C}ourant-\textsc{F}riedrichs-\textsc{L}ewy (CFL) restriction \cite{Courant_1928} is relaxed. Interested readers may consult \cite{Gasparin_2017a,Gasparin_2017b,Berger_2019b} for example of its applications for one-dimensional heat and moisture transfer in building porous materials.

\subsubsection{Stability}
\label{sec:DF_stability}

To proof the unconditional stability of the numerical scheme, a standard \textsc{von} \textsc{Neumann} analysis is carried out. Assuming constant diffusion coefficient, the solution is decomposed according to:
\begin{align}
\label{eq:decomposed_solution}
u_{\,ji}^{\,n} \egal \rho^{\,n} \cdot \exp \bigl(\, -\ii \, \beta \, x_{\,j} \,) \cdot \exp \bigl(\, -\,\ii \, \gamma \, y_{\,i} \,) \,,
\end{align}
where $\ii \egal \sqrt{- 1} \,$, $\gamma$ and $\beta$ are real numbers and $\rho$ is a complex one. Substituting Eq.~\eqref{eq:decomposed_solution} into Eq.~\eqref{eq:DF_num_scheme}, one obtains: 
\begin{align*}
\rho \egal 
\Sigma_{\,x} \cdot \Bigl(\, \exp \bigl(\, \ii \, \beta \, \Delta x \,) \plus \exp \bigl(\, -\ii \, \beta \, \Delta x \,)\, \,\Bigr) 
\plus
\Sigma_{\,y} \cdot \Bigl(\, \exp \bigl(\, \ii \, \gamma \, \Delta y \,) \plus \exp \bigl(\, -\ii \, \gamma \, \Delta y \,)\, \,\Bigr) 
\plus 
\frac{\Sigma_{\,xy} }{\rho} \,.
\end{align*}
It leads to the following second-order polynomials in $\rho$ equation:
\begin{align}
\label{eq:second_order_polyn}
\rho^{\,2} \moins B \, \rho \plus C \egal 0 \,,
\end{align}
with 
\begin{align*}
B \, \eqdef \,  2 \, \Bigl(\,
\Sigma_{\,x} \, \cos \bigl(\, \beta \, \Delta x \,)
\plus
\Sigma_{\,y} \, \cos \bigl(\, \gamma \, \Delta y \,)
 \,\Bigr) \,, \qquad
 C \, \eqdef \,  - \, \Sigma_{\,xy} \,.
\end{align*}
The general solution of Eq.~\eqref{eq:second_order_polyn} is:
\begin{align*}
\rho_{\,\pm} \egal - \, \frac{1}{2} \, \bigl(\, B \, \pm \, \sqrt{D} \,\bigr) \,, \qquad
D \, \eqdef \, B^{\,2} \moins 4 \, C \,.
\end{align*}
The modulus $\bigl| \,\rho_{\,\pm} \,\bigr|$ verifies:
\begin{align*}
\bigl| \, \rho_{\,\pm} \,\bigr| \, \leqslant \, \frac{1}{2} \, \Bigl(\, \bigl|\, B\,\bigr| \, \pm \, \bigl|\, \sqrt{D}\,\bigr| \,\Bigr) \,.
\end{align*}
It is straightforward that $\frac{1}{2} \, \bigl|\, B\,\bigr| \, \leqslant \, \Sigma_{\,x} \plus \Sigma_{\,y} \,$. 
Given the expression of $D\,$, we have:
\begin{align*}
\frac{1}{2} \, \bigl|\, \sqrt{D} \,\bigr| \egal
\frac{1}{1 \plus \lambda_{\,x} \plus \lambda_{\,y}} \cdot \Biggl(\, 
\Bigl(\, \lambda_{\,x} \, \cos \bigl(\, \beta \, \Delta x \,)
\plus
\lambda_{\,y} \, \cos \bigl(\, \gamma \, \Delta y \,) \,\Bigr)^{\,2}
\plus 1
\moins
\Bigl(\, \lambda_{\,x} \plus \lambda_{\,y} \,\Bigr)^{\,2}
\,\Biggr)^{\,\half} \,.
\end{align*}
Thus, 
\begin{align*}
\frac{1}{2}\, \bigl|\, \sqrt{D} \,\bigr| \, \leqslant \, \frac{1}{1 \plus \lambda_{\,x} \plus \lambda_{\,y}} \,,
\end{align*}
and
\begin{align*}
\bigl| \, \rho_{\,\pm} \,\bigr| \, \leqslant \, \Sigma_{\,x} \plus \Sigma_{\,y} \plus \frac{1}{1 \plus \lambda_{\,x} \plus \lambda_{\,y}} \,.
\end{align*}
One can note that the right hand side is equal to 
\begin{align*}
\Sigma_{\,x} \plus \Sigma_{\,y} \plus \frac{1}{1 \plus \lambda_{\,x} \plus \lambda_{\,y}} \egal 1 \,.
\end{align*}
Therefore,  the $\bigl| \, \rho_{\,\pm} \,\bigr| \, \leqslant \, 1$ always holds and the scheme is unconditionally stable. 

\subsubsection{Accuracy}

The consistence analysis of the scheme~\eqref{eq:DF_num_scheme}, using \textsc{Taylor} expansion, gives the following result: 
\begin{align*}
& u_{\,j\,i}^{\,n+1} \moins
\Sigma_{\,x} \cdot \Bigl(\, u_{\,j-1\,i}^{\,n} \plus u_{\,j+1\,i}^{\,n} \,\Bigr)
\moins
\Sigma_{\,y} \cdot  \Bigl(\, u_{\,j\,i-1}^{\,n} \plus u_{\,j\,i+1}^{\,n} \,\Bigr)
\moins 
\Sigma_{\,xy} \cdot u_{\,j\,i}^{\,n-1} \\[4pt]
\egal 
& c \cdot \pd{u}{t}  \moins \pd{}{x} \biggl(\, k_{\,x} \cdot \pd{u}{x}\,\biggr) \moins \pd{}{y} \biggl(\, k_{\,y} \cdot \pd{u}{y}\,\biggr) 
\plus
\Biggl(\, 
\biggl(\, \frac{k_{\,x}}{\Delta x^{\,2}} \plus \frac{k_{\,y}}{\Delta y^{\,2}}\,\biggr) \cdot \pd{^{\,2}u}{t^{\,2}} 
\plus \frac{1}{6} \, \, \pd{^{\,3}u}{t^{\,3}} 
\,\Biggr) \cdot \Delta t^{\,2} \\[4pt]
& \plus \mathcal{O}\bigl(\, \Delta x^{\,2} \plus \Delta y^{\,2} \plus \Delta t^{\,4}  \,\bigr) \,.
\end{align*}
Thus, the scheme is second-order accurate in space $\mathcal{O}\bigl(\, \Delta x^{\,2} \plus \Delta y^{\,2}\,\bigr)\,$. However, the accuracy in time depends on the quantity $\tau$ defined as: 
\begin{align*}
\tau \, \eqdef \, \biggl(\, \frac{k_{\,x}}{\Delta x^{\,2}} \plus \frac{k_{\,y}}{\Delta y^{\,2}}\,\biggr) \cdot \Delta t^{\,2} \,.
\end{align*}
If $\tau \, \ll \, 1\,$, then the scheme is second-order accurate in time $\mathcal{O}\bigl(\, \Delta t^{\,2}\,\bigr)\,$. If the condition $\tau \, \ll \, 1$ is not respected, then the scheme is not consistent with the discretized equation. For practical applications, in the case $\Delta x \, \approx \, \Delta y$, then the second-order accuracy is obtained when $\Delta t \egal \mathcal{O}\bigl(\, \Delta x^{\,2} \,\bigr)\,$.

\subsection{Metrics of efficiency and reliability of a model}

To evaluate the efficiency of a numerical model, one criterion is the computational (CPU) run time required to compute the solution. It is measured using the \texttt{Matlab\texttrademark} environment with a computer equipped with \texttt{Intel} i$7$ CPU ($2.7 \ \mathsf{GHz}$ $6^{\,th}$ generation) and $32$ GB of RAM. Hence the following ratio is defined: 
\begin{align*}
R_{\,\cpu} \ \eqdef \ \frac{t_{\,\mathrm{cpu}}}{\tref} \,,
\end{align*}
where $t_{\,\cpu} \ \unit{s}$ is the measured CPU time and $\tref$ is a chosen reference time. 

The accuracy of the numerical model is assessed by comparing the results to a reference solution denoted by the superscript $\mathrm{ref}$. The error can be applied to a certain time varying output $\Phi\,$, as for instance the temperature, the flux or the thermal loads, that may depend on the space coordinates $x$ and $y$. Then, the error for this output is defined by the compound function:
\begin{align*}
\Bigl(\, \varepsilon_{\,2} \, \circ \, \Phi \,\Bigr) \,(\,t\,) & \eqdef \ 
\Biggl(\, \frac{1}{L \ H} 
\int_{\,\Omega_{\,x}} \ 
\int_{\,\Omega_{\,y}} \ 
\Bigl(\,
\Phi(\,x \,,\, y \,, t \,) \moins \Phi^{\mathrm{\, ref}}(\,x \,,\, y \,, t \,) 
\,\Bigr)^{\,2}
\mathrm{d}x \
\mathrm{d}y
\,\Biggr)^{\,\half}\,.
\end{align*}
A normalized version of the error is also used:
\begin{align*}
\displaystyle
\Bigl(\, \varepsilon_{\,2} \, \circ \, \Phi \,\Bigr) \,(\,t\,) & \eqdef \ 
\Biggl(\, \frac{1}{L \ H} 
\int_{\,\Omega_{\,x}} \ 
\int_{\,\Omega_{\,y}} \ 
\biggl(\,
\frac{\Phi(\,x \,,\, y \,, t \,) \moins \Phi^{\mathrm{\, ref}}(\,x \,,\, y \,, t \,) }
{\displaystyle \max_{\,x\,y} \Phi^{\mathrm{\, ref}}(\,x \,,\, y \,, t \,) \moins \min_{\,x\,y} \Phi^{\mathrm{\, ref}}(\,x \,,\, y \,, t \,)}
\,\biggr)^{\,2}
\mathrm{d}x \
\mathrm{d}y
\,\Biggr)^{\,\half}\,.
\end{align*}
For investigations of the physical phenomena, the relative error can also be relevant:
\begin{align*}
\Bigl(\, \varepsilon_{\,r} \, \circ \, \Phi \,\Bigr) \,(\,t\,) & \eqdef \ 
\Biggl(\, \frac{1}{L \ H} 
\int_{\,\Omega_{\,x}} \ 
\int_{\,\Omega_{\,y}} \ 
\biggl(\,\frac{
\Phi(\,x \,,\, y \,, t \,) \moins \Phi^{\mathrm{\, ref}}(\,x \,,\, y \,, t \,) }
{\Phi^{\mathrm{\, ref}}(\,x \,,\, y \,, t \,)}
\,\biggr)^{\,2}
\mathrm{d}x \
\mathrm{d}y
\,\Biggr)^{\,\half}\,.
\end{align*}

\section{Validation of the numerical model}
\label{sec:validation_case}

\subsection{Case study}

To validate the implementation and verify the theoretical features, the model results are compared with an analytical solution from the \texttt{EXACT}\footnote{http://exact.unl.edu/exact/home/home.php} toolbox, namely X$33$B$00$Y$33$B$00$Tx$5$y$5$ \cite{Cole_2014}. Since the objective of this section is the validation, the problem is described in dimensionless formulation. The domain is defined for $(\,x^{\,\star} \,,\, y^{\,\star} \,) \, \in \, \bigl[\,0\,,\,1\,\bigr] \times \bigl[\,0\,,\,1\,\bigr]\,$. The initial condition is piece-wise defined on the sub-domain $\Omega_{\,ab} \, \eqdef \, \bigl[\,0 \,,\, L_{\,a}^{\,\star} \,\bigr] \, \times \,  \bigl[\,0 \,,\, L_{\,b}^{\,\star} \,\bigr] $:
\begin{align*}
u_{\,0}(\,x^{\,\star}\,,\,y^{\,\star}\,) \egal 
\begin{cases}
& 1 \,, \qquad (\,x^{\,\star}\,,\,y^{\,\star}\,) \, \in \, \Omega_{\,ab} \,,\\[4pt]
& 0 \,, \qquad (\,x^{\,\star}\,,\,y^{\,\star}\,) \, \notin \, \Omega_{\,ab} \,,
\end{cases} 
\end{align*}
where $L_{\,a}^{\,\star} \egal 0.6$ and $L_{\,b}^{\,\star} \egal 0.5\,$. The \textsc{Fourier} numbers are set to unity $\Fo_{\,x} \egal \Fo_{\,y} \egal 1\,$. The plate is composed of one material so the distortion coefficients are equal to the unity $k^{\,\star} \egal c^{\,\star} \egal 1\,$. At the boundaries, the imposed \textsc{Robin} condition is homogeneous, so $u_{\,\infty\,,\,i} \egal q_{\,\infty\,,\,i} \egal 0 \,, \forall \quad  i \, \in \, \bigl\{\, 1 \,,\, \ldots \,,\, 4 \,\bigr\}\,$. The \textsc{Biot} numbers are equal to:
\begin{align*}
\Bi_{\,1} \egal 3 \,, \quad 
\Bi_{\,2} \egal 0.5 \,, \quad 
\Bi_{\,3} \egal 1.5 \,, \quad 
\Bi_{\,4} \egal 4 \,.
\end{align*}
The time horizon is $\tf^{\,\star} \egal 0.04 \,$.

\subsection{Results}

\begin{figure}
\centering
\subfigure[\label{fig:u_fy}]{\includegraphics[width=.45\textwidth]{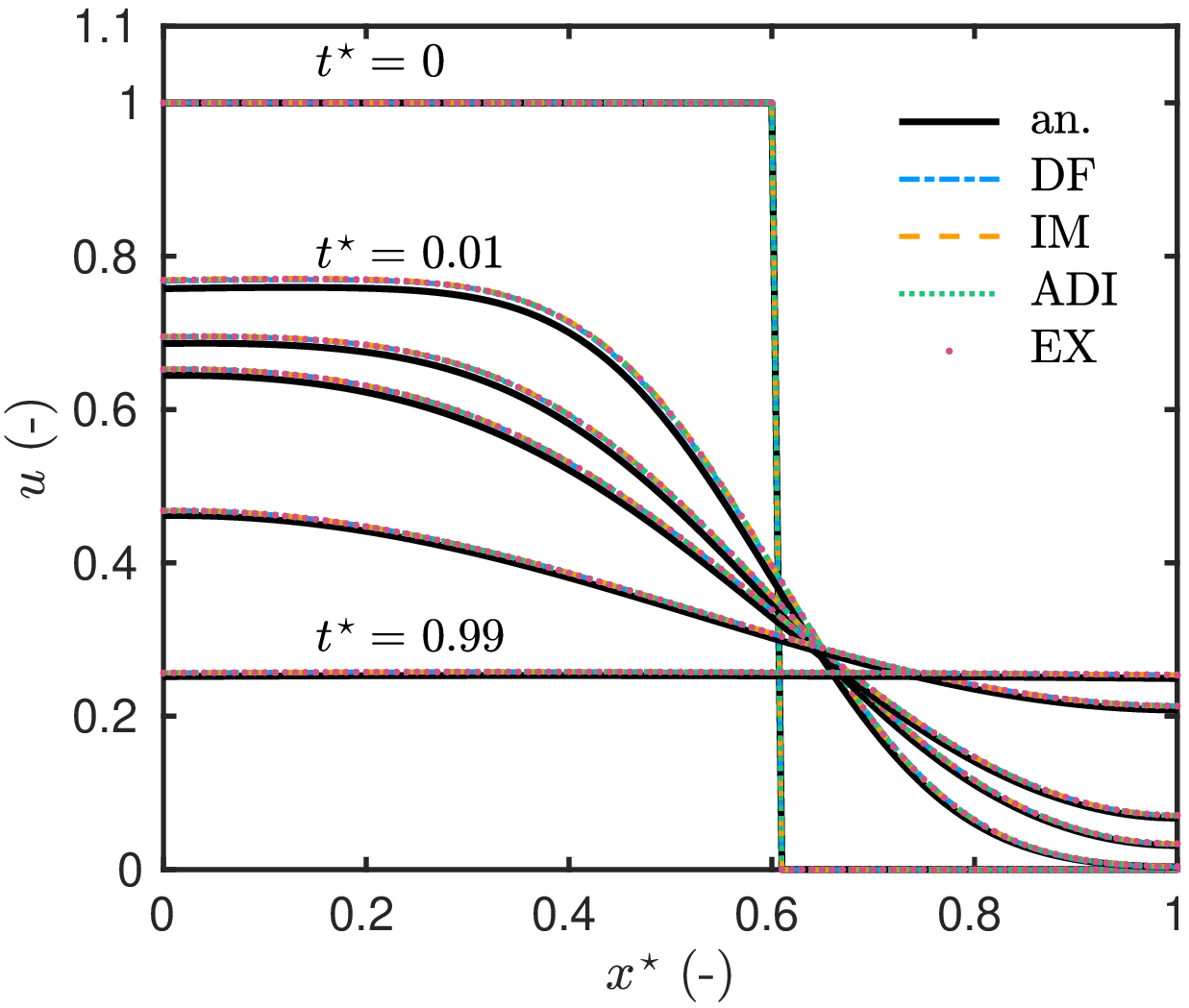}}  \hspace{0.2cm}
\subfigure[\label{fig:u_fx}]{\includegraphics[width=.45\textwidth]{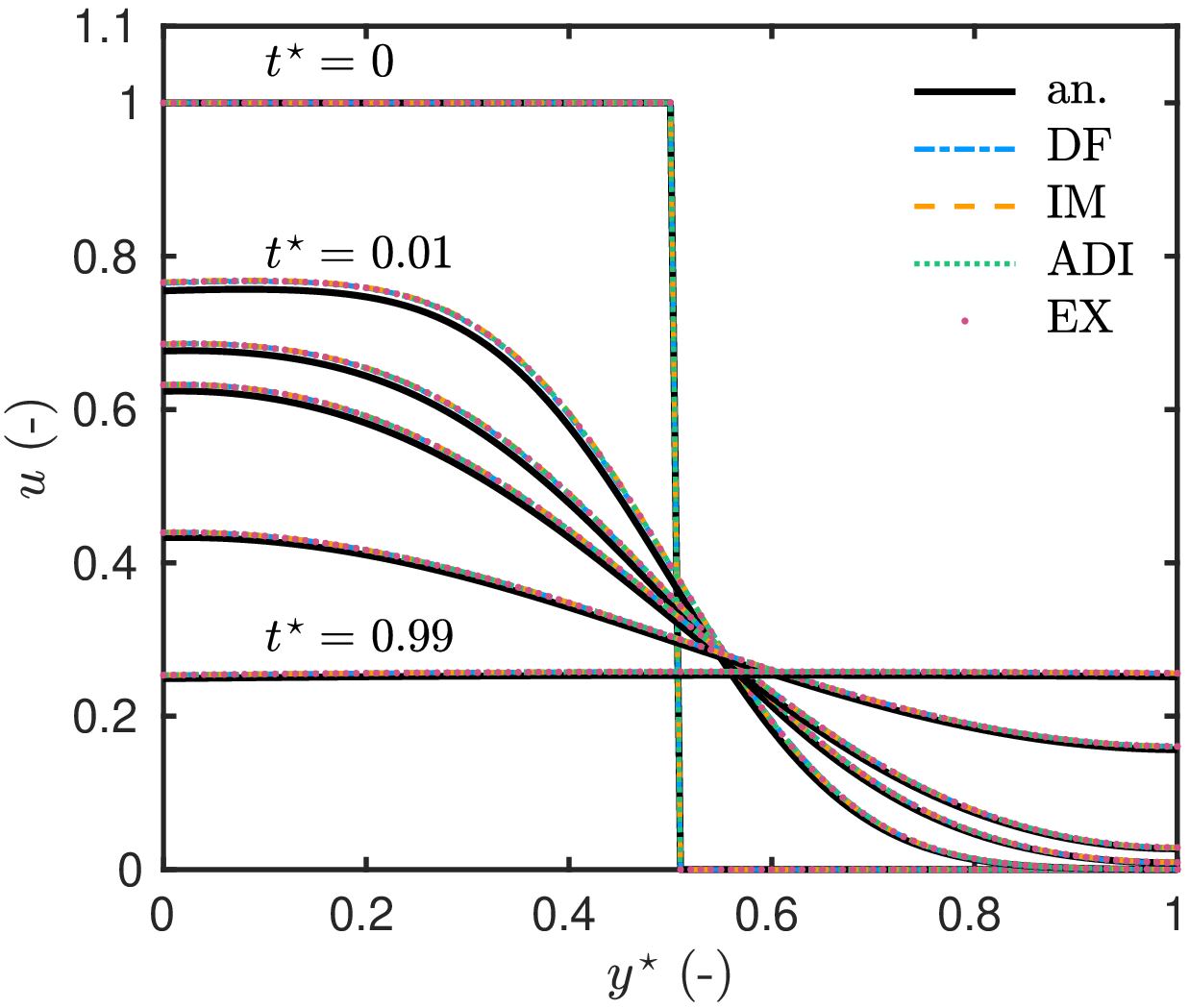}} \\
\subfigure[\label{fig:u_ft}]{\includegraphics[width=.45\textwidth]{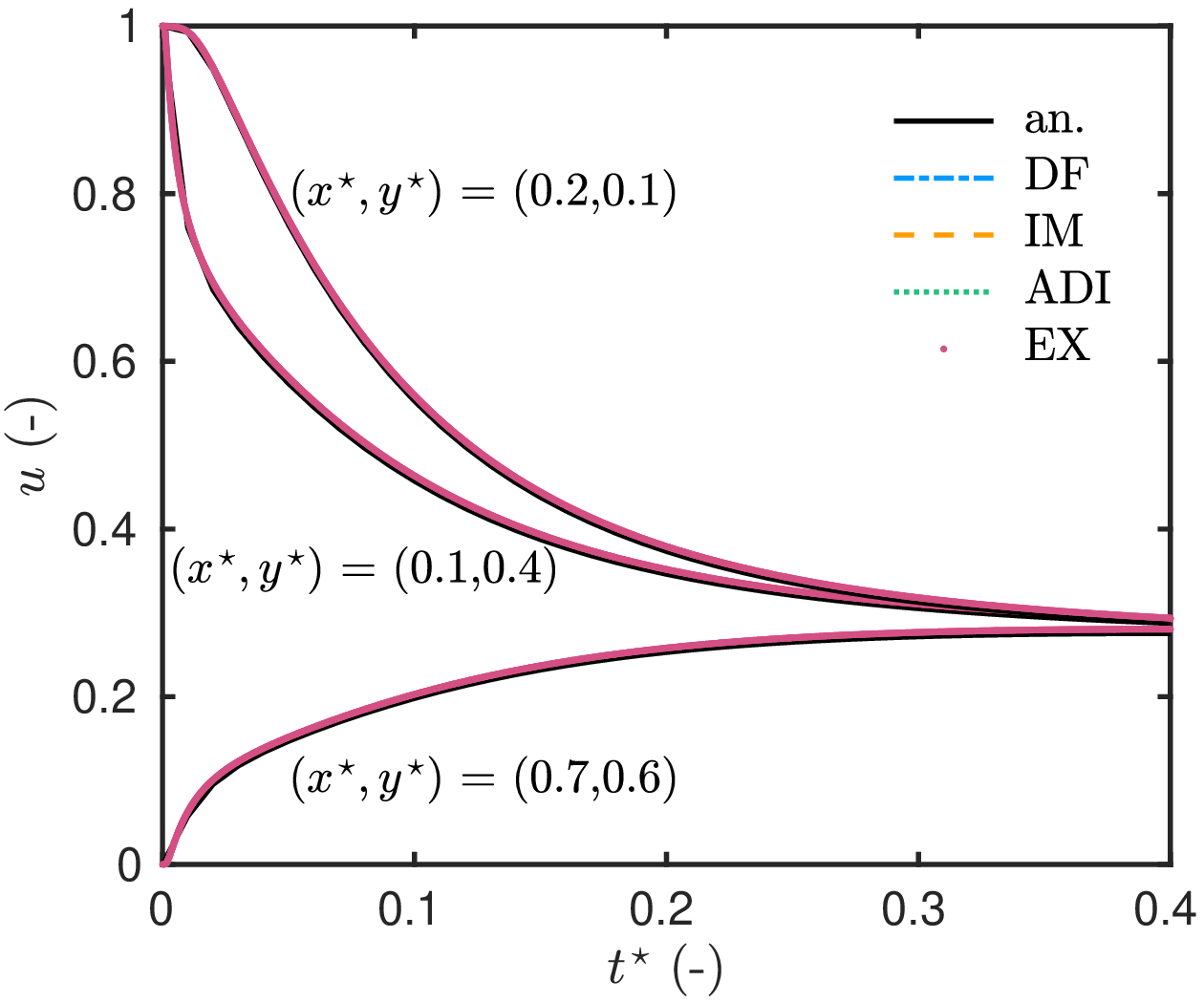}}  \hspace{0.2cm}
\subfigure[\label{fig:u_fxy}]{\includegraphics[width=.45\textwidth]{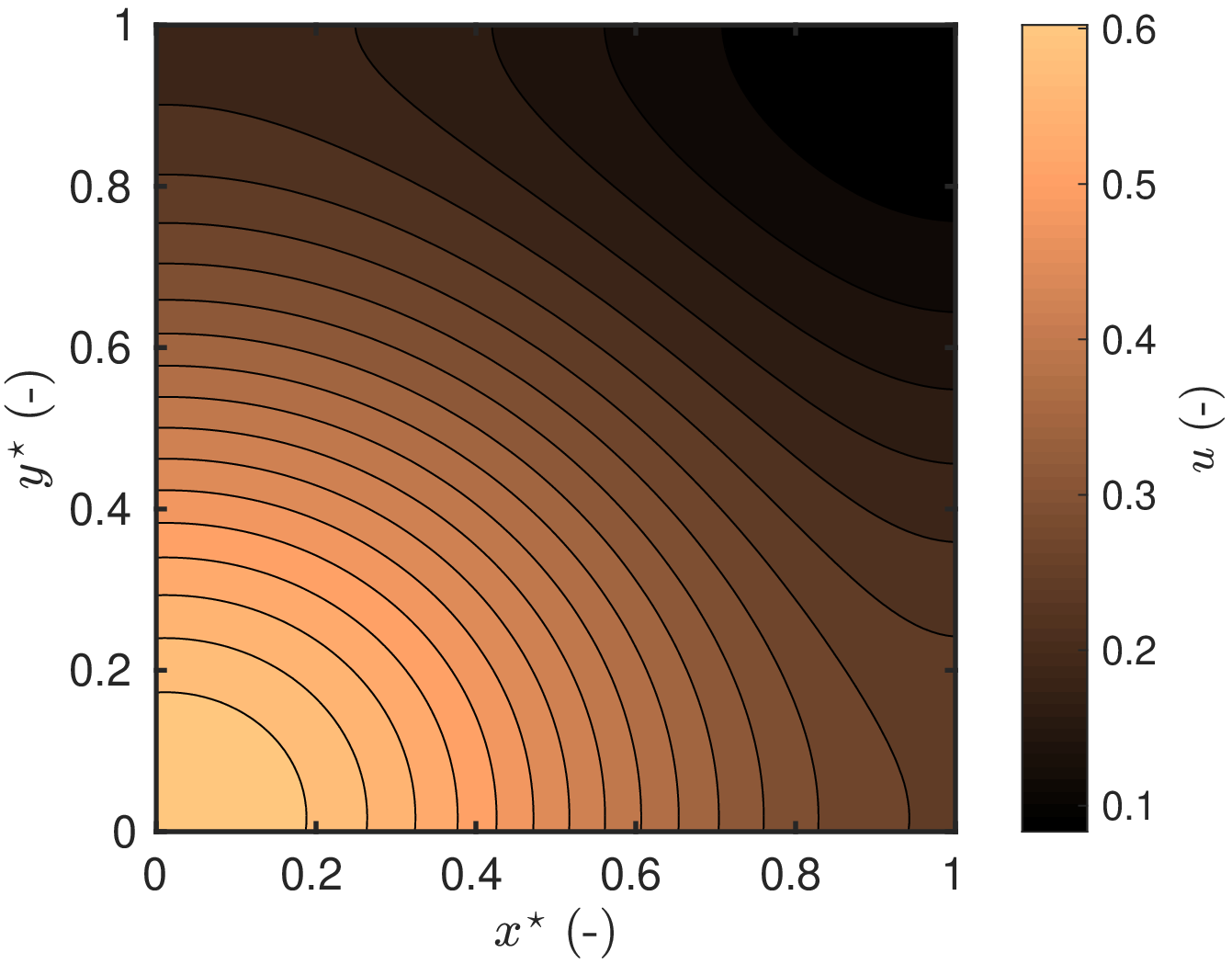}} 
\caption{Variation of the field according to $x^{\,\star}$ at $y^{\,\star} \egal 0.5$ \emph{(a)}, according to $y^{\,\star}$ at $x^{\,\star} \egal 0.4$  \emph{(b)} and according to $t^{\,\star}$ \emph{(c)}. Slice of the solution from the \DF ~model at $t^{\,\star} \egal 0.04$ (\emph{d}).}
\end{figure}

The solution is computed using four numerical models, namely the \DF ~(denoted DF), the implicit \textsc{Euler} (denoted IM), the explicit \textsc{Euler} (denoted EX) and the \ADI ~(denoted ADI). The second and third models use central finite difference approaches for the space discretisation. The ADI method is described in \cite{Peaceman_1955} with details in Appendix~\ref{sec:ADI}. First, all models except \Eu ~explicit considers the same space and time discretisations $\Delta t^{\,\star} \egal 10^{\,-4}$ and $\Delta x^{\,\star} \egal \Delta y^{\,\star} \egal 10^{\,-2}\,$. For the \Eu ~explicit approach, the discretization parameters are required to satisfy the following stability condition:
\begin{align}
\label{eq:cfl_condition}
\displaystyle \Delta t^{\,\star} \, \leqslant \, \frac{\bigl(\,\Delta x^{\,\star} \ \Delta y^{\,\star} \,\bigr)^{\,2}}{2 \ \Fo \ \Bigl(\, \Delta \bigl(\,x^{\,\star}\,\bigr)^{\,2} \plus \Delta \bigl(\,y^{\,\star}\,\bigr)^{\,2} \,\Bigr)}\,,
\end{align}
which corresponds to $\Delta t^{\,\star} \, \leqslant \, 2.5 \e{-5} $ for this case study. Thus, a smaller time step is used $\Delta t^{\,\star} \egal 10^{\,-5}$ for this model.  Figures~\ref{fig:u_fy} to \ref{fig:u_ft} compare the solutions. All of them are overlapped, highlighting the validation of the numerical models compared to the analytical solution. Figure~\ref{fig:u_fxy} enhances the two-dimensional aspect of the heat transfer through the domain. Table~\ref{tab:res_eps_cpu} provides a synthesis of the efficiency of the numerical models. All models have a satisfying error with the same order of accuracy $\varepsilon_{\,2} \egal \mathcal{O}(\,10^{\,-3}\,)\,$. It also justifies why the solutions are overlapped in Figures~\ref{fig:u_fy} to \ref{fig:u_ft}. However, the computational time ratio is very different among the models. The \Eu ~explicit requires only $10\%$ of the computational time of the \Eu ~implicit, even for a time step one order lower. However, this model is not reliable for predicting the phenomena in building materials due to its conditional stability Eq.~\eqref{eq:cfl_condition}. The \ADI ~and \DF ~approaches require only $5\%$ and $0.7\%$ of the implicit computational time. The important differences are due to the computational efforts to inverse the matrix in the implicit method. It represents $99.5\%$ of the total computational time. Note that the problem is linear in parameters. The differences in computational time should increase when considering nonlinear problems due to the requirement of subiterations to treat the nonlinearities.

\begin{figure}
\centering
\subfigure[\label{fig:eps_fdt}]{\includegraphics[width=.9\textwidth]{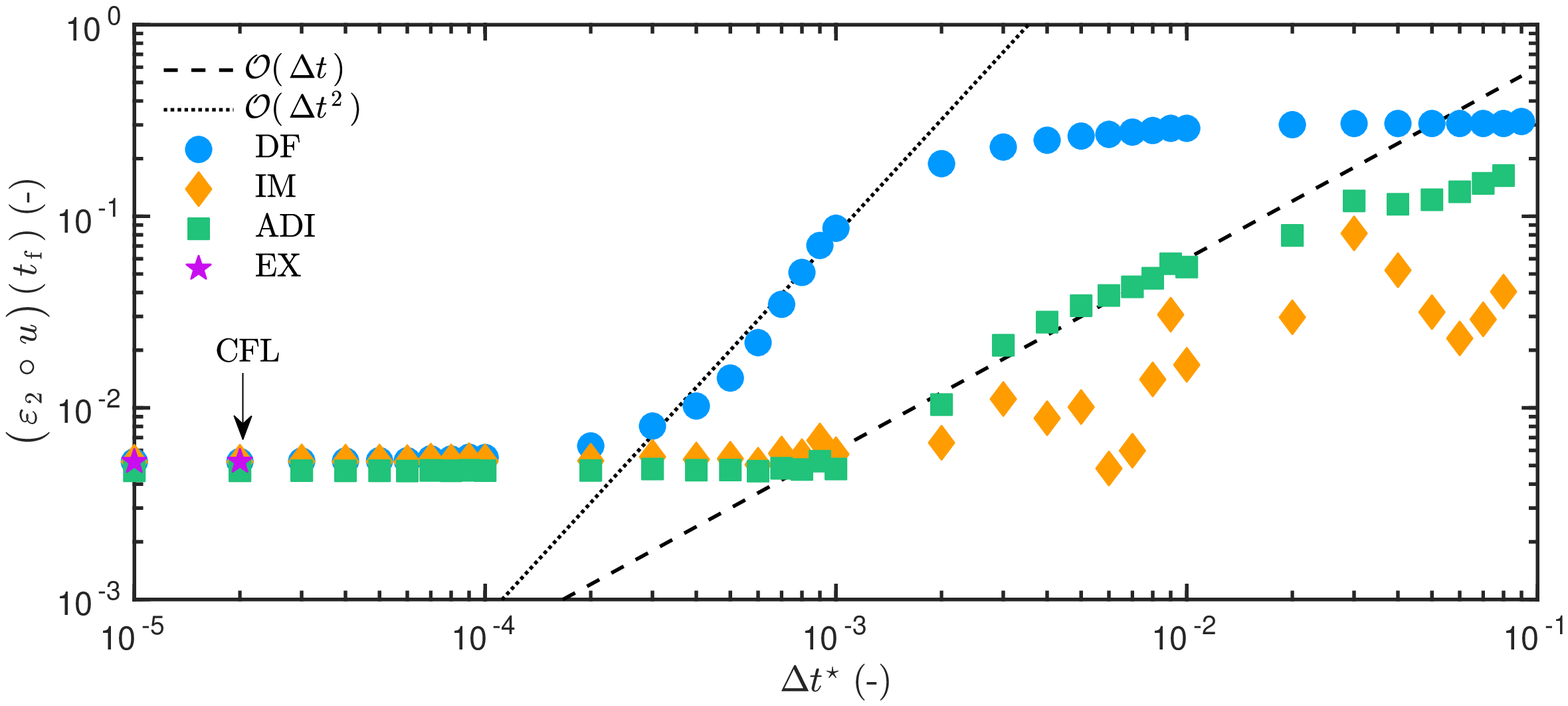}} \\
\subfigure[\label{fig:cpu_feps}]{\includegraphics[width=.9\textwidth]{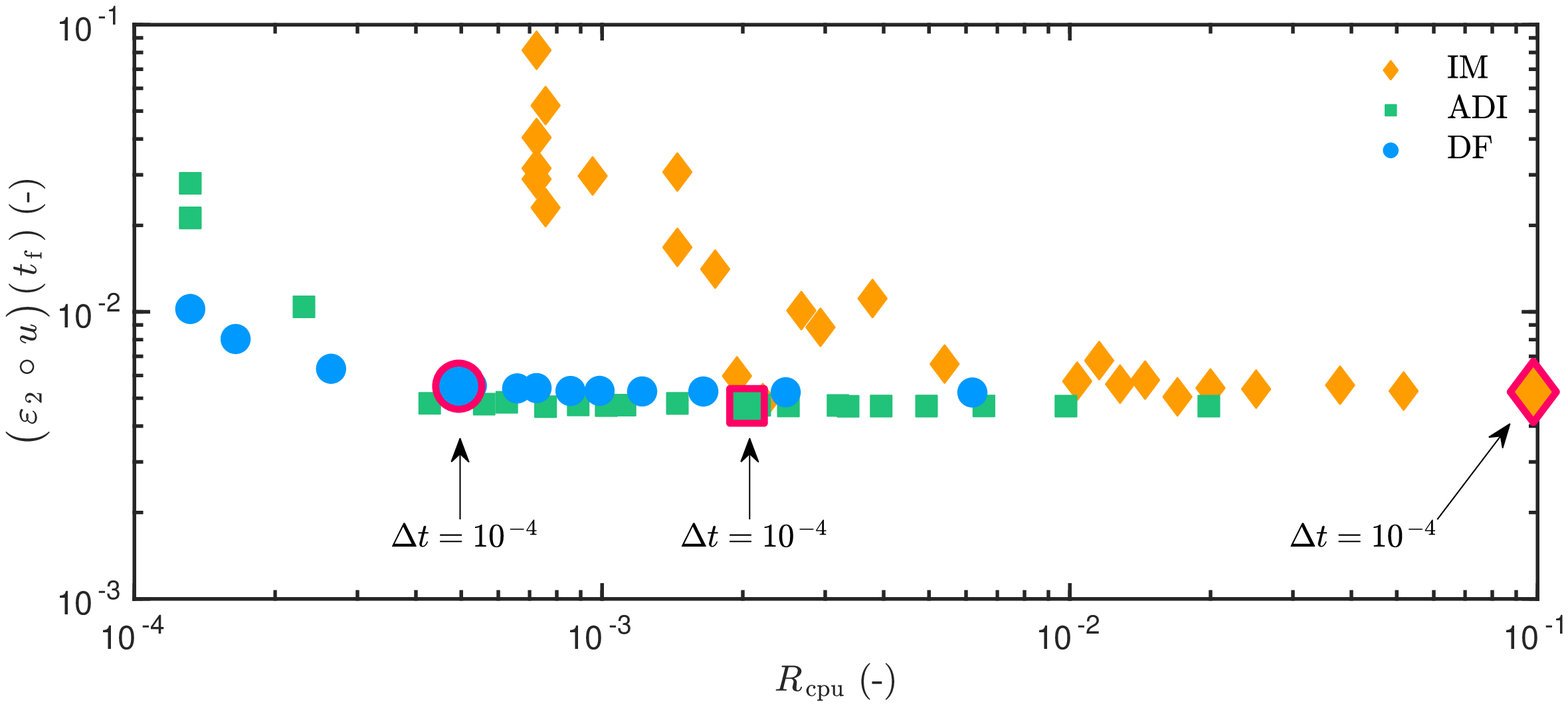}} 
\caption{Variation  of the numerical models errors according to the time discretisation \emph{(a)}. Variation of the numerical models errors according to the computational effort (\emph{b}). For the last one, the reference computational time is $t_{\,0} \egal 294 \ \mathsf{s}\,$, the one of the \Eu ~implicit model.}
\end{figure}

Further investigations are carried out by setting the space mesh to $\Delta x^{\,\star} \egal \Delta y^{\,\star} \egal 10^{\,-2}\,$ and performing computations for several values of time discretisation $\Delta t^{\,\star}\,$. For each computation, the error with the analytical solution and the computational time of the four numerical models are evaluated. Figure~\ref{fig:eps_fdt} shows the variation of the error according to the time discretisation. Several theoretical results can be confirmed. First, the \Eu ~explicit scheme enables to compute the solution only until the CFL restriction $\Delta t^{\,\star} \, \leqslant \, 2.5 \e{-5}\,$. Then, it can be remarked that the \DF, the \Eu ~implicit and the \ADI ~approaches are unconditionally stable, as proven theoretically in Section~\ref{sec:DF_stability} for the primer. However, some differences are observed between those models. The \DF ~scheme is second-order accurate in time while the two others are only first order. Figure~\ref{fig:cpu_feps} gives the variation of the accuracy with the computational ratio. The \DF ~model is always faster than the other approaches. For $\Delta t^{\,\star} \egal 10^{\,-4}\,$, it can be remarked that the \DF ~approach is as accurate as the others. However, it computes more than a thousand times faster than the \Eu ~implicit model. 

\begin{table}
\centering
\caption{Efficiency of the models for the validation case.}
\label{tab:res_eps_cpu}
\setlength{\extrarowheight}{.5em}
\begin{tabular}[l]{@{} c cc cc}
\hline
\hline
\multirow{2}{*}{\textit{Model}}
& \textit{Time step} 
& \textit{Space mesh} 
& \textit{Error}  
& \textit{Computational time}$^{\,\dagger}$ \\

& $\Delta t^{\,\star}$
& $\Delta x^{\,\star} \egal \Delta y^{\,\star}$
& $\bigl(\,\varepsilon_{\,2} \, \circ\, u \,\bigr) \,(\,\tf^{\,\star}\,)$
& $R_{\,\cpu}$ \\
\Eu ~implicit
& $10^{\,-4}$
& $10^{\,-2}$
& $4.63 \e{-3}$
& $1$ \\
\Eu ~explicit
& $10^{\,-5}$
& $10^{\,-2}$
& $4.63 \e{-3}$
& $0.11$ \\
\ADI
& $10^{\,-4}$
& $10^{\,-2}$
& $4.63 \e{-3}$
& $0.05$ \\
\DF
& $10^{\,-4}$
& $10^{\,-2}$
& $4.64 \e{-3}$
& $0.007$ \\
\hline
\hline
\multicolumn{5}{l}{$^{\,\dagger}$ the reference computational time is $t_{\,0} \egal 773 \ \mathsf{s}\,$,the one of the \Eu ~implicit model.} 
\end{tabular}
\end{table}

\newpage

\section{Real case study}
\label{sec:real_case_study}

\subsection{Description}
\label{sec:description_case_study}
 
\begin{figure}
\centering
\subfigure[\label{fig:vBC_ft}]{\includegraphics[width=.45\textwidth]{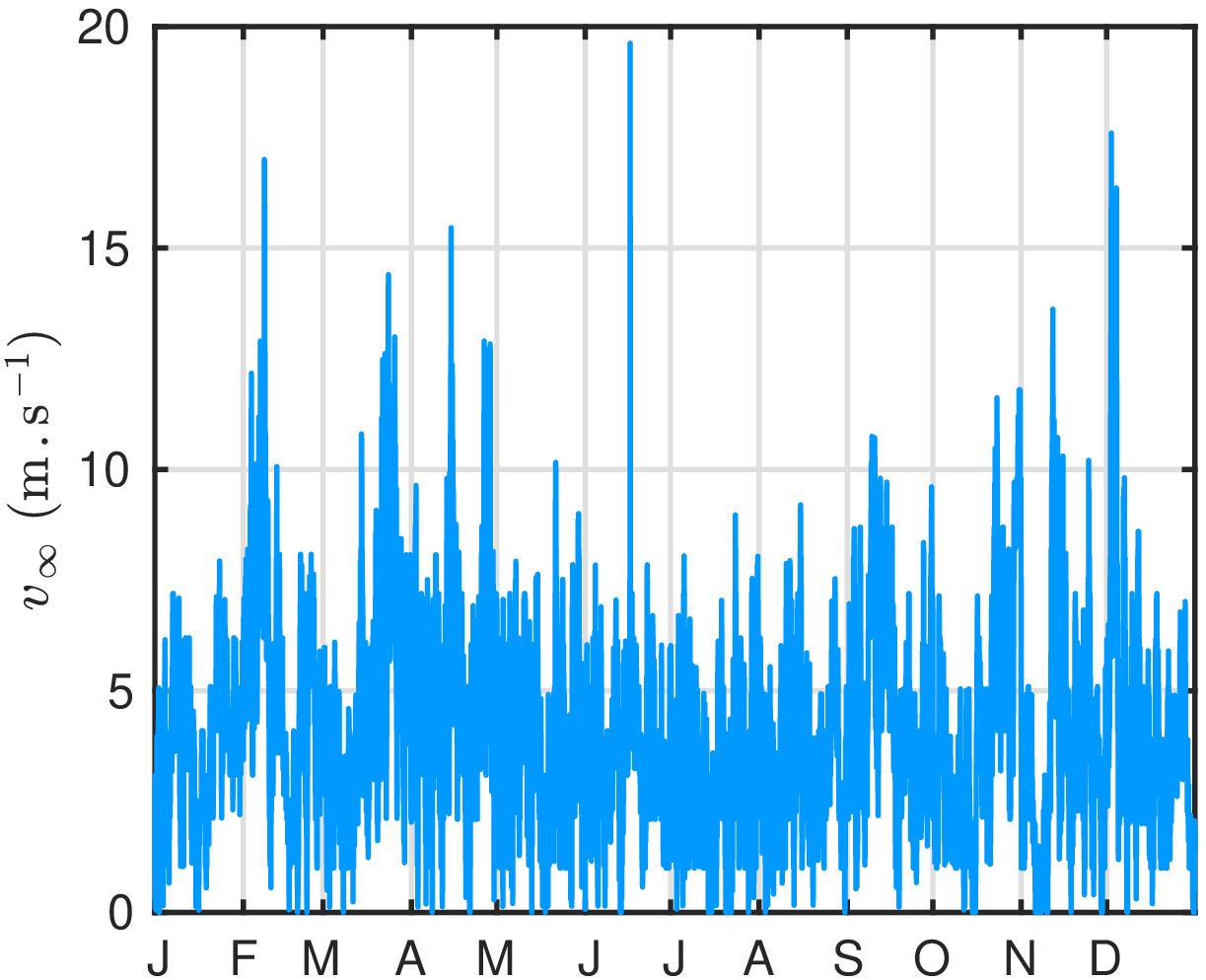}}  \hspace{0.2cm}
\subfigure[\label{fig:pdf_h1}]{\includegraphics[width=.44\textwidth]{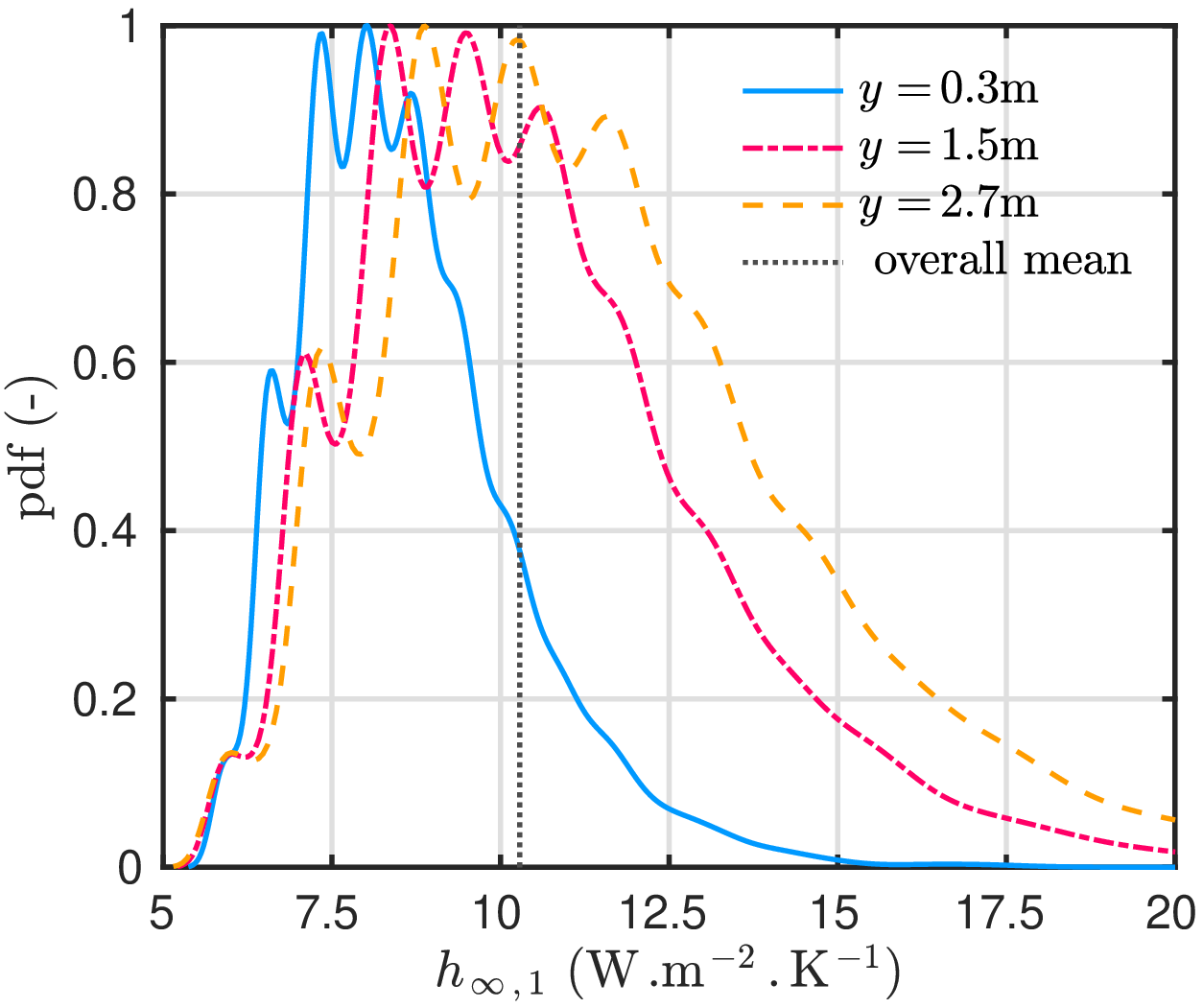}} \\
\subfigure[\label{fig:hy1_ft}]{\includegraphics[width=.45\textwidth]{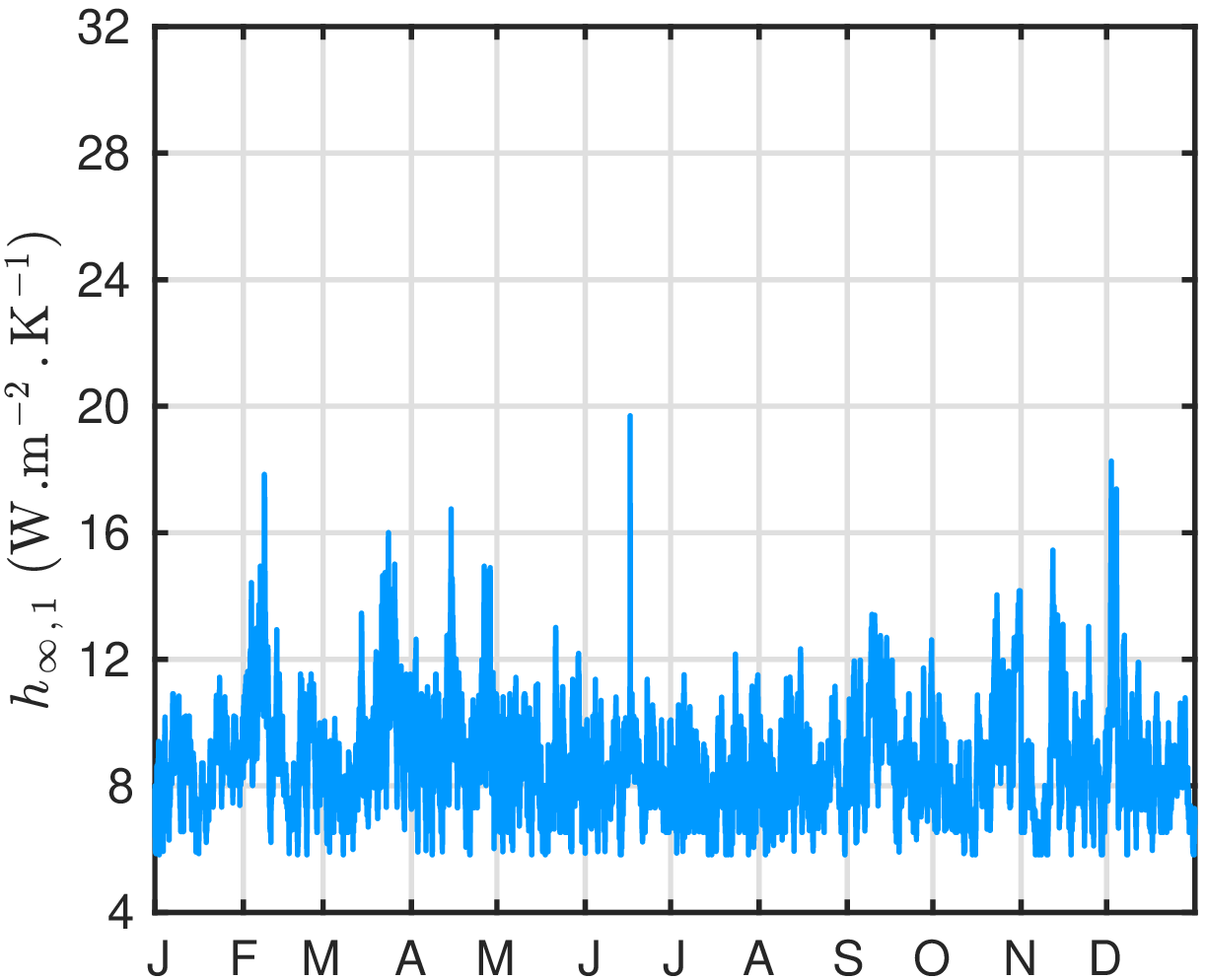}}  \hspace{0.2cm}
\subfigure[\label{fig:hy3_ft}]{\includegraphics[width=.45\textwidth]{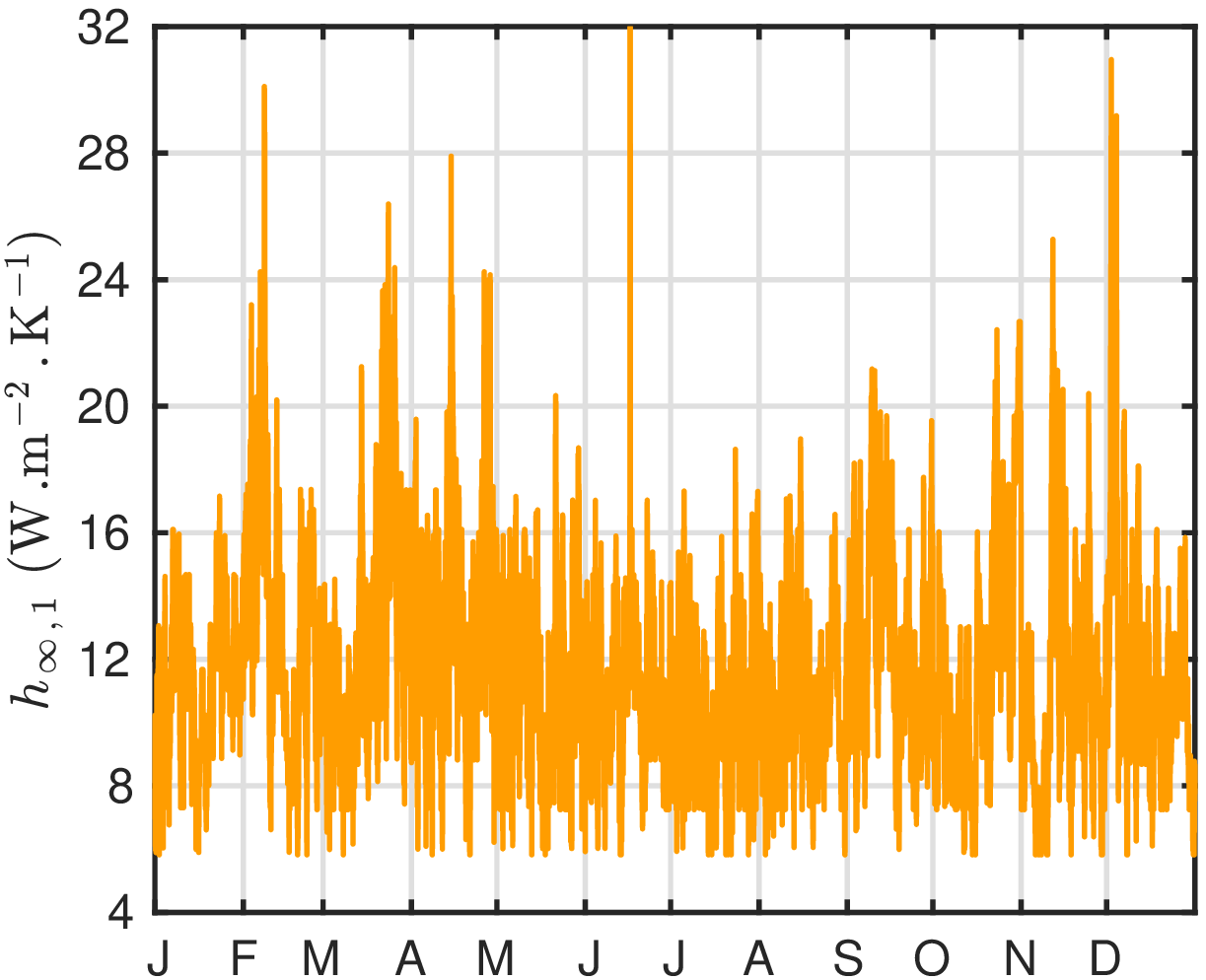}} 
\caption{Variation of the outside velocity \emph{(a)}. Probability density function of the outside surface heat transfer coefficient $h_{\,1}$ \emph{(b)}. Variation of the outside surface heat transfer coefficient $h_{\,1}$ at $y \egal 0.3 \ \mathsf{m}$  \emph{(c)} and $y \egal 2.7 \ \mathsf{m}$  \emph{(d)}.}
\end{figure}

The case study considers a south-oriented facade of a house located in Paris, France. The wall is composed of three layers: concrete (outside part), wood fiber insulation and gypsum board (inside part). The material properties are taken from the French standards \cite{RT_2010} and given in Table~\ref{tab:mat_prop}.  This configuration corresponds to a building with improved energy efficiency. As illustrated in Figure~\ref{fig:domain}, the height  and width of the wall are $H \egal 3 \ \mathsf{m}$ and $L \egal 37 \ \mathsf{cm}\,$. The facade is located in an urban area so it is facing other buildings. The latter is located at a distance $D \egal 5 \ \mathsf{m}$ and has a height $F \egal 3 \ \mathsf{m}\,$, which induces a shadow on the studied facade. The height of the shadow varies according to time. As a consequence, the outside incident radiation flux $q_{\,\infty\,,\,1}$ varies according to height and time. The diffusivity is set as $\alpha \egal 0.6 \,$. The outside surface heat transfer coefficient $h_{\,1}$ depends on height position $y$ and time varying climate wind velocity as defined in Eq.~\eqref{eq:h1}. The following parameters are used $h_{\,10} \egal 5.82 \ \mathsf{W\,.\,m^{\,-2}\,.\,K^{\,-1}}\,$, $h_{\,11} \egal 3.96 \ \mathsf{W\,.\,m^{\,-2}\,.\,K^{\,-1}}\,$, $v_{\,0} \egal 1 \ \mathsf{m\,.\,s^{\,-1}}\,$,  $y_{\,0} \egal 1 \ \mathsf{m}$ and $\beta \egal 0.32\,$. The outside wind velocity is shown in Figure~\ref{fig:vBC_ft}. It varies around a mean of $4 \ \mathsf{m\,.\,s^{\,-1}}\,$. The occurrences of the surface heat transfer coefficient $h_{\,\infty\,,\,1}$ are shown in Figure~\ref{fig:pdf_h1}. The surface heat transfer coefficient increases according to the height, with a mean around $8.6 \ \mathsf{W\,.\,m^{\,-2}\,.\,K^{\,-1}}$ at $y \egal 0.3 \ \mathsf{m}\,$, $10.5 \ \mathsf{W\,.\,m^{\,-2}\,.\,K^{\,-1}}$ at $y \egal 1.5 \ \mathsf{m}$ and $11.5 \ \mathsf{W\,.\,m^{\,-2}\,.\,K^{\,-1}}$ at $y \egal 2.7 \ \mathsf{m}\,$. The mean over the whole year and height gives a coefficient of $\overline{h}_{\,\infty\,,\,1} \egal 10.28 \ \mathsf{W\,.\,m^{\,-2}\,.\,K^{\,-1}}\,$. Figures~\ref{fig:hy1_ft} and \ref{fig:hy3_ft} enable to compare the time variation of the coefficient between the bottom and the top of the facade. Higher magnitudes of variation are observed at the top of the facade. Moreover, the discrepancy with the value of $12\ \mathsf{W\,.\,m^{\,-2}\,.\,K^{\,-1}} $ used for standard computation is locally important.

\begin{figure}
\centering
\includegraphics[width=.95\textwidth]{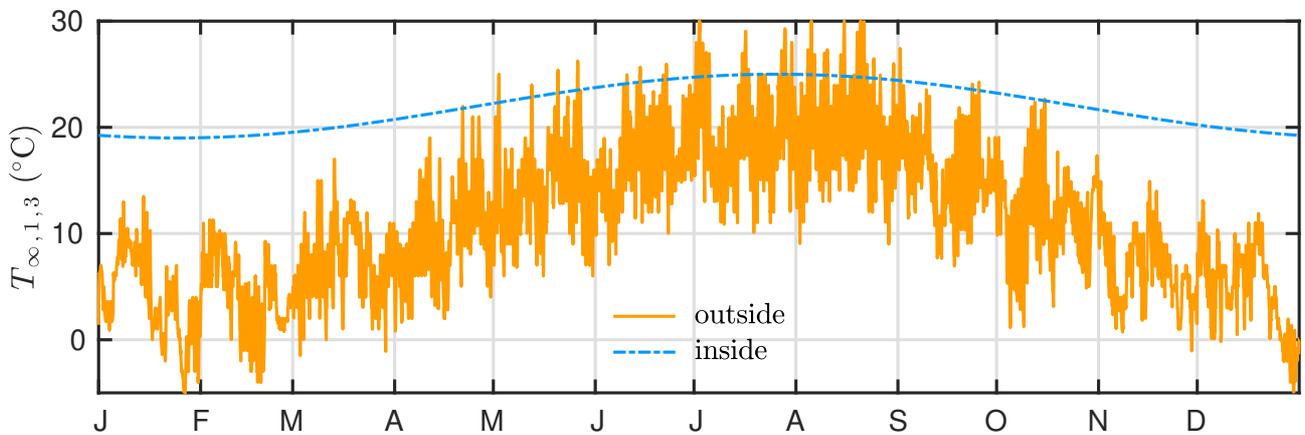}
\caption{Variation of the inside and outside temperatures.}
\label{fig:TBC_ft}
\end{figure}

The outside temperature is also given by weather data file. The inside temperature is controlled and defined according to sinusoidal variations depending on the winter and summer seasons. The time variation of inside and outside temperature is shown in Figure~\ref{fig:TBC_ft}. The inside surface transfer coefficient is set as constant to $h_{\,3} \egal 10 \ \mathsf{W\,.\,m^{\,-2}\,.\,K^{\,-1}}\,$. The top and bottom boundaries of the facade $\Gamma_{\,2}$ and $\Gamma_{\,4}$ are set as adiabatic. Indeed, the investigations focus on the influence of the space and time variations of outside boundary conditions on the thermal efficiency of the facade. The simulation horizon is of one year so $\tf \egal 365 \ \mathsf{d}\,$.

\begin{table}
\centering
\caption{Thermal properties of the material composing the facade.}
\label{tab:mat_prop}
\setlength{\extrarowheight}{.5em}
\begin{tabular}[l]{@{} c ccc}
\hline
\hline
\multirow{2}{*}{\textit{Layer}}
& \textit{Thermal conductivity}
& \textit{Volumetric  heat capacity}
& \textit{length}  \\
&  $\unit{W\,.\,m^{\,-1}\,.\,K^{\,-1}}$ 
&  $\unit{MJ\,.\,m^{\,-3}\,.\,K^{\,-1}}$ 
& $\unit{m}$ \\
concrete
& $1.4$
& $2$ 
& $0.2$ \\
wood fiber
& $0.05$
& $0.85$ 
& $0.15$ \\
gypsum
& $0.25$
& $0.85$ 
& $0.02$ \\
\hline
\hline
\end{tabular}
\end{table}

\subsection{Generating the incident radiation flux}

\begin{figure}
\centering
\subfigure[\label{fig:scene_0512}]{\includegraphics[width=.95\textwidth]{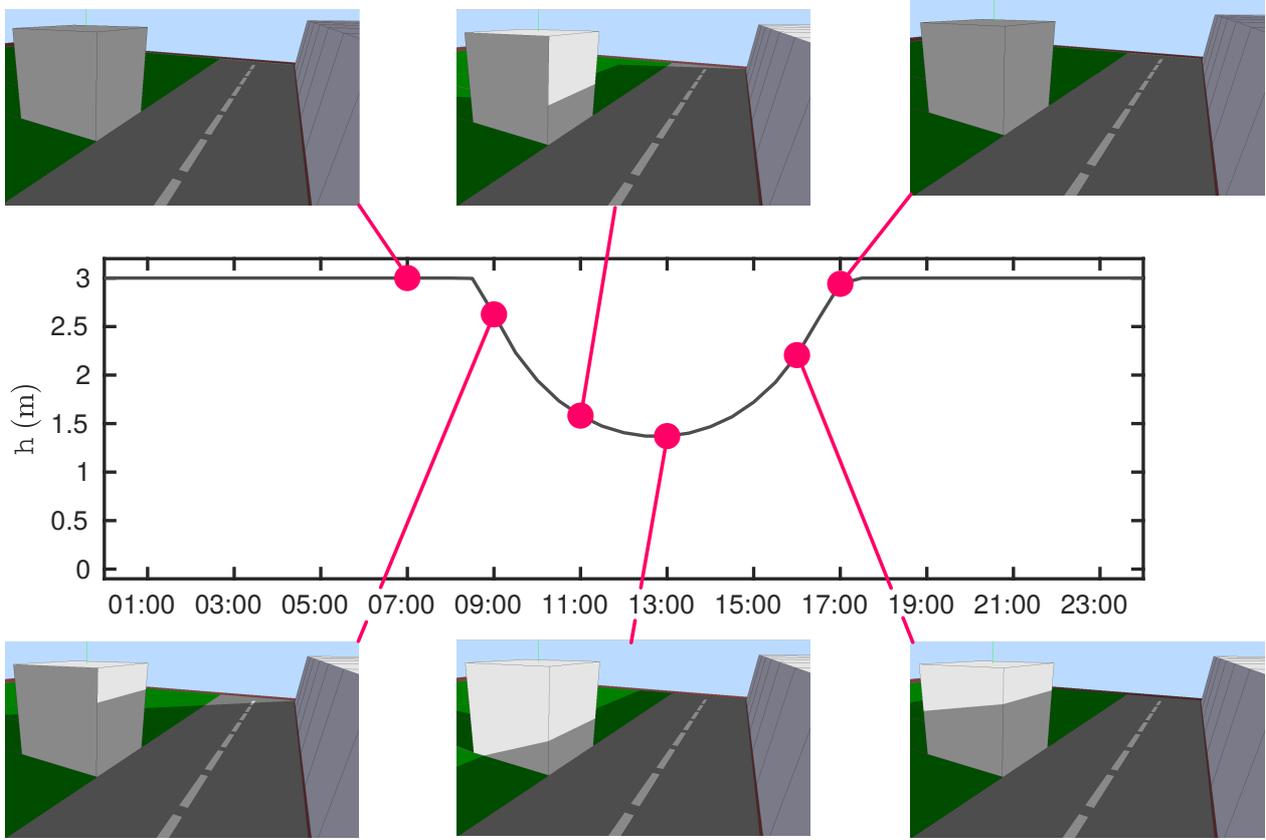}}   \\
\subfigure[\label{fig:chi_0512_1day}]{\includegraphics[width=.45\textwidth]{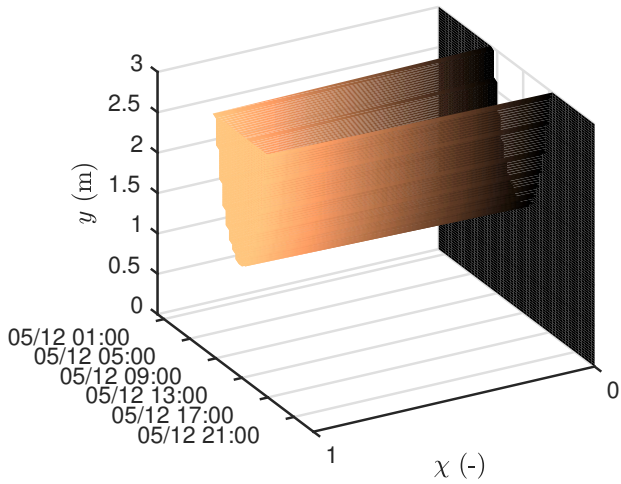}}  \hspace{0.2cm}
\subfigure[\label{fig:q_0512_1day}]{\includegraphics[width=.44\textwidth]{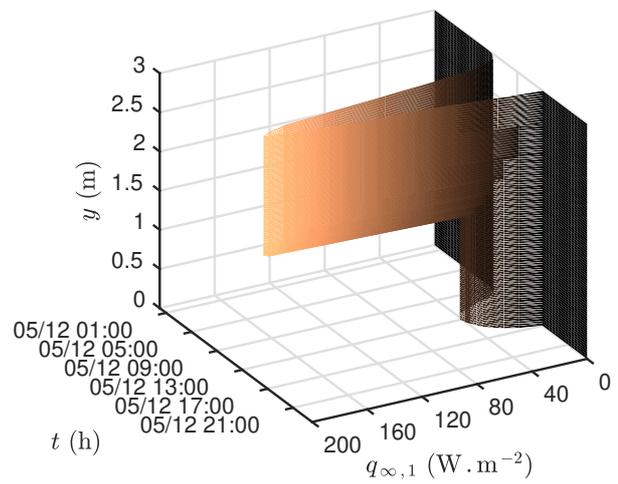}} 
\caption{Results of the pixel counting technique compared with the height of the shadow on the facade for December $5^{\,th}$  \emph{(a)}. Variation of the indicator function \emph{(b)} and the corresponding incident radiation heat flux \emph{(c)}}
\end{figure}

The pixel counting technique is employed to determine the outside incident radiation heat flux and the variation of the shadow height. A time step of $6 \ \mathsf{min}$ is used to provide the data. Figure~\ref{fig:scene_0512} compares the results of the pixel counting technique with the shadow height. From these results, the indicator function $\chi$ defined in Eq.~\eqref{eq:chi} can be computed as illustrated in Figure~\ref{fig:chi_0512_1day}. For this winter day, the bottom of the facade remains in the shadow. Around midday, the top of the facade receives the direct sun. During this time, the indicator function is equal to $1$ and thus this part of the facade receives the direct heat flux added to the reflected and diffuse ones. As remarked in Figure~\ref{fig:q_0512_1day}, the magnitude of the flux is three times higher on the top of facade around midday. 

\begin{figure}
\centering
\subfigure[\label{fig:hsummer_ft} summer]{\includegraphics[width=.45\textwidth]{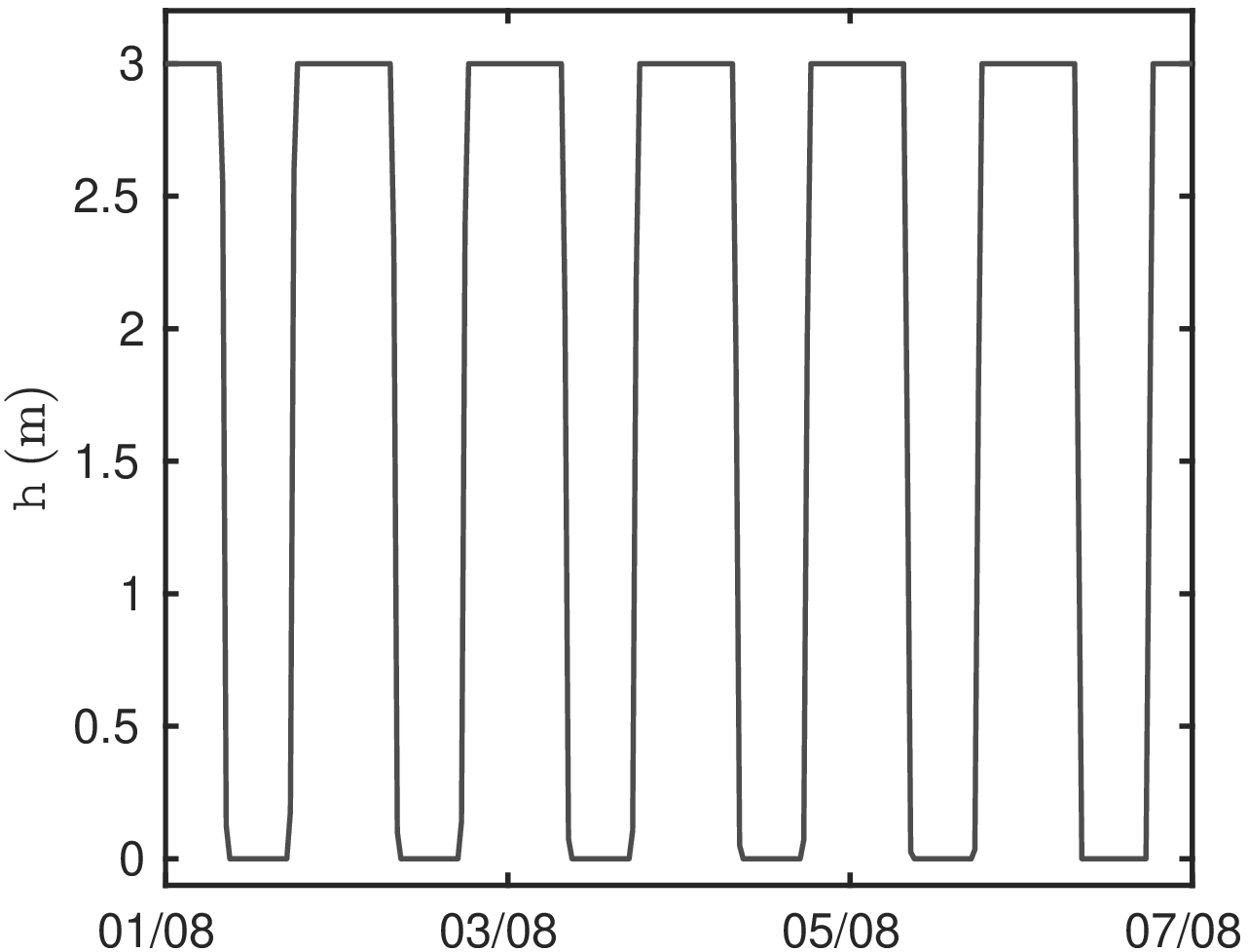}}  \hspace{0.2cm}
\subfigure[\label{fig:hwinter_ft} winter]{\includegraphics[width=.45\textwidth]{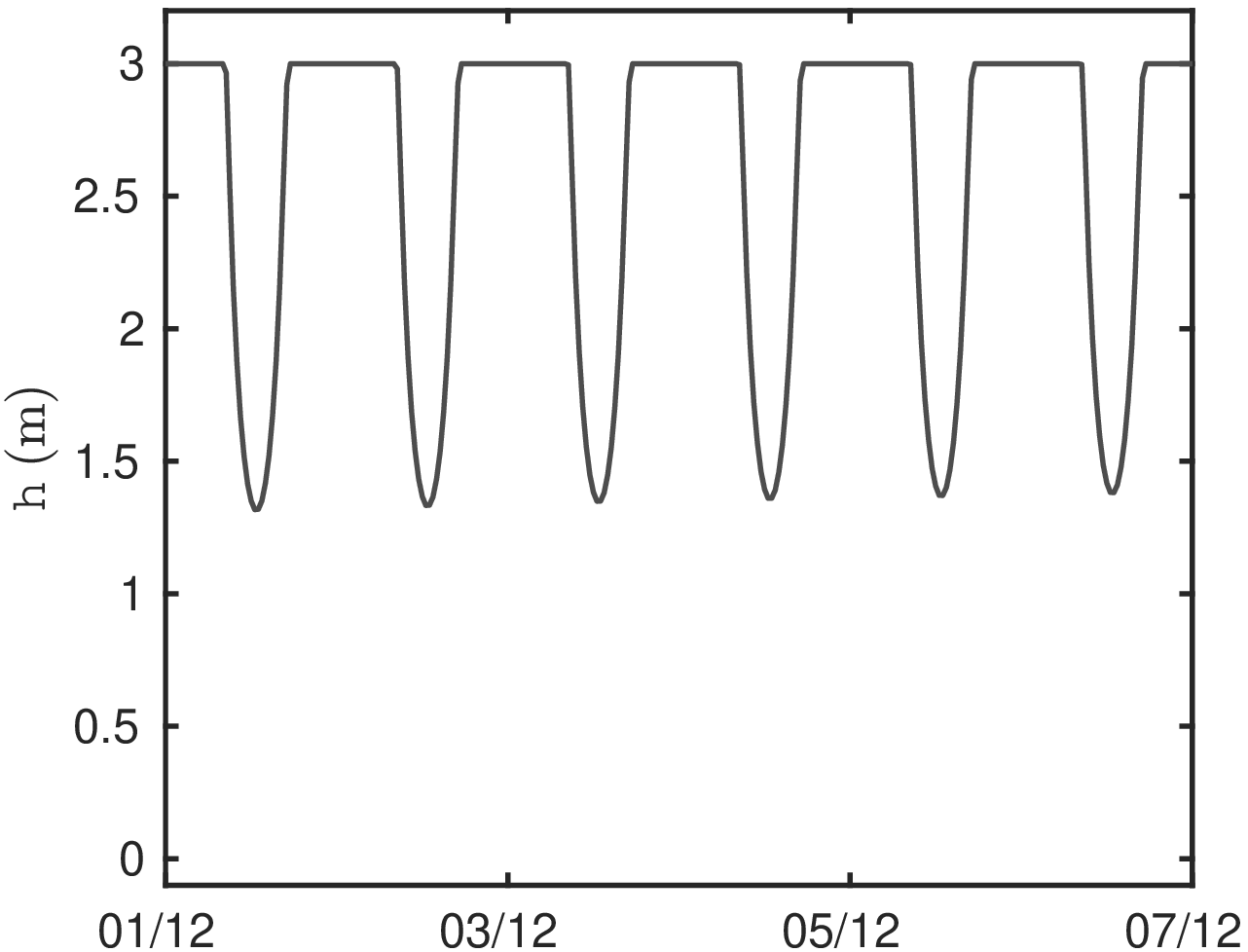}} \\
\subfigure[\label{fig:qsummer_fy} summer]{\includegraphics[width=.45\textwidth]{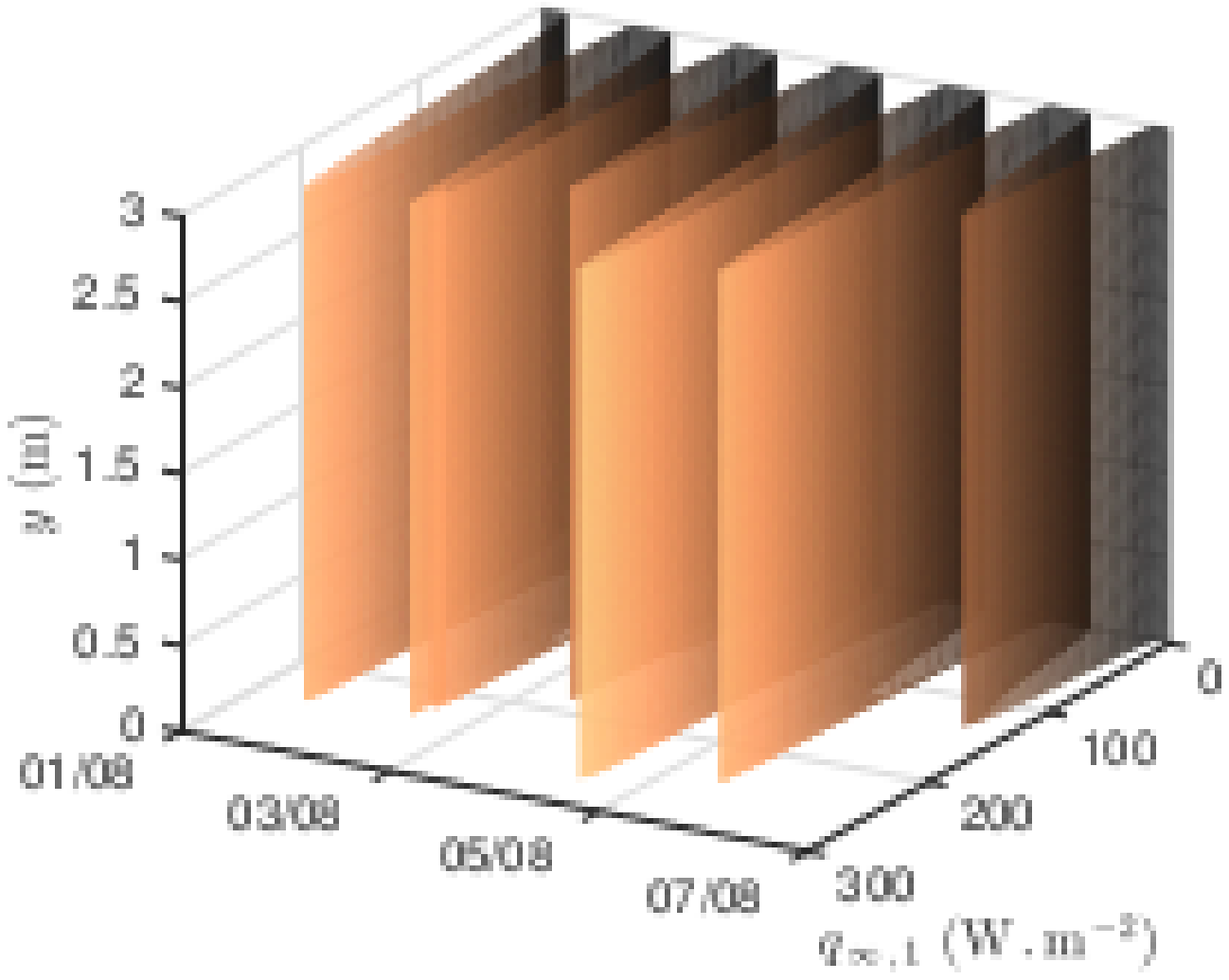}}  \hspace{0.2cm}
\subfigure[\label{fig:qwinter_fy} winter]{\includegraphics[width=.45\textwidth]{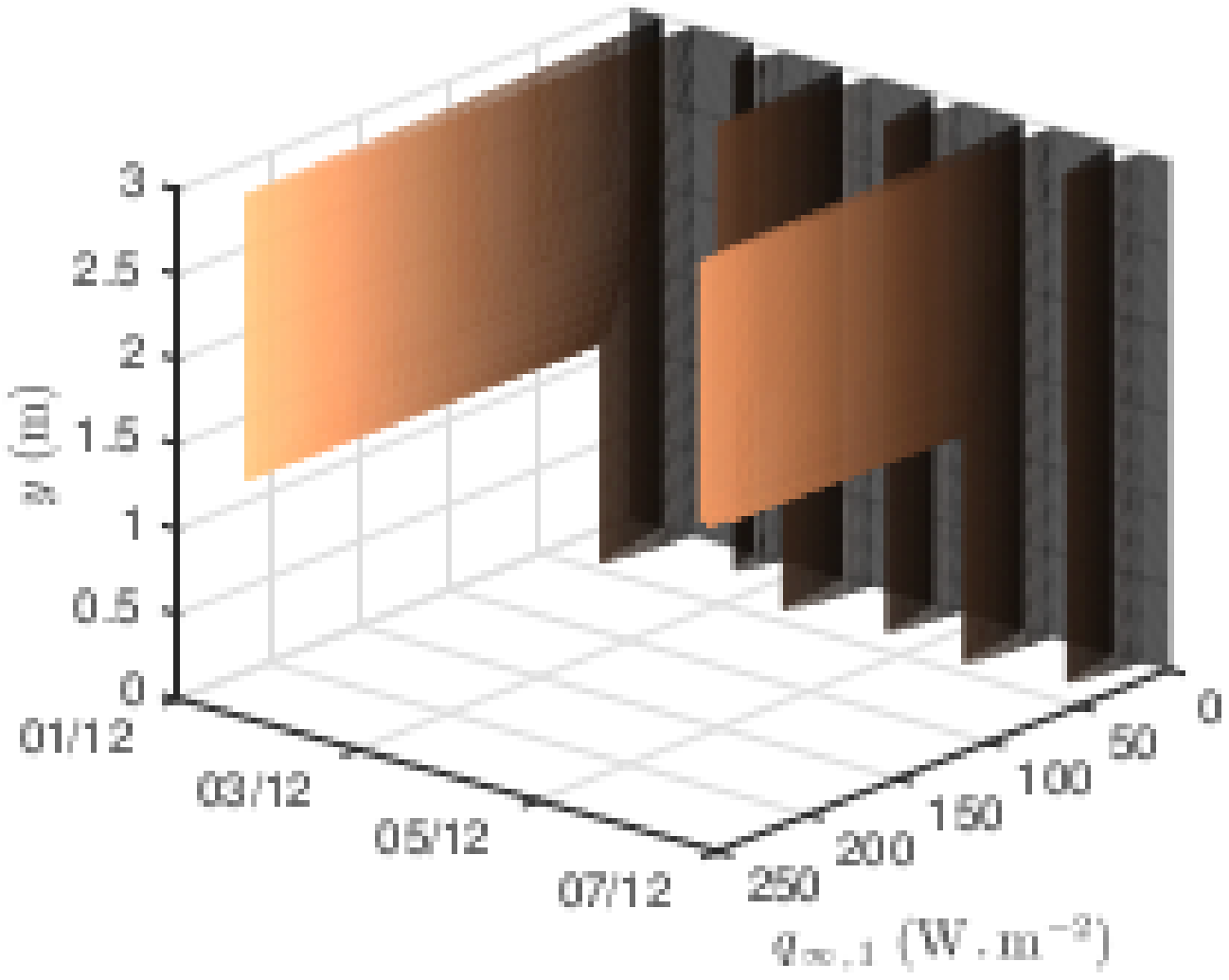}} 
\caption{Variation of the shadow height \emph{(a,b)} and incident short-wave radiation heat flux \emph{(c,d)}.}
\end{figure}

Figures~\ref{fig:qsummer_fy} and \ref{fig:qwinter_fy} illustrates the variation of the incident shot-wave radiation flux for two different weeks. In summer, the sunlit area covers the whole facade, as shown in Figures~\ref{fig:hsummer_ft}. Thus, there is no significant difference in terms of incident flux on the top and bottom surfaces. However, in winter the contrast is more noticeable since the height of the shadow reaches almost $y \egal 1.5 \ \mathsf{m}$ each day at midday. As a consequence, the bottom of the facade receives less flux. Note that when the direct flux is negligible compared to the diffuse one, the magnitude of the total incident flux is homogeneous over the whole facade. This can be remarked on December $6^{\,th}$  in Figure~\ref{fig:qwinter_fy}.

\begin{figure}
\centering
\subfigure[\label{fig:qy1_ft} $y \egal 0.3 \ \mathsf{m}$ ]{\includegraphics[width=.45\textwidth]{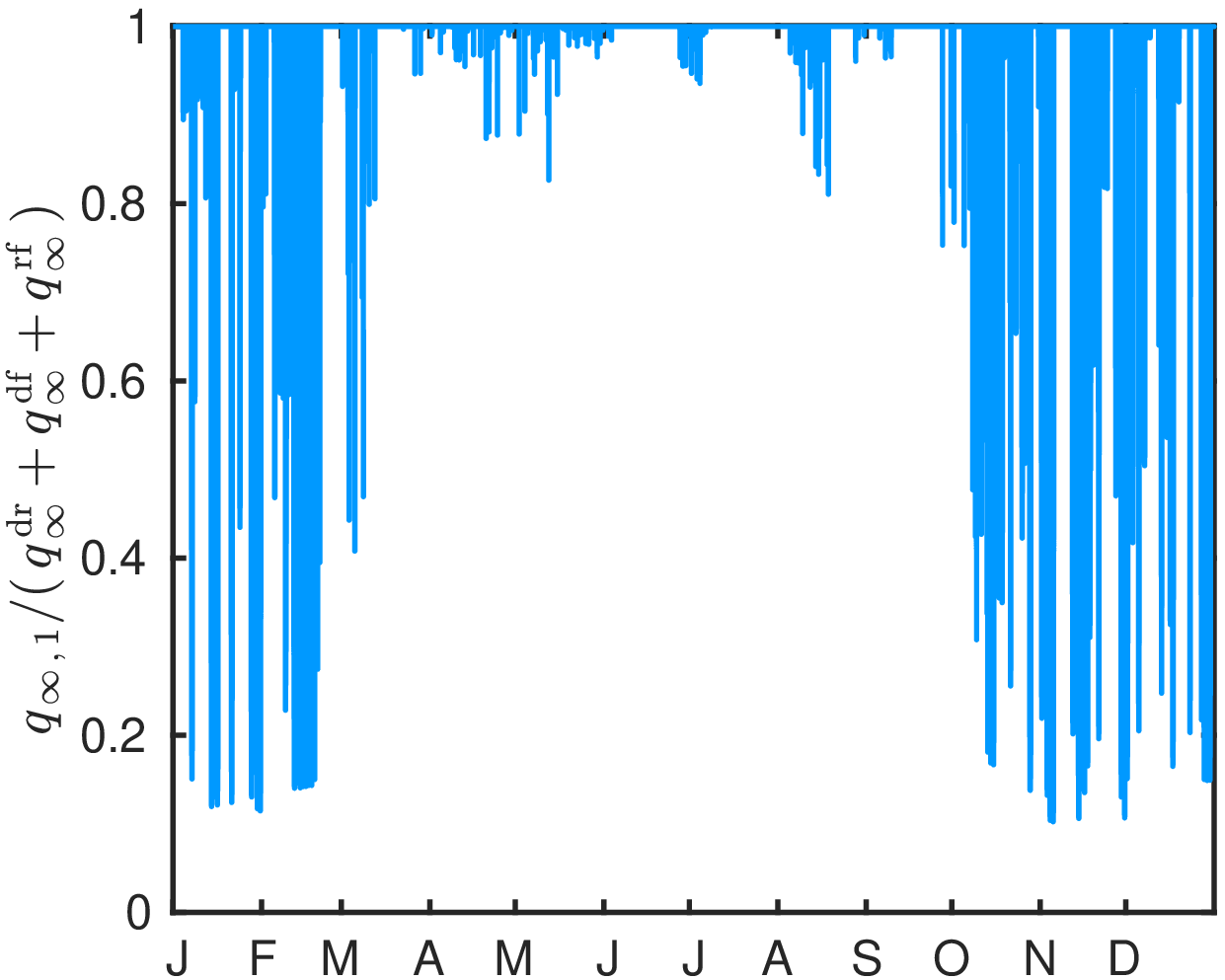}}  \hspace{0.2cm}
\subfigure[\label{fig:qy3_ft} $y \egal 2.7 \ \mathsf{m}$ ]{\includegraphics[width=.44\textwidth]{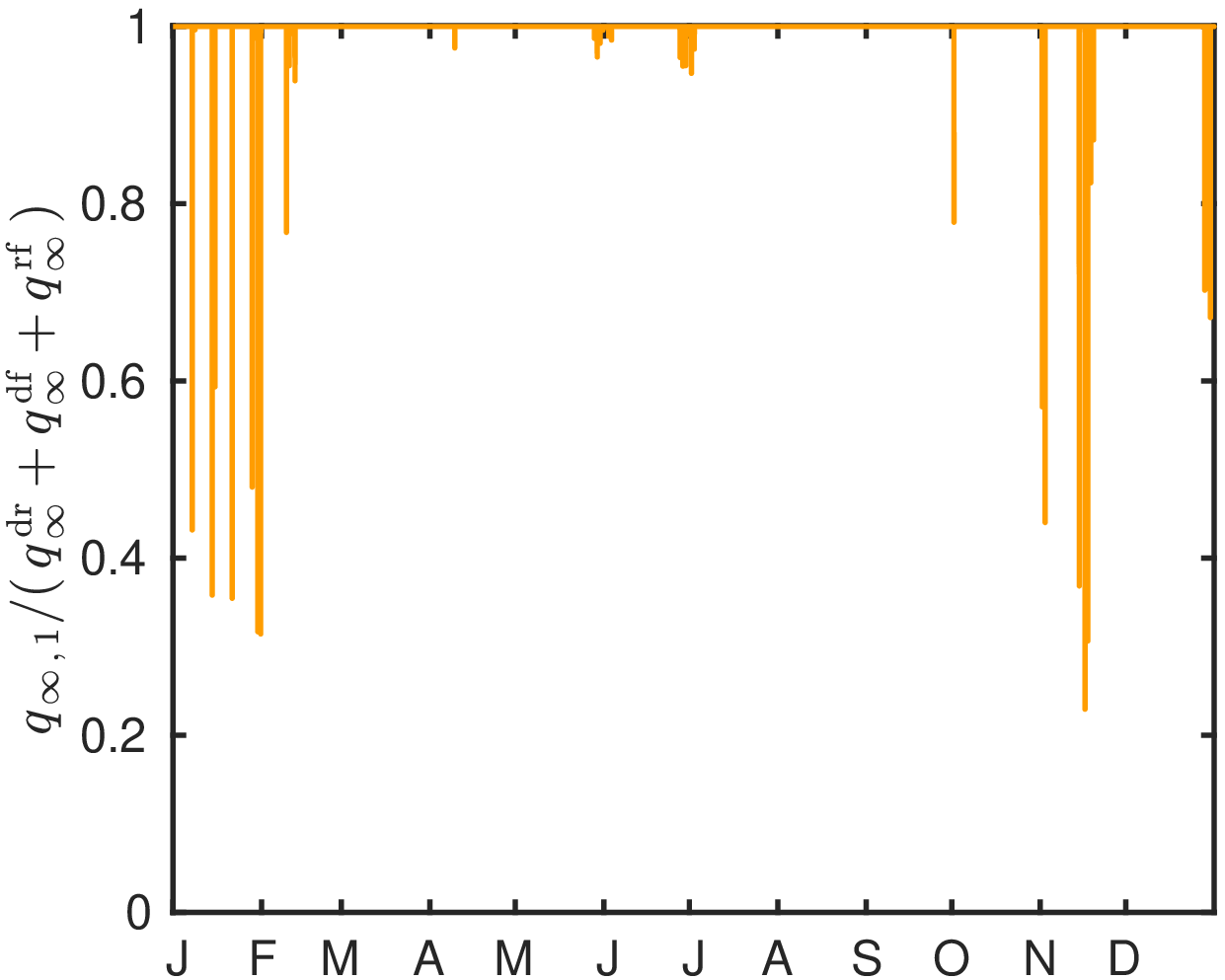}} 
\caption{Ratio between the effective and total incident radiation heat flux at two different heights.}
\end{figure}

The incident flux on the facade varies according to the height. The ratio between the effective and total incident radiation flux on the facade is illustrated for two different heights in Figures~\ref{fig:qy1_ft} and \ref{fig:qy3_ft}. For the top of the facade, the ratio is almost always equal to $1\,$. In other words, the top of the facade is not affected by the shadow and it receives the total incident radiation flux. However, at the bottom, the ratio can reach $20 \%$. During the winter period, the effective incident radiation flux is particularly reduced compared to the total one. Note that the sunlight exposure of the facade is shorter in winter than in summer. The average time of daily sunlight exposure is $7.1 \ \mathsf{h}$ in December and $8.9 \ \mathsf{h}$ n August.


\subsection{Assessing the thermal efficiency}
\label{sec:thermal_efficiency}

After generating the variation of the boundary conditions according to space and time, the numerical model is used to evaluate the thermal performance of the wall. The discretisation parameters are $\Delta t \egal 36 \ \mathsf{s}$ and $\Delta x \egal \Delta y \egal 3.7 \ \mathsf{mm}\,$. The computational time of this simulation is of $t_{\,\cpu} \egal 5.6 \ \mathsf{min}\,$, corresponding to a ratio $R_{\,\cpu} \egal 0.93 \ \mathsf{s\,/\,days}$ of physical simulation. Comparatively, the same simulation with the one-dimensional \DF ~model and same discretisation parameters has a ratio of $R_{\,\cpu} \egal 0.54 \ \mathsf{s\,/\,days}$ of physical simulation. The increase of computational time is moderate compared to the one-dimensional approach. Considering the CFL stability condition from Eq.~\eqref{eq:cfl_condition} and the parameters of the problem, the computational ratio with the \Eu ~explicit approach is estimated to $R_{\,\cpu} \egal 13 \ \mathsf{\%}\,$. The \DF ~numerical model enables to save significant computational efforts compared to standard approaches. 

\begin{figure}
\centering
\subfigure[\label{fig:Tsummer_y270_fx} summer, $y \egal 2.7 \ \mathsf{m}$ ]{\includegraphics[width=.45\textwidth]{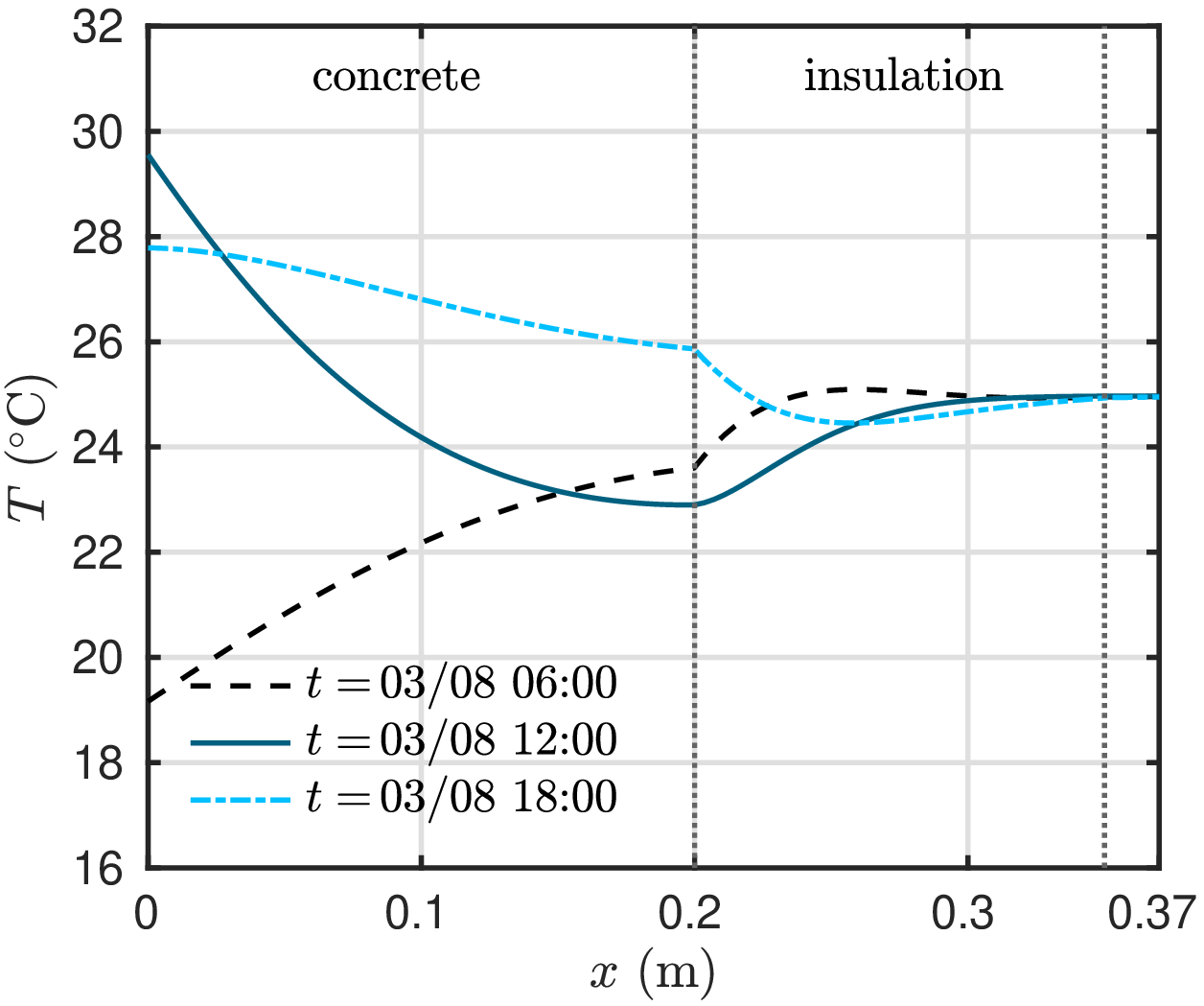}}  \hspace{0.2cm}
\subfigure[\label{fig:Twinter_y270_fx} winter, $y \egal 2.7 \ \mathsf{m}$ ]{\includegraphics[width=.45\textwidth]{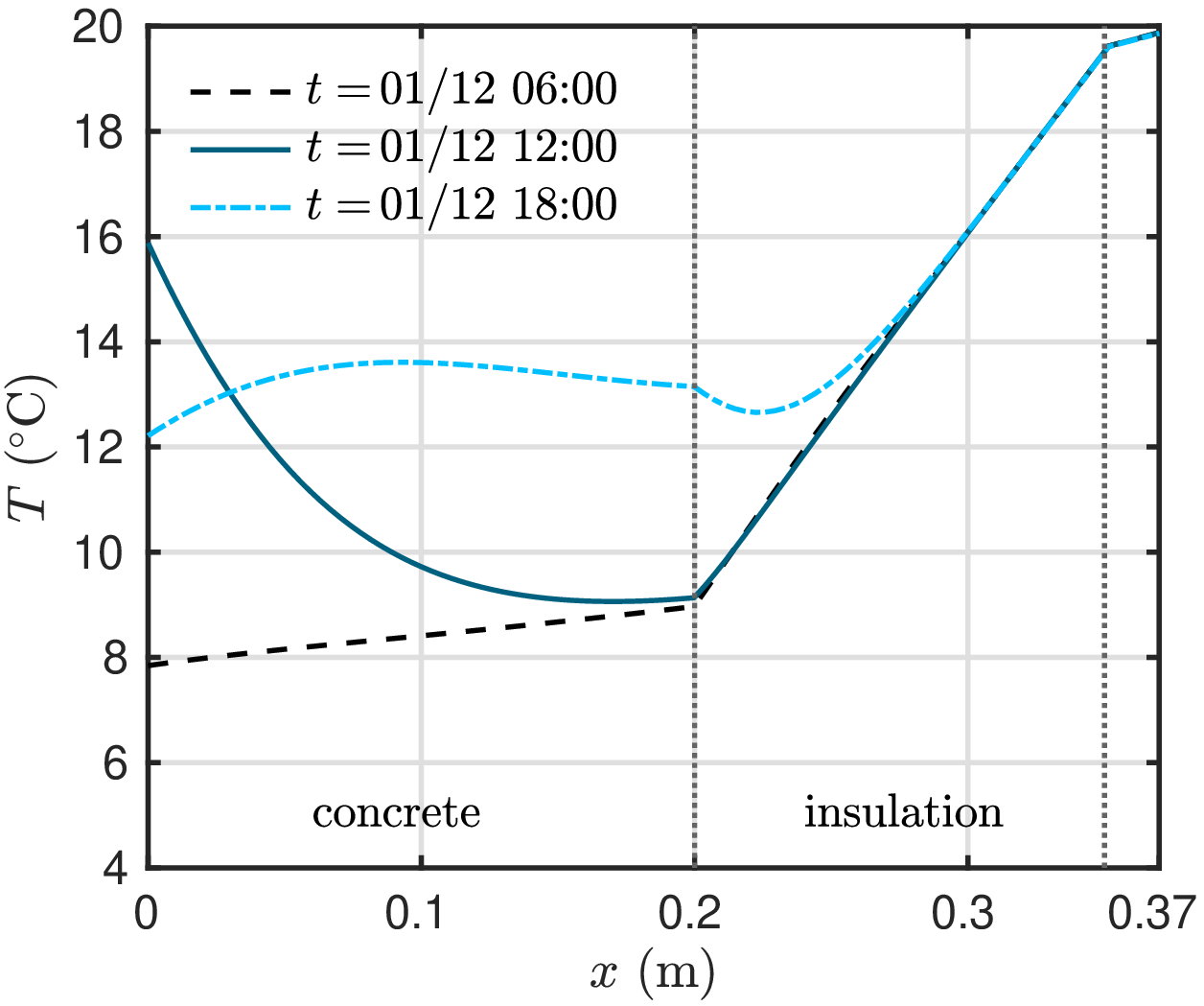}} \\
\subfigure[\label{fig:Tsummer_y30_fx} summer, $y \egal 0.3 \ \mathsf{m}$ ]{\includegraphics[width=.45\textwidth]{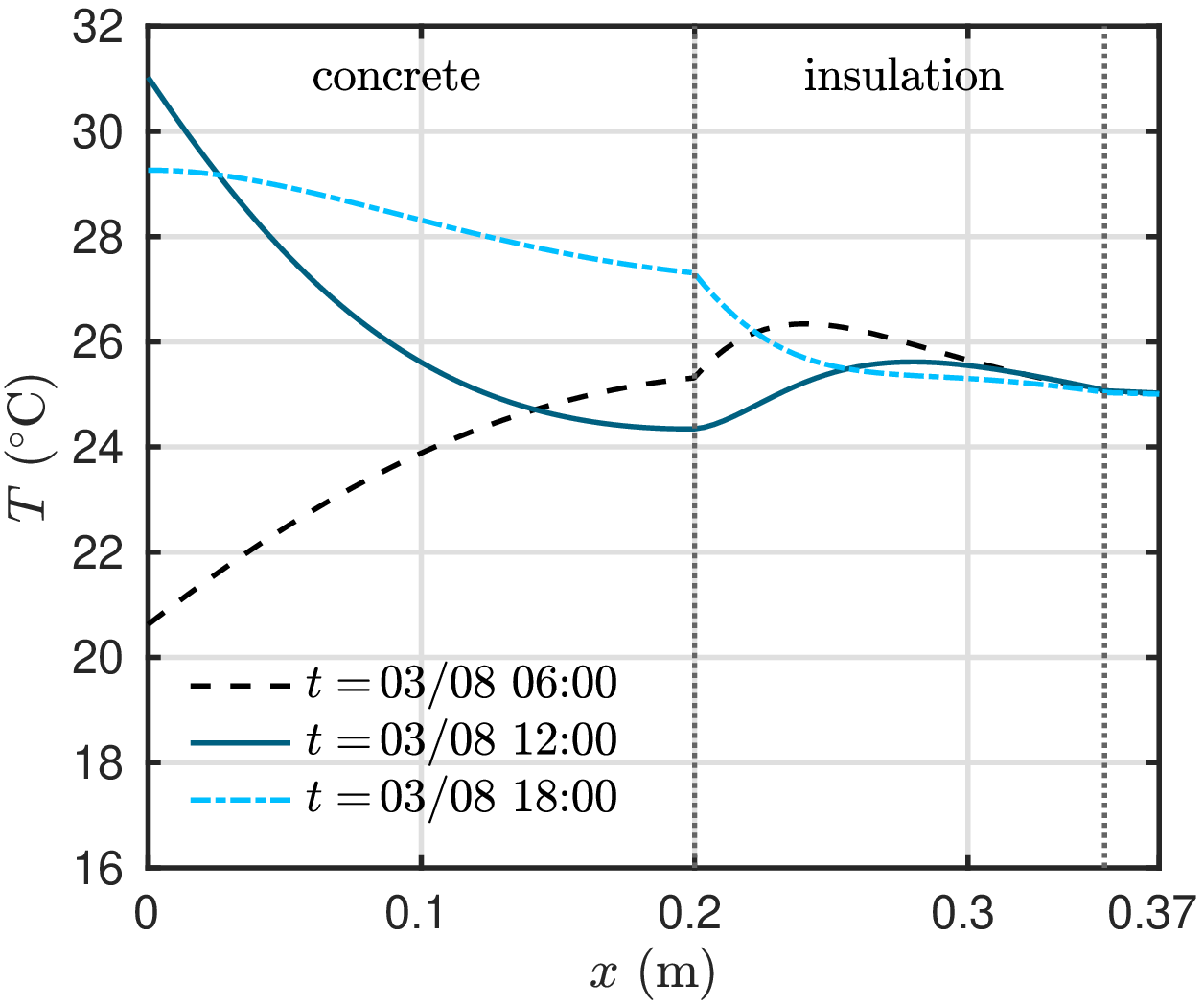}}  \hspace{0.2cm}
\subfigure[\label{fig:Twinter_y30_fx} winter, $y \egal 0.3 \ \mathsf{m}$ ]{\includegraphics[width=.45\textwidth]{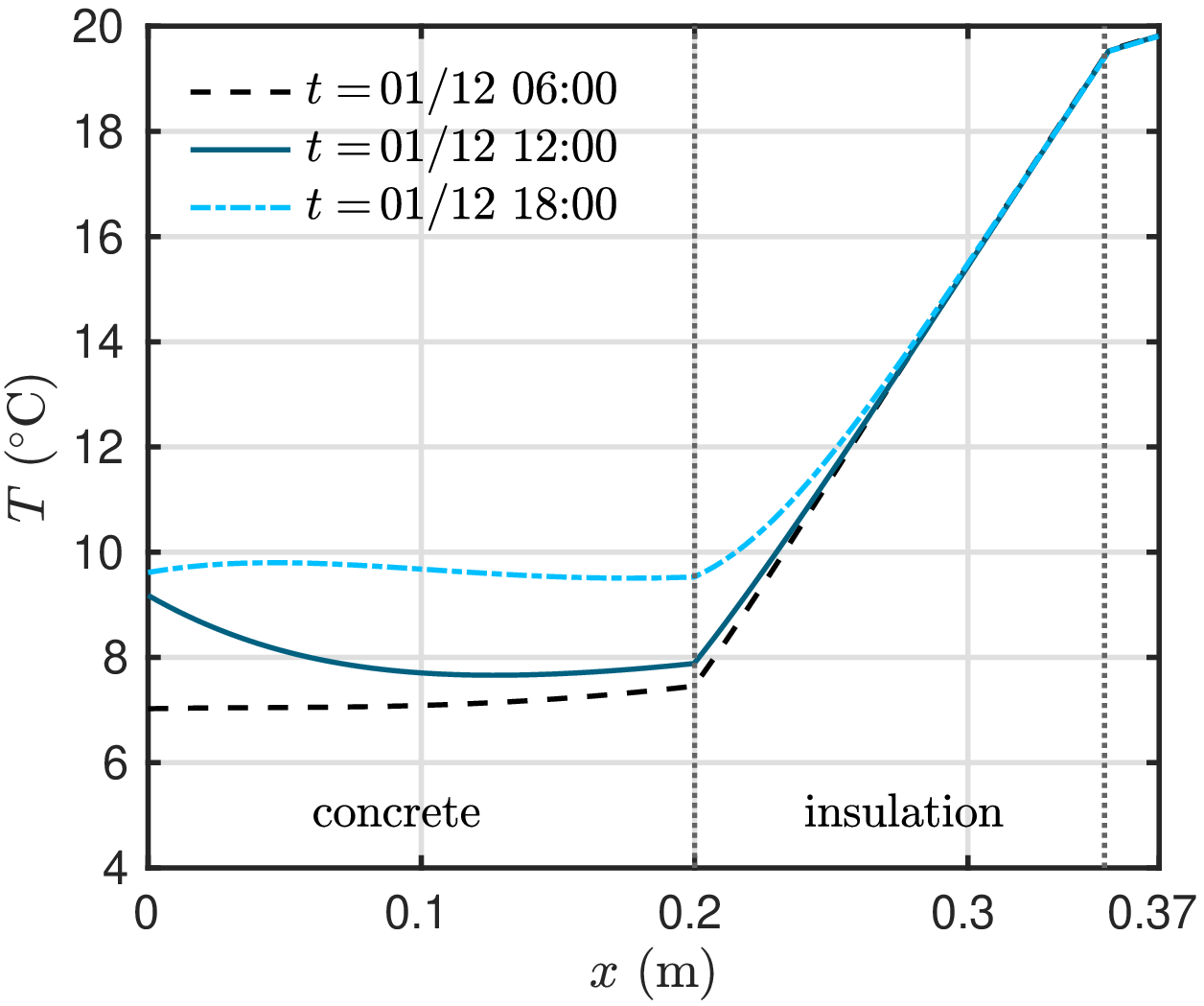}} \\
\subfigure[\label{fig:Tsummer_x0_fy} summer, $x \egal 0 \ \mathsf{m}$ ]{\includegraphics[width=.45\textwidth]{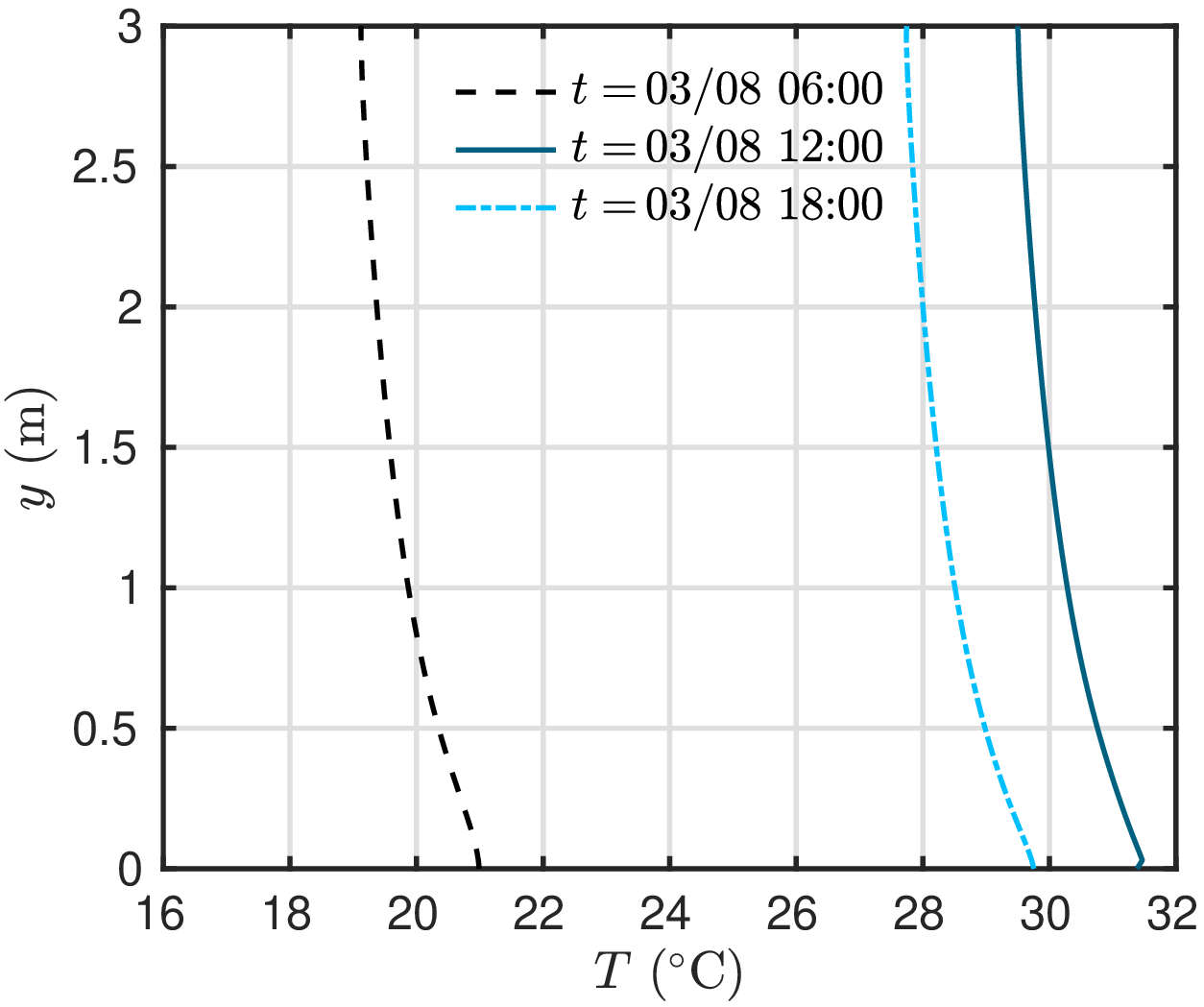}}  \hspace{0.2cm}
\subfigure[\label{fig:Twinter_x0_fy} winter, $x \egal 0 \ \mathsf{m}$ ]{\includegraphics[width=.45\textwidth]{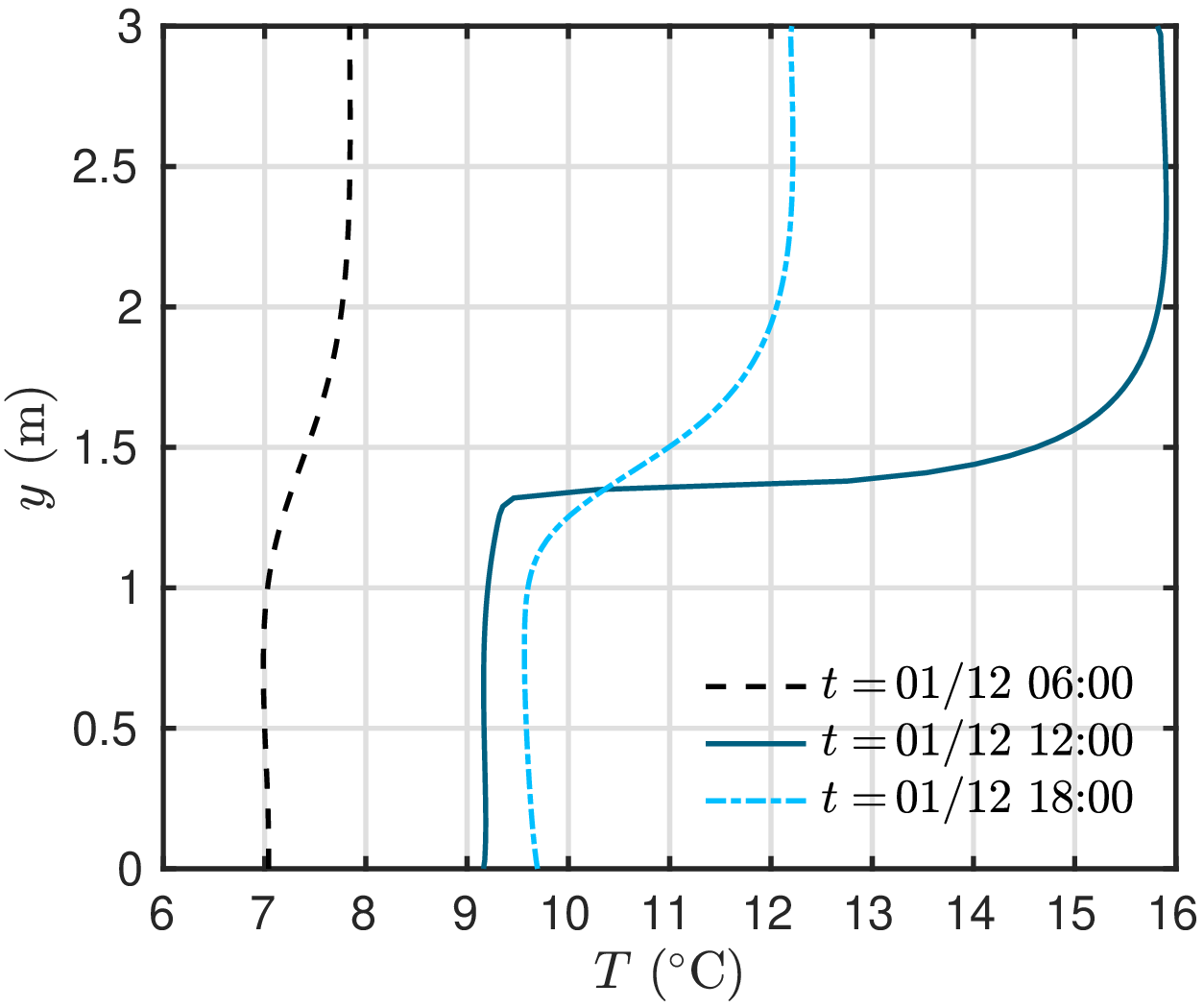}} 
\caption{Temperature variation according to the height \emph{(a,b,c,d)} and length \emph{(e,f)} of the wall.}
\end{figure}

\begin{figure}
\centering
\subfigure[\label{fig:Tsummer0308_13h_fxy} summer, $03/08$ ]{\includegraphics[width=.45\textwidth]{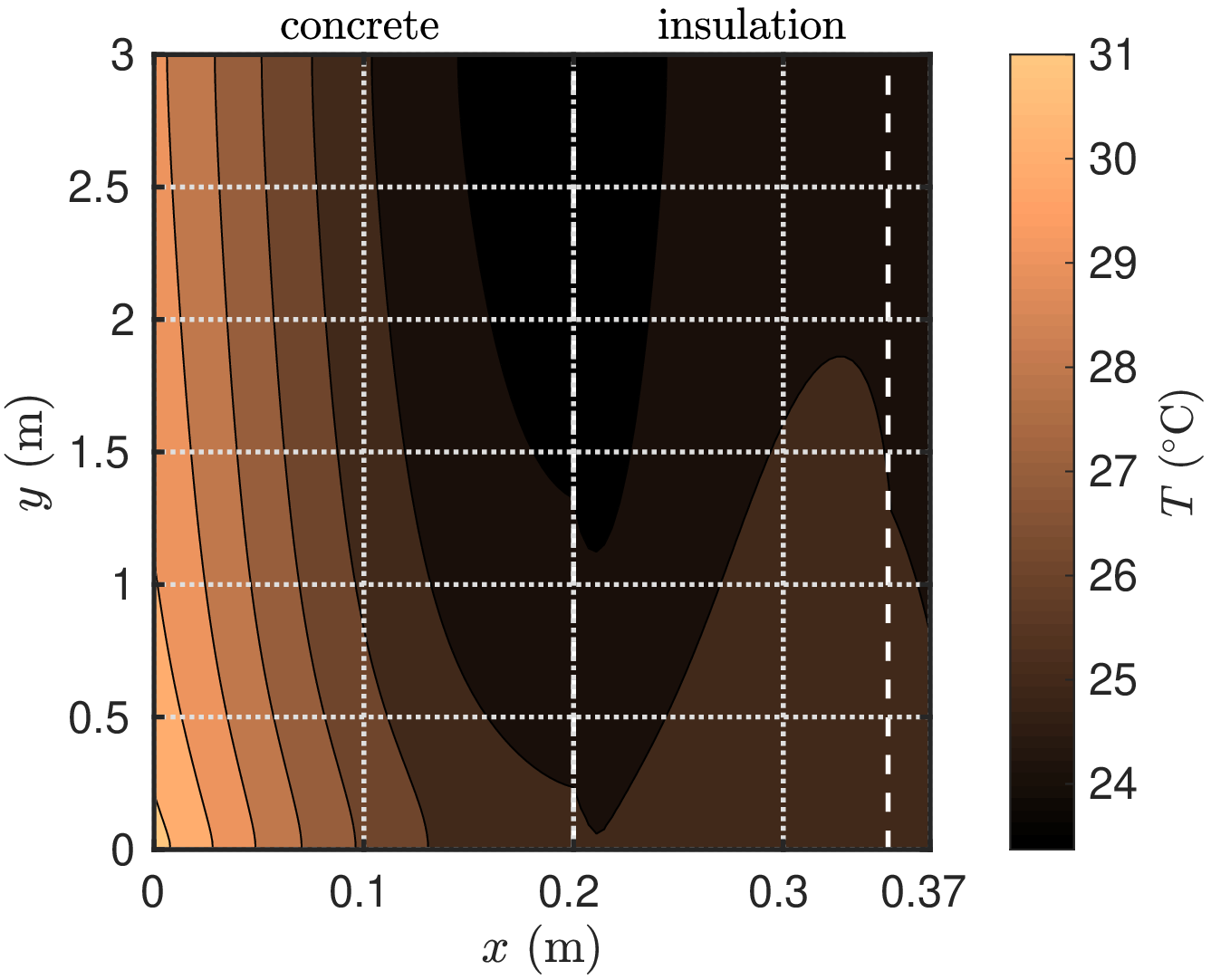}}  \hspace{0.2cm}
\subfigure[\label{fig:Twinter0112_13h_fxy} winter, $01/12$ ]{\includegraphics[width=.45\textwidth]{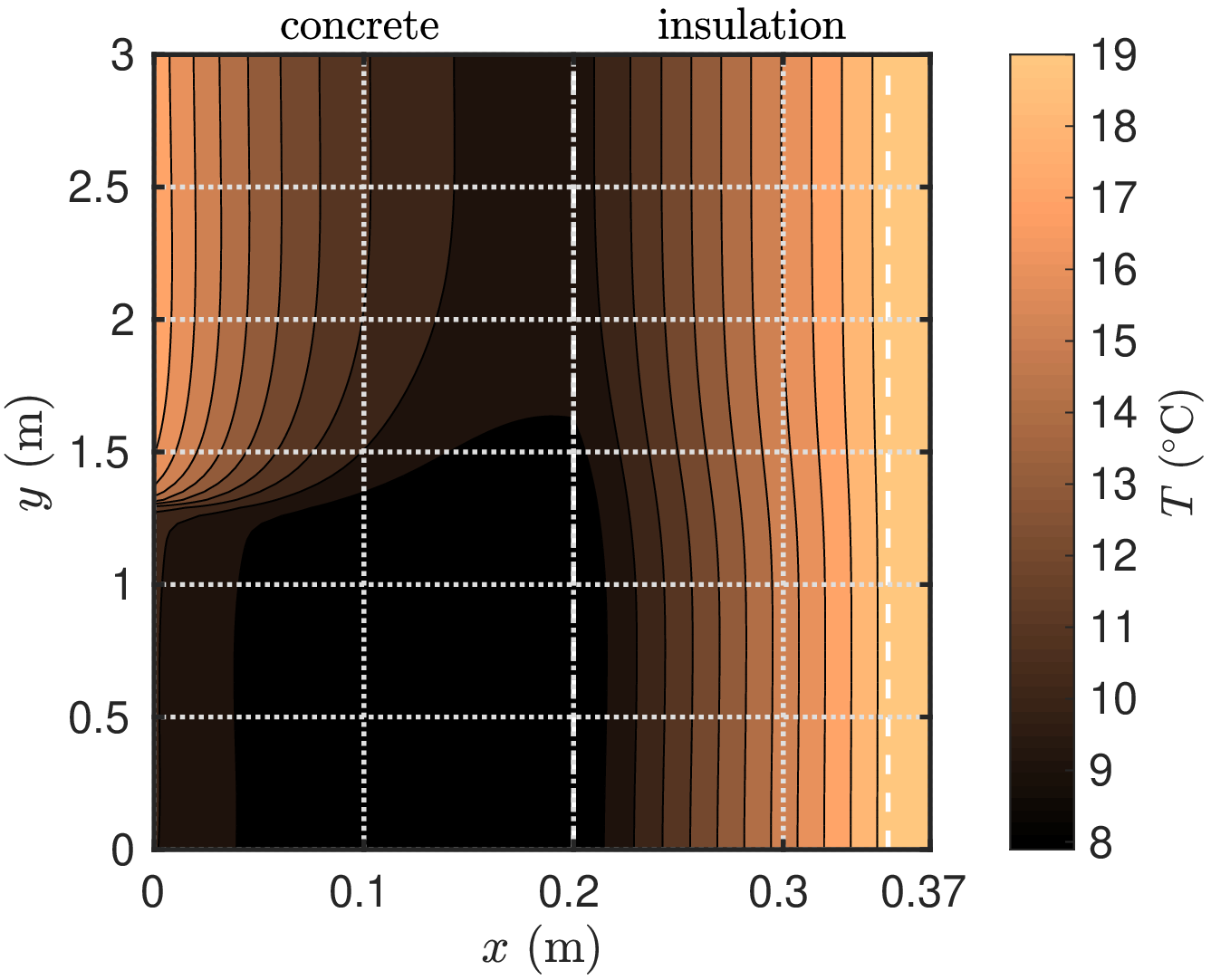}} 
\caption{Section of the temperature according to $x$ and $y$ at $13:00$ in summer \emph{(a)} and winter \emph{(b)}.}
\label{fig:sectionT}
\end{figure}

The temperature variation according to $x$ and $y$ is provided in Figures~\ref{fig:Tsummer_y270_fx} to \ref{fig:Twinter_x0_fy} for summer and winter periods. Since the incident heat flux is more homogeneous on the facade in summer, there are no important differences between the temperature at the bottom and top. As remarked in Figure~\ref{fig:Tsummer_x0_fy}, the differences scale with $2 \ \mathsf{^{\,\circ}C}\,$. Those small differences are mainly due to the variation of the surface heat transfer coefficient according to the height $y\,$.  In winter, there are more discrepancies in the surface temperature between the top and bottom of the facade. Figure~\ref{fig:Twinter_x0_fy} highlights those contrasts. At 06:00, the temperature is relatively homogeneous along the facade since there is no incident flux. However, at 12:00 the incident flux induces a variation of almost $7 \ \mathsf{^{\,\circ}C}$ on the temperature between the top and bottom. Note that on the inside surface at $ x \egal 0.37 \ \mathsf{m}\,$, the temperature does not vary with the height of the facade. Those results are confirmed by the sections of temperature illustrated in Figure~\ref{fig:sectionT}. The influence of the boundary conditions is mostly remarkable in the first concrete layer of the facade. The insulation layer reduces significantly the temperature gradients along with the height. 

\begin{figure}
\centering
\subfigure[\label{fig:jRsummer_fy} summer]{\includegraphics[width=.45\textwidth]{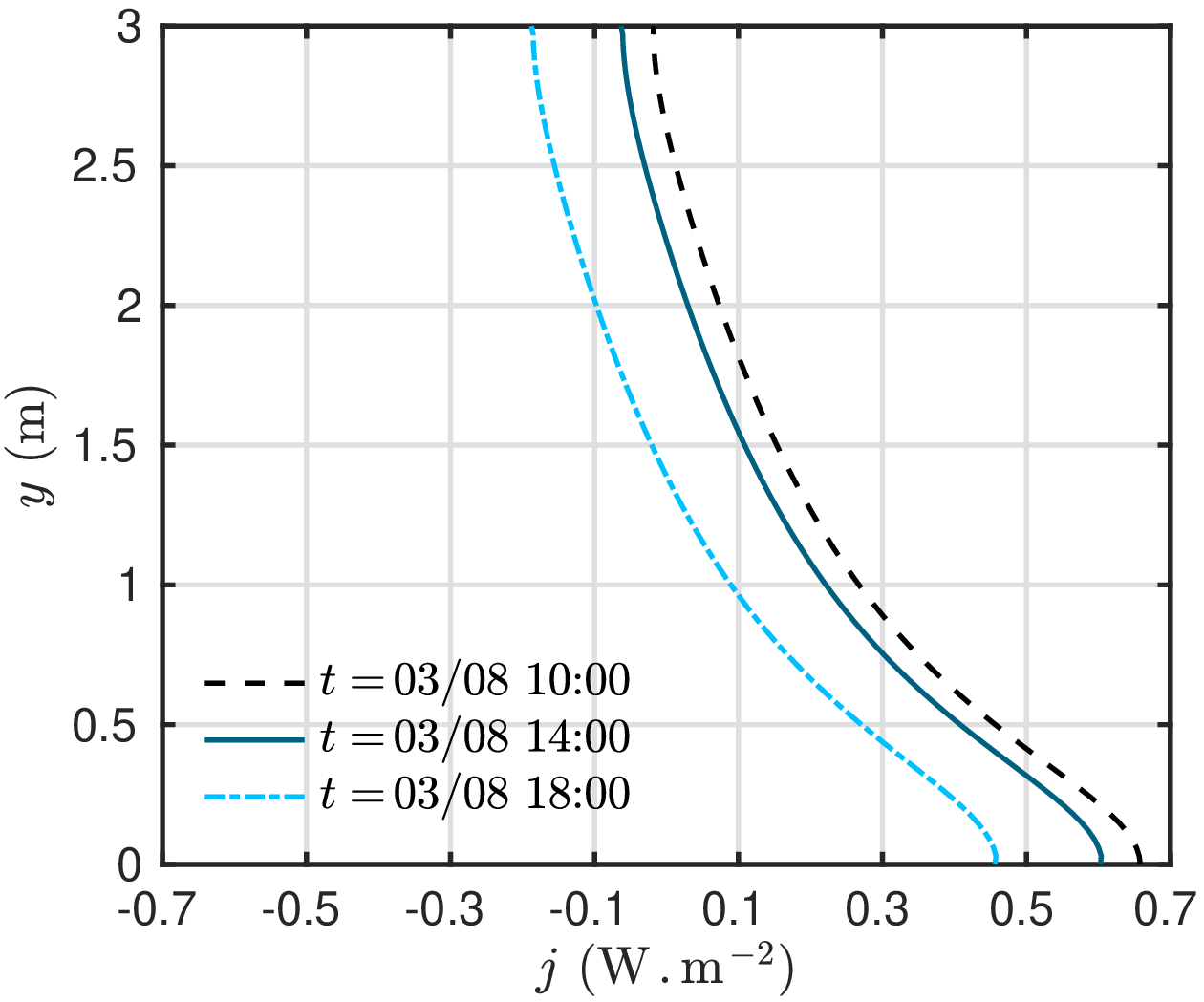}}  \hspace{0.2cm}
\subfigure[\label{fig:jRwinter_fy} winter]{\includegraphics[width=.45\textwidth]{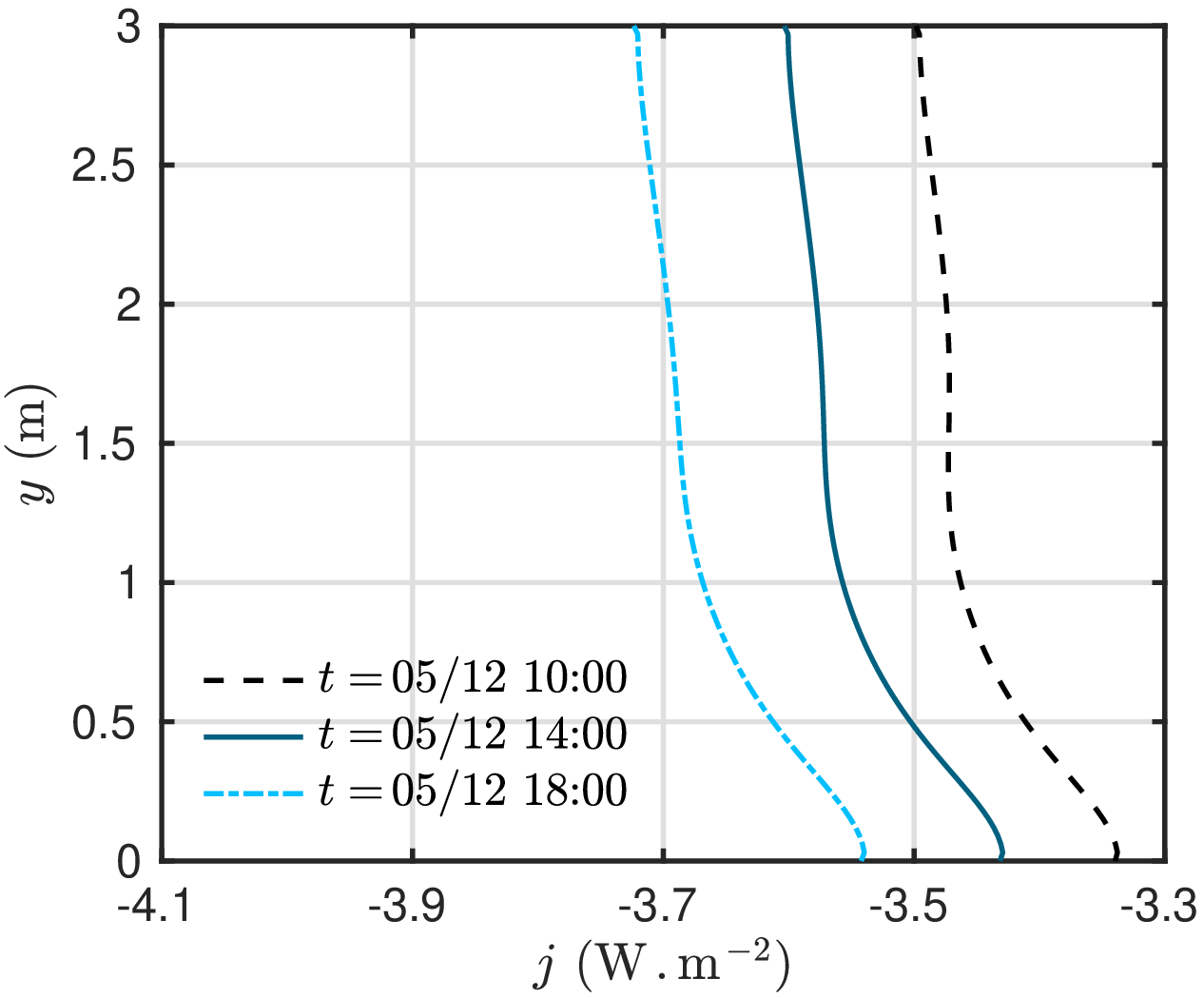}} 
\caption{Variation of the heat flux on the inside boundary $x \egal 0$ in summer \emph{(a)} and winter \emph{(b)}.}
\label{fig:jR}
\end{figure}

Even if the variation of temperature is small at the inner surface, it still induces a variation of the heat flux along $y$ as remarked in Figure~\ref{fig:jR}. It is important to note that in summer the flux changes of sign between the bottom and top of the wall. At  18:00, below $y \egal 1.25 \ \mathsf{m}$ the flux is positive so the bottom of the wall is heating the inside zone. Inversely, the top of the wall is cooling the inside zone. This effect is not due to the incident radiation but to the variation of the heat surface coefficient with the height. The latter is higher at the top of the facade. It increases the heat transfer at the top surface. In winter, the flux on the inside surface is entirely negative. Thus, the inside zones are losing energy through the wall. 

\begin{figure}
\centering
\subfigure[\label{fig:jRtot_ft}]{\includegraphics[width=.95\textwidth]{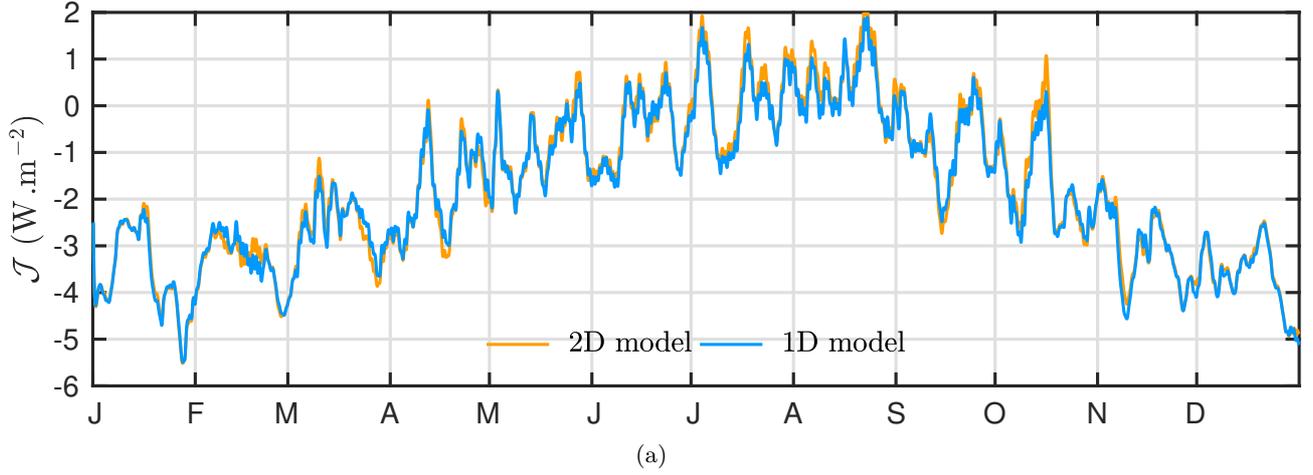}}
\caption{Variation of the total heat flux on the inside boundary computed with the standard one dimensional approach and with the two-dimensional modeling.} 
\end{figure}

The time and space variation of the climatic boundary conditions induces two-dimensional heat transfer through the facade. It is important to evaluate the thermal efficiency of the wall compared to standard approaches. The building simulation program \texttt{Domus} is used to assess standard building energy efficiency. Within this approach, the heat transfer is modeled in one-dimension and the outside surface heat transfer coefficient is considered as constant. The latter is set to the mean $\overline{h}_{\,\infty\,,\,1} \egal 10.28 \ \mathsf{W\,.\,m^{\,-2}\,.\,K^{\,-1}}\,$. Moreover, the outside incident short-wave radiation flux includes the shading effects by evaluating the sunlit area ratio $S$ on the facade using the pixel counting technique:
\begin{align}
\label{eq:average_value_qinf}
q_{\,\infty\,,\,1}^{\,\mathrm{st}} (\,t\,) \egal  \alpha \cdot \Bigl(\, q_{\,\infty}^{\,\mathrm{dr}}(\,t\,) \cdot S (\,t\,)
\plus q_{\,\infty}^{\,\mathrm{df}}(\,t\,) 
\plus q_{\,\infty}^{\,\mathrm{rf}}(\,t\,) \, \Bigr) \,.
\end{align}
The results from the standard approach are compared to the one obtained with the two-dimensional modeling. Figure~\ref{fig:jRtot_ft} compares the heat flux at the inside surface. Small discrepancies are noted in both winter and summer periods. The magnitude of the flux is higher with the two-dimensional modeling. In the mid-season, the two approaches have similar predictions. Table~\ref{tab:thermal_loads} gives the thermal loads per month for the two-dimension model. The relative error with the one-dimensional modeling is also presented. The error is higher for the summer period. In July, the one-dimensional model underestimated by $80\%$ the thermal loads. In winter the error is lower by around $1\%\,$.

A parametric comparison is carried out to analyze the influence of the shadow on the incident short-wave radiation flux and the variation of the surface heat transfer coefficient with the height and the wind speed. Three additional simulations are performed with the two-dimensional model. The first one considers both a constant outside surface heat transfer coefficient $\overline{h}_{\,\infty\,,\,1} \egal 10.28 \ \mathsf{W\,.\,m^{\,-2}\,.\,K^{\,-1}}$ and no shadow modeling on the facade. Thus, the sunlit area is set to unity $S \egal 1\,$. The second computation deals only with a constant coefficient while the third one only does not take into account the shadow modeling. 
Results are shown in Table~\ref{tab:thermal_loads}. The shadow modeling does not impact the prediction of thermal loads during the summer period (April to September). It is consistent with the analysis of Figures~\ref{fig:qy1_ft} and \ref{fig:qy3_ft} since the wall is always exposed to the sunlited. However, the modeling of the heat surface coefficient with height and wind velocity significantly influence the predictions during this period. In July, the relative error reaches $-80\%$ when considering a constant surface transfer coefficient. In the winter period (October to February), both shadow modeling and varying surface transfer coefficients have remarkable effects on the predictions. For the month of December, the effects are counter-balanced. Not including the shadow modeling induces a relative error of $3.65 \%$. On the contrary, not taking into account the time variation of the surface transfer coefficient implies a relative error of $-1\%\,$.

A similar study is performed by computing the relative error on the thermal loads with the one-dimensional model. The first computation is the standard one described above. It considers a constant surface transfer coefficient and a direct radiation flux computed according to the sunlit area ratio. The second simulation combines the same approach for the direct radiation flux and a time varying surface transfer coefficient compute using Eq.~\eqref{eq:h1}, for the middle height $y \egal 1.5 \ \mathsf{m}\,$. The last simulation includes a constant surface transfer coefficient. For the radiation, it is assumed that the facade is always exposed to the sunlit. The results are presented in Table~\ref{tab:thermal_loads}. 
For the summer period, the 1D model with ratio radiation flux and time varying coefficient has a lower error compared to the 2D modeling. For the winter, there is no particular tendency. It can be highlighted that the 1D model with no sunlit ratio and constant coefficient lacks of accuracy to predict the thermal loads in this case.

Those results highlight the importance of modeling the two-dimensional transfer induced by time and space variations of the incident short-wave radiation heat flux and of the heat surface coefficient to accurately predict the wall energy efficiency. Both the shadow and the surface transfer coefficient modelings have an important effect on the thermal loads. The one-dimensional model cannot predict the phenomena with a reduced relative error over the whole year.

\begin{table}
\centering
\addtolength{\leftskip} {-2cm}
\addtolength{\rightskip}{-2cm}
\caption{Influence of several hypotheses on the thermal loads of the wall.}
\label{tab:thermal_loads}
\setlength{\extrarowheight}{.5em}
\begin{tabular}[l]{@{} c cc c cccccccccccc }
\hline
\hline

& \multicolumn{2}{c}{\textit{Hypothesis}}
& \multicolumn{12}{c}{Months} \\
\textit{Model} & \textit{Shad.} & \textit{Conv.} & \textit{Output} &
J & F & M & A & M & J & J & A & S & O & N & D \\ \hline
2D & Yes & $f(\,y\,,\,t\,)$  & $E \, [\mathsf{MJ}]$ &
$-9.68$ & $-7.99$ & $-7.06$ & $-4.98$ & $-2.41$ & $-1.30$ & $0.67$ & $1.52$ & $-1.99$ & $-4.36$ & $-7.92$ & $-9.69$ \\
2D & No & $f(\,y\,,\,t\,)$ & $\bigl(\,\varepsilon_{\,r} \, \circ\, E \,\bigr)\, [\mathsf{\%}]$ &
$3.84$ & $4.62$ & $0.08$ & $0.01$ & $0.05$ & $0.16$ & $0.39$ & $0.14$ & $0.02$ & $2.52$ & $5.87$ & $3.65$ \\
2D & Yes & Const. & $\bigl(\,\varepsilon_{\,r} \, \circ\, E \,\bigr) \, [\mathsf{\%}]$ &
$-0.69$ & $0.45$ & $0.13$ & $4.61$ & $-5.52$ & $-15.2$ & $-80.5$ & $-31.8$ & $-1.64$ & $-8.48$ & $-2.24$ & $-1$ \\
2D & No & Const. & $\bigl(\,\varepsilon_{\,r} \, \circ\, E \,\bigr)\, [\mathsf{\%}]$ &
$2.61$ & $4.5$ & $0.19$ & $4.62$ & $-5.46$ & $-15.1$ & $-80.2$ & $-31.7$ & $-1.62$ & $-6.34$ & $2.74$ & $2.32$ \\\hline
1D & Ratio & Const. & $\bigl(\,\varepsilon_{\,r} \, \circ\, E \,\bigr) \, [\mathsf{\%}]$ &
$-0.38$ & $0.44$ & $0.09$ & $4.56$ & $-5.62$ & $-15.3$ & $-80.6$ & $-31.7$ & $-1.54$ & $-8.43$ & $-2.21$ & $-0.98$ \\
1D & Ratio & $f(\,t\,)$ & $\bigl(\,\varepsilon_{\,r} \, \circ\, E \,\bigr) \, [\mathsf{\%}]$ &
$0.11$ & $-0.42$ & $-1.41$ & $-2.66$ & $-5.61$ & $-10.5$ & $-18.4$ & $-9.22$ & $-6.08$ & $-1.97$ & $-0.19$ & $-0.21$ \\
1D & No & Const & $\bigl(\,\varepsilon_{\,r} \, \circ\, E \,\bigr) \, [\mathsf{\%}]$ &
$2.92$ & $4.49$ & $0.16$ & $4.57$ & $-5.56$ & $-15.2$ & $-80.3$ & $-31.6$ & $-1.51$ & $-6.28$ & $2.77$ & $2.34$ \\
\hline
\hline
\end{tabular}
\end{table}

\subsection{Results for other cities}

Further investigations are carried for $119$ cities in France, according to weather data filed taken from \texttt{Meteonorm} \cite{Meteonorm}. The purpose is to evaluate the difference between the two- and the one-dimensional modeling for the given case study. For that, the procedure is as follows: \emph{(i)} compute the incident radiation and shadow height using \texttt{Domus} software, \emph{(ii)} compute the fields in the facade using the proposed two-dimensional model, \emph{(iii)} compute the fields in the facade using the one-dimensional model considering the mean value $\overline{h}_{\,\infty\,,\,1}$ and the average value of radiation flux given by Eq.~\eqref{eq:average_value_qinf} \emph{(iv)} compute the errors on the inside heat flux $\mathcal{J}$ and the thermal loads $E\,$. The discretisation parameters and the inside boundary conditions are equal to the one described in Section~\ref{sec:description_case_study}. The results are then projected on the France map using the geographic information system QGIS \cite{QGIS}. The interpolation between the results is carried using the inverse distance weight method.

\begin{figure}
\centering
\subfigure[\label{fig:eps2J}]{\includegraphics[width=.7\textwidth]{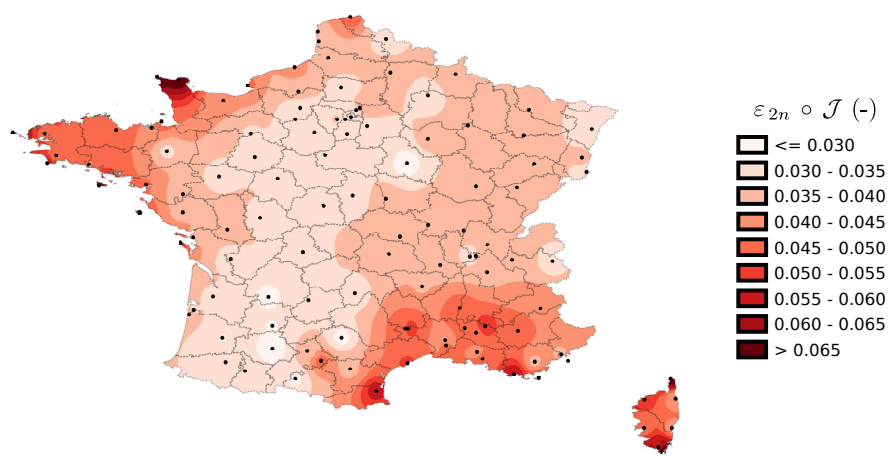}}  \\
\subfigure[\label{fig:eps2E}]{\includegraphics[width=.7\textwidth]{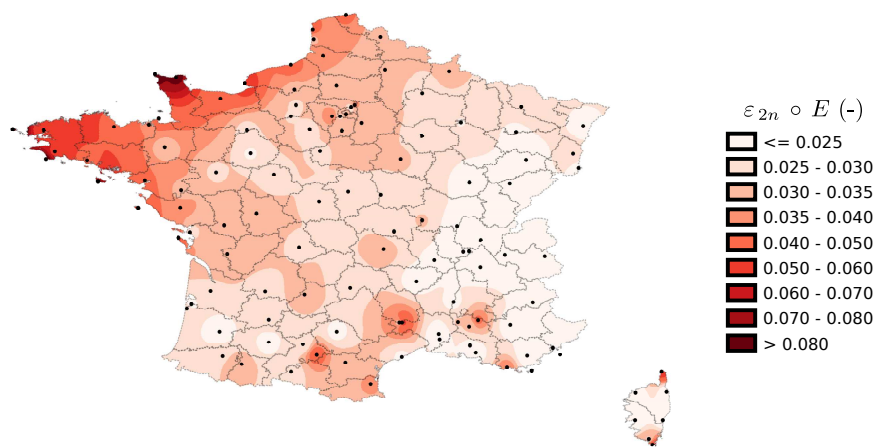}} 
\caption{Error between one- and two-dimensional modeling for the inside heat flux \emph{(a)} and the thermal loads \emph{(b)} across France.}
\end{figure}

\begin{figure}
\centering
\subfigure[\label{fig:meanQ}]{\includegraphics[width=.7\textwidth]{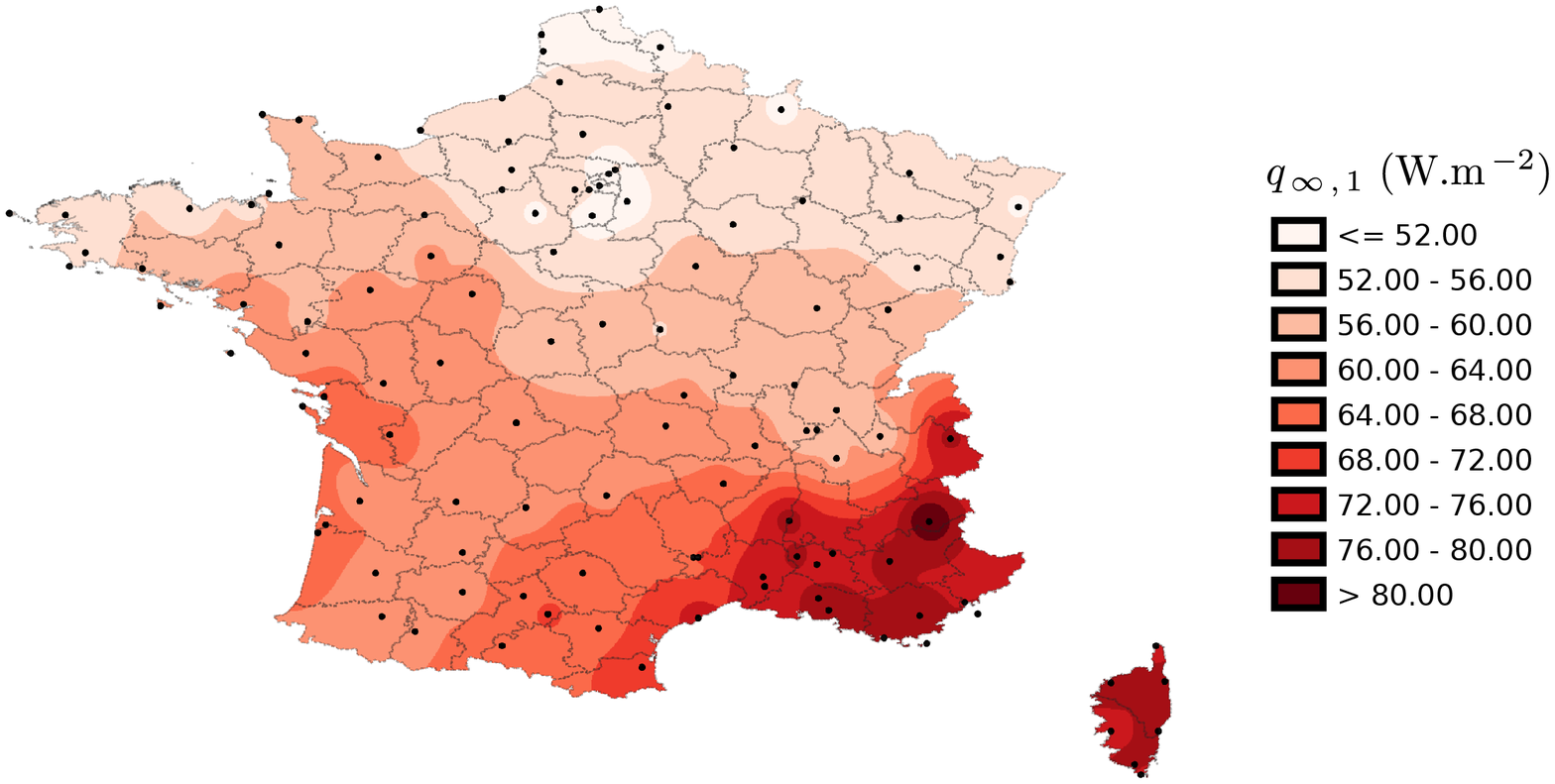}} \\
\subfigure[\label{fig:meanV}]{\includegraphics[width=.7\textwidth]{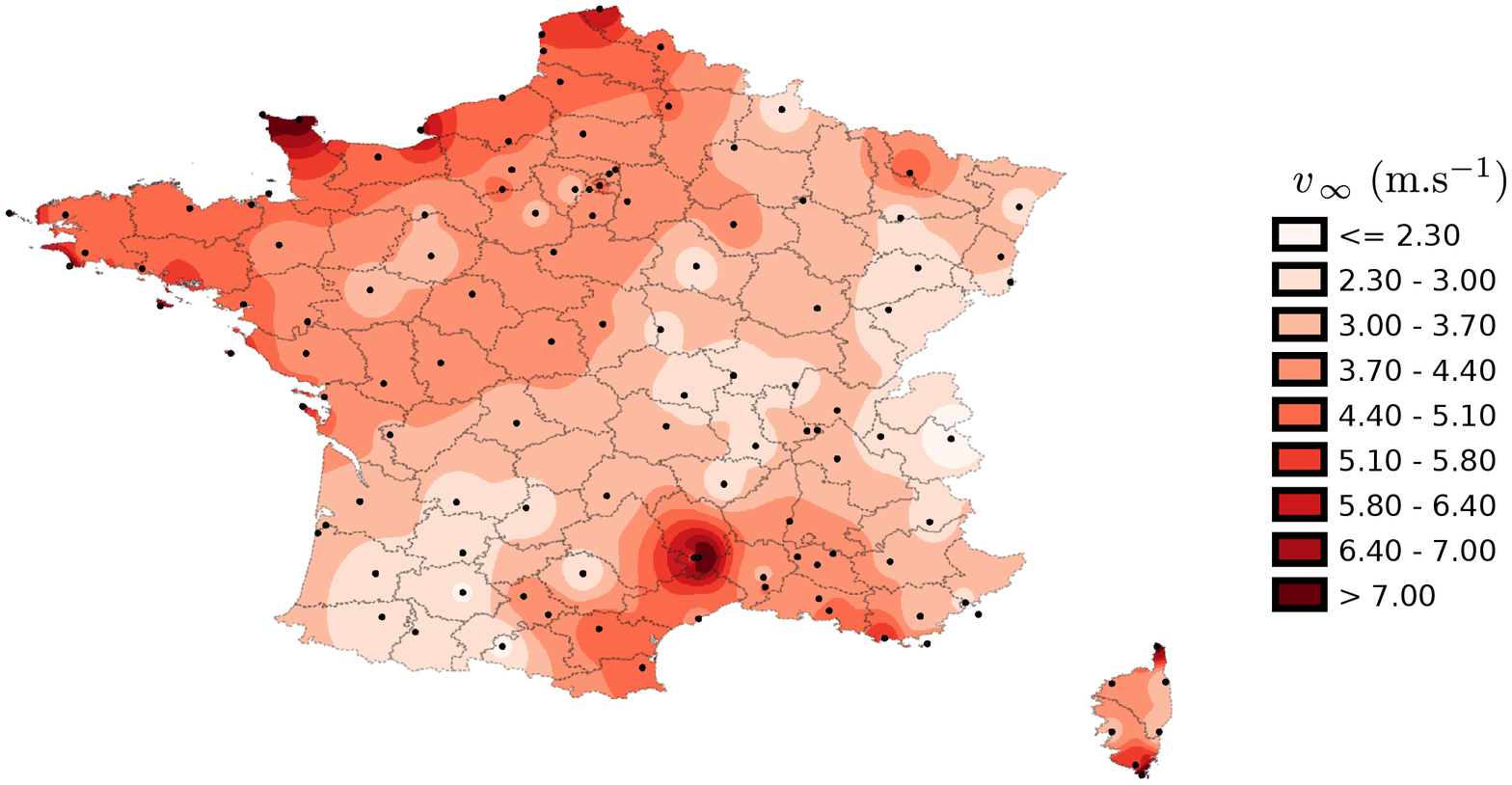}} 
\caption{Time mean total incident radiation flux \emph{(a)} and air velocity \emph{(b)} across France.}
\end{figure}

Figures~\ref{fig:eps2J} and \ref{fig:eps2E} show the error between the two modeling approaches for the inside heat flux and for the thermal loads, respectively. The maximal errors $\varepsilon_{\,2n}$ are $0.069$ and $0.088 $ for the inside flux and thermal loads, respectively. For the inside heat flux, the South-East and North-West regions of France are zones of higher errors. Indeed, as remarked in Figures~\ref{fig:meanQ} and \ref{fig:meanV}, it corresponds to a geographic area of high radiation flux and wind velocity. Thus, the accurate description of the outside boundary condition in the two-dimensional model is essential to predict accurately the inside heat flux. The diagonal South-West North-East is a region of a relatively small error. It corresponds to the same diagonal region with lower wind velocity. For the thermal loads, the error is maximal in the North-West region. On the contrary to the inside thermal flux, the error has a low magnitude in the South-East region. This is probably due to the fact that the magnitude of the thermal loads is lower.


\subsection{Local sensitivity analysis of the two-dimensional boundary conditions}

As remarked with the previous results, the modeling of the boundary conditions outside the facade can impact significantly the assessment of the energy efficiency. To investigate this influence, the \textsc{Taylor} expansion~\eqref{eq:Taylor_dev} is used. It enables to assess the sensitivity of the important fields, \emph{i.e.} the temperature, the total heat flux on the inside part and the thermal loads according to variation of the input parameters. The latter includes the first order surface heat transfer coefficient $h_{\,11}$ and the coefficient $\beta$ used in Eq.~\eqref{eq:h1}. Both parameters influence the empirical model of the outside surface heat transfer coefficient and its modification with the height $y$ and the wind velocity. Two additional parameters are considered, the front building height position $F$ and distance $D\,$. Those two parameters are related to the height of the shadow on the outside facade and thus, the incident direct flux. Those four parameters enable to
analyze the uncertainty in the two-dimensional modeling, \emph{i.e.} in space height $y$ and in time $t$, boundary conditions. The variations are $\delta h_{\,11} \egal \pm \, 0.198 \ \mathsf{W\,.\,m^{\,-2}\,.\,K^{\,-1}}\,$, $\delta \beta \egal \pm \, 0.016\,$, $\delta F \egal \pm \, 0.15 \ \mathsf{m}$ and $ \delta D \egal \pm \, 0.15 \ \mathsf{m}\,$. It corresponds to an equal relative variation for each parameter:
\begin{align*}
\frac{\delta h_{\,11}}{h_{\,11}} \egal \frac{\delta \beta}{\beta} \egal \frac{\delta F}{H} \egal \frac{\delta D}{H} \egal 0.05 \,.
\end{align*} 
For the sake of clarity, we denote by:
\begin{align*}
\delta p \, \eqdef \, \bigl(\,\delta h_{\,11} \,,\, -\,\delta \beta \,,\, \delta F \,,\, -\, \delta D \,\bigr) \,,
\end{align*}
which by convention induces a decrease of entering heat flux on the outside surface of the facade. The numerical model composed of the governing and the four sensitivity equations is computed with the following discretisation parameters: $\Delta t \egal 36 \ \mathsf{s}$ and $\Delta x \egal \Delta y \egal 3.7 \ \mathsf{mm}\,$. The climatic data from Marseille is selected. The computational ratio is $R_{\,\cpu} \egal 3.77 \ \mathsf{s\,/\,days}\,$. As expected, the computational time increases compared to the one in Section~\ref{sec:thermal_efficiency} since there are additional equations to solve.

\begin{figure}
\centering
\subfigure[\label{fig:vinffeb7d_ft}]{\includegraphics[width=.45\textwidth]{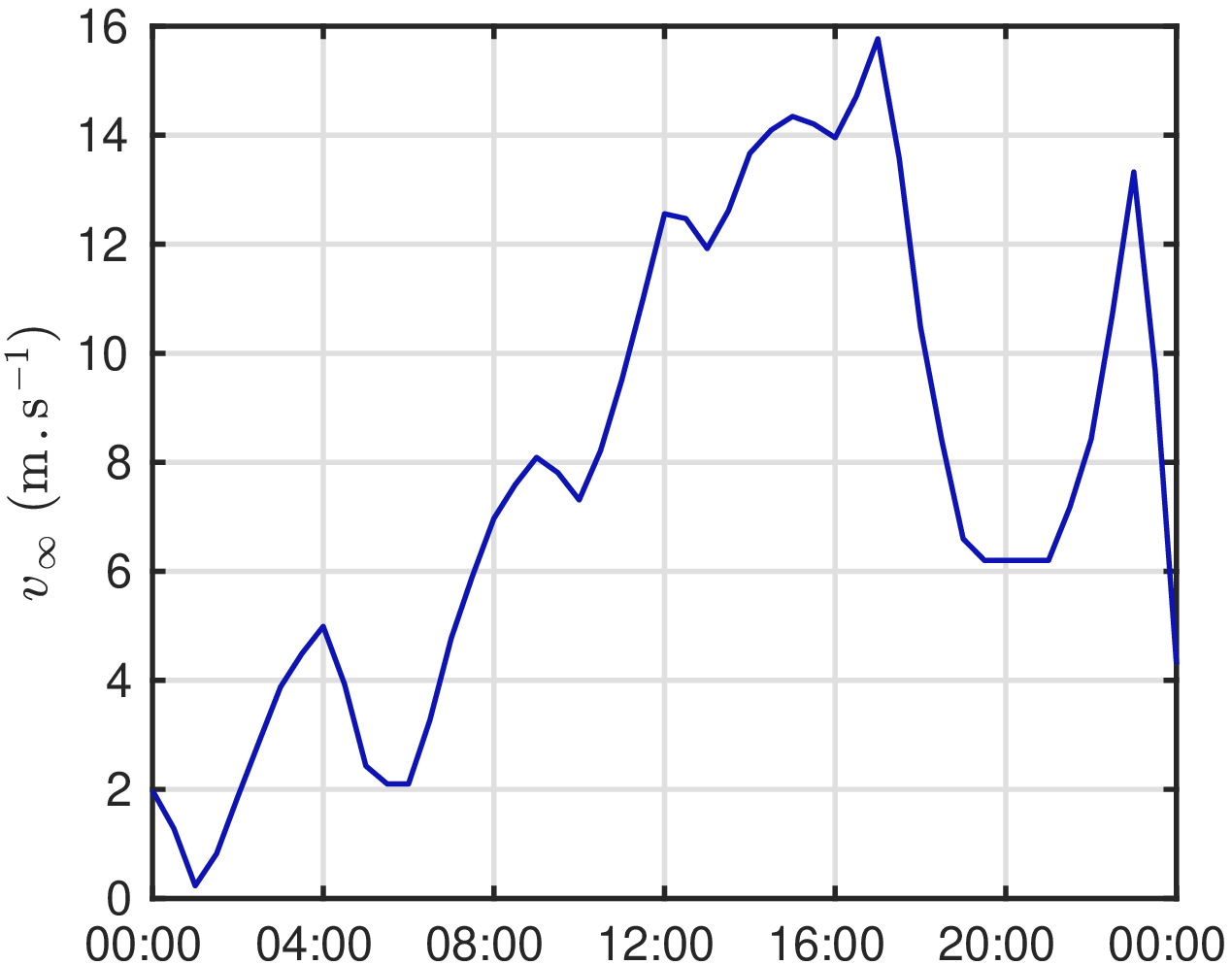}} \hspace{0.5cm}
\subfigure[\label{fig:h1inf_ft}]{\includegraphics[width=.45\textwidth]{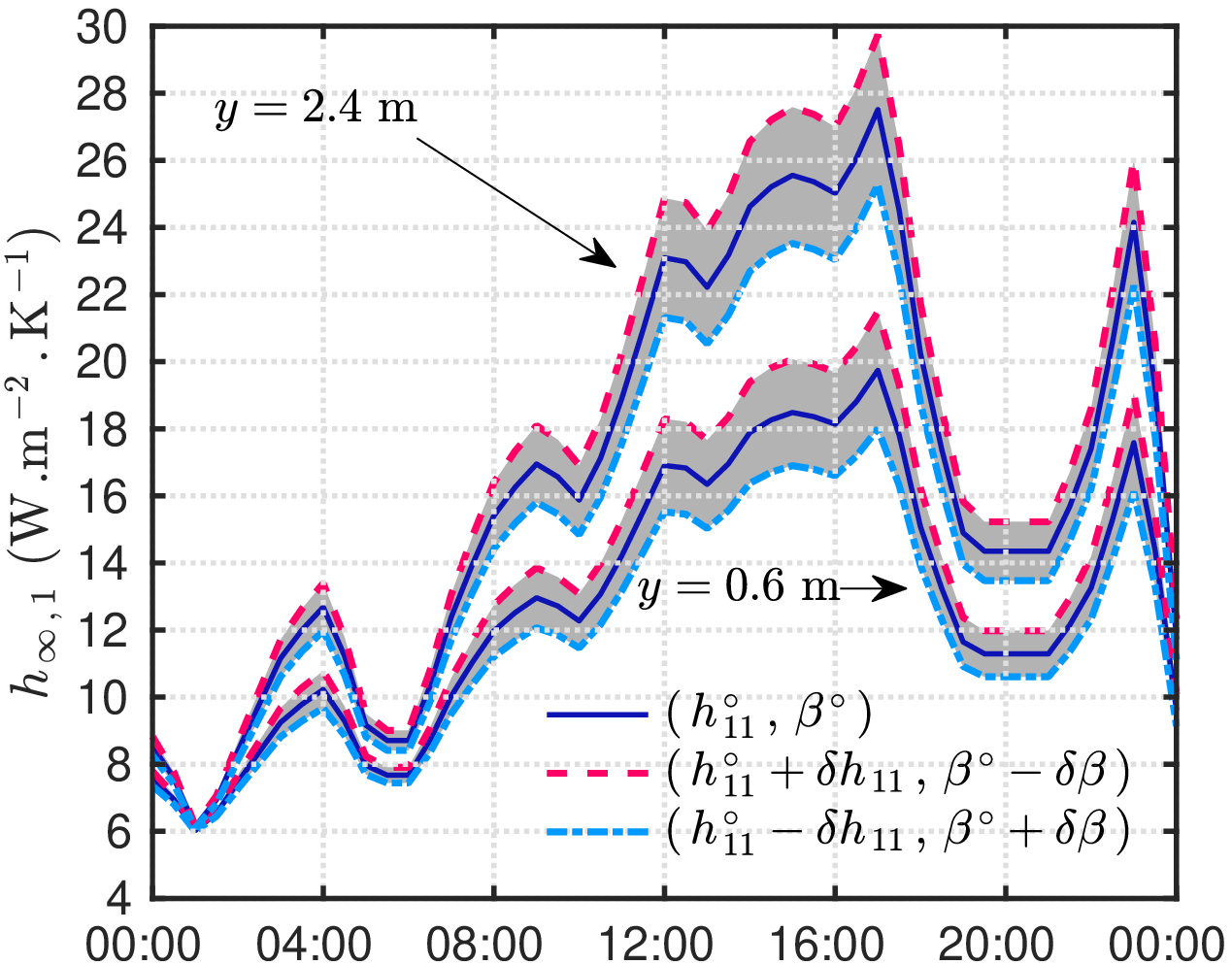}} 
\caption{Time evolution of the wind velocity \emph{(a)}. Taylor extension of the outside surface transfer coefficient at $y \egal 0.6 \ \mathsf{m}$ according to a slight variation of parameters $h_{\,11}^{\,\circ}$ and $\beta^{\,\circ}$ \emph{(b)}.
All results are plotted for February $7^{\,th}\,$.}
\label{fig:h1inf}
\end{figure}

\begin{figure}
\centering
\subfigure[\label{fig:hF_ft}]{\includegraphics[width=.45\textwidth]{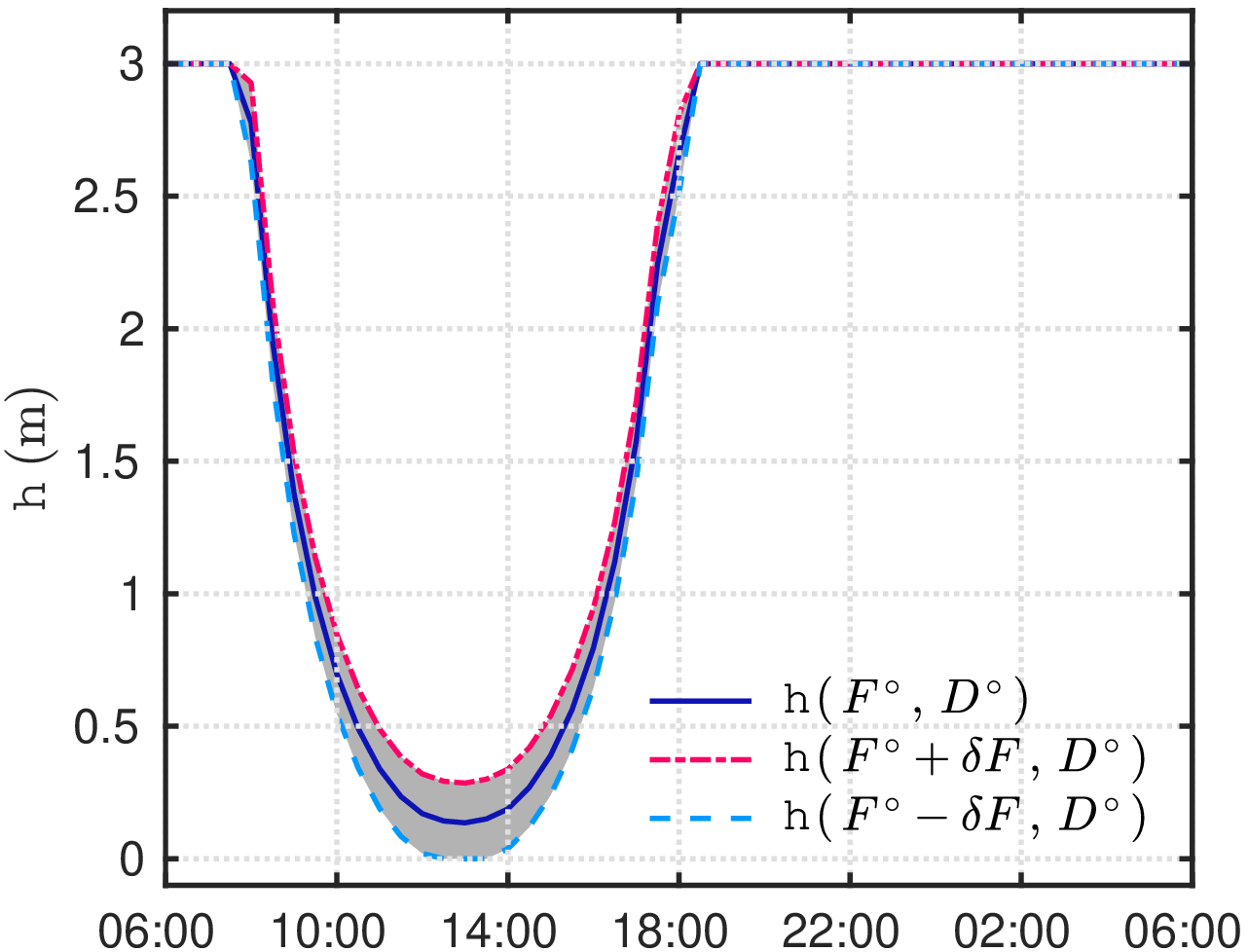}} \hspace{0.5cm}
\subfigure[\label{fig:hD_ft}]{\includegraphics[width=.45\textwidth]{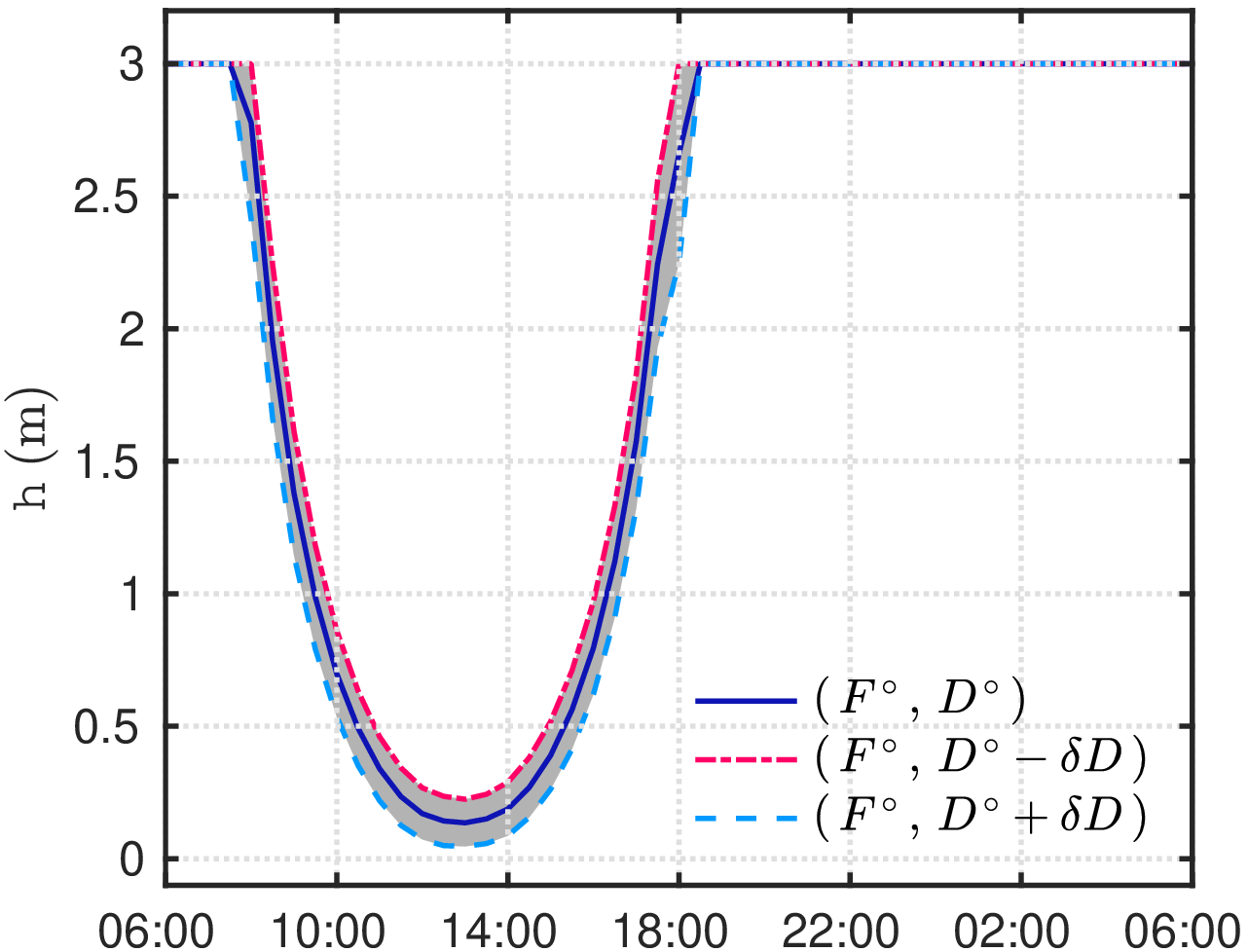}} \\
\subfigure[\label{fig:q1infF_ft}]{\includegraphics[width=.45\textwidth]{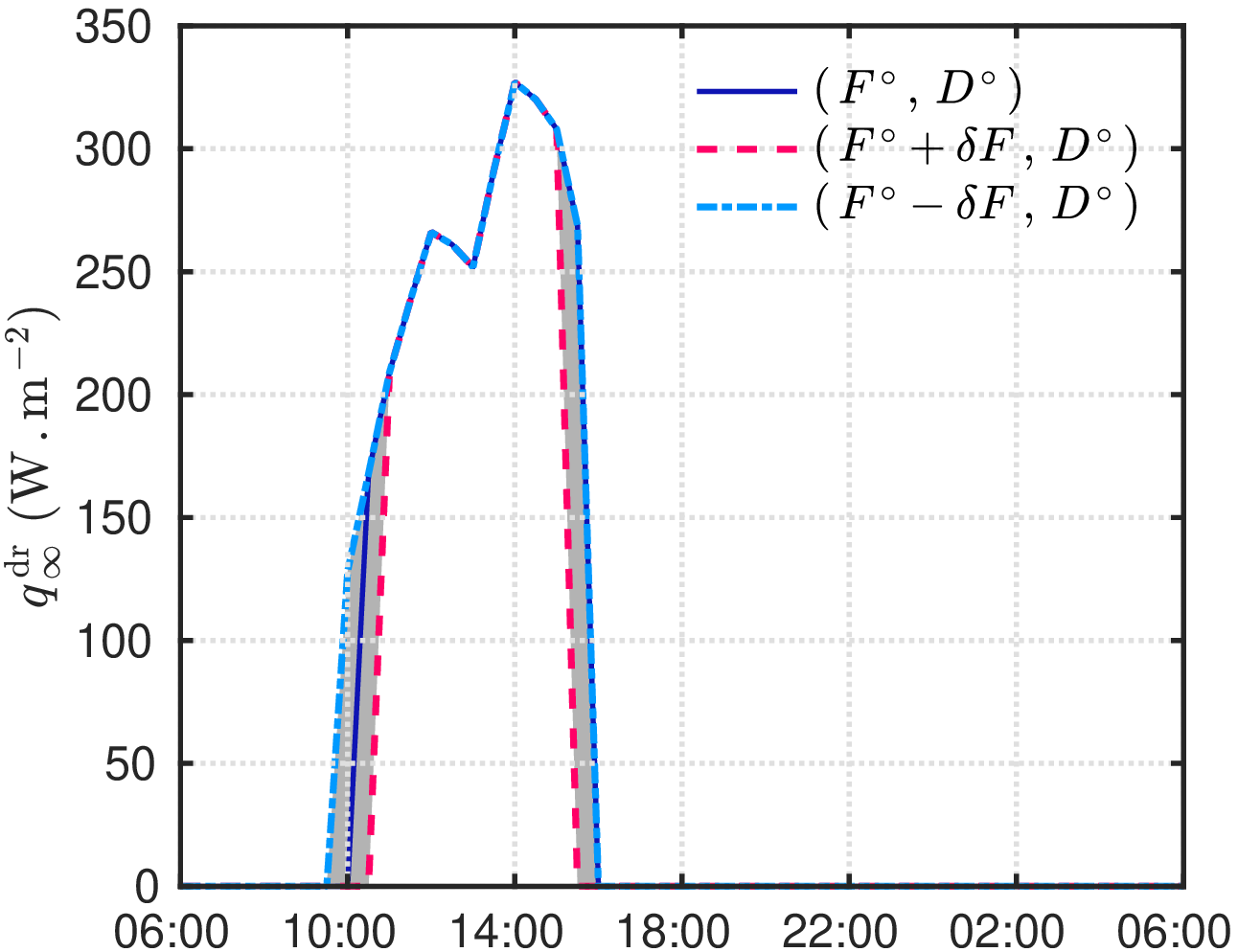}} \hspace{0.5cm}
\subfigure[\label{fig:q1infD_ft}]{\includegraphics[width=.45\textwidth]{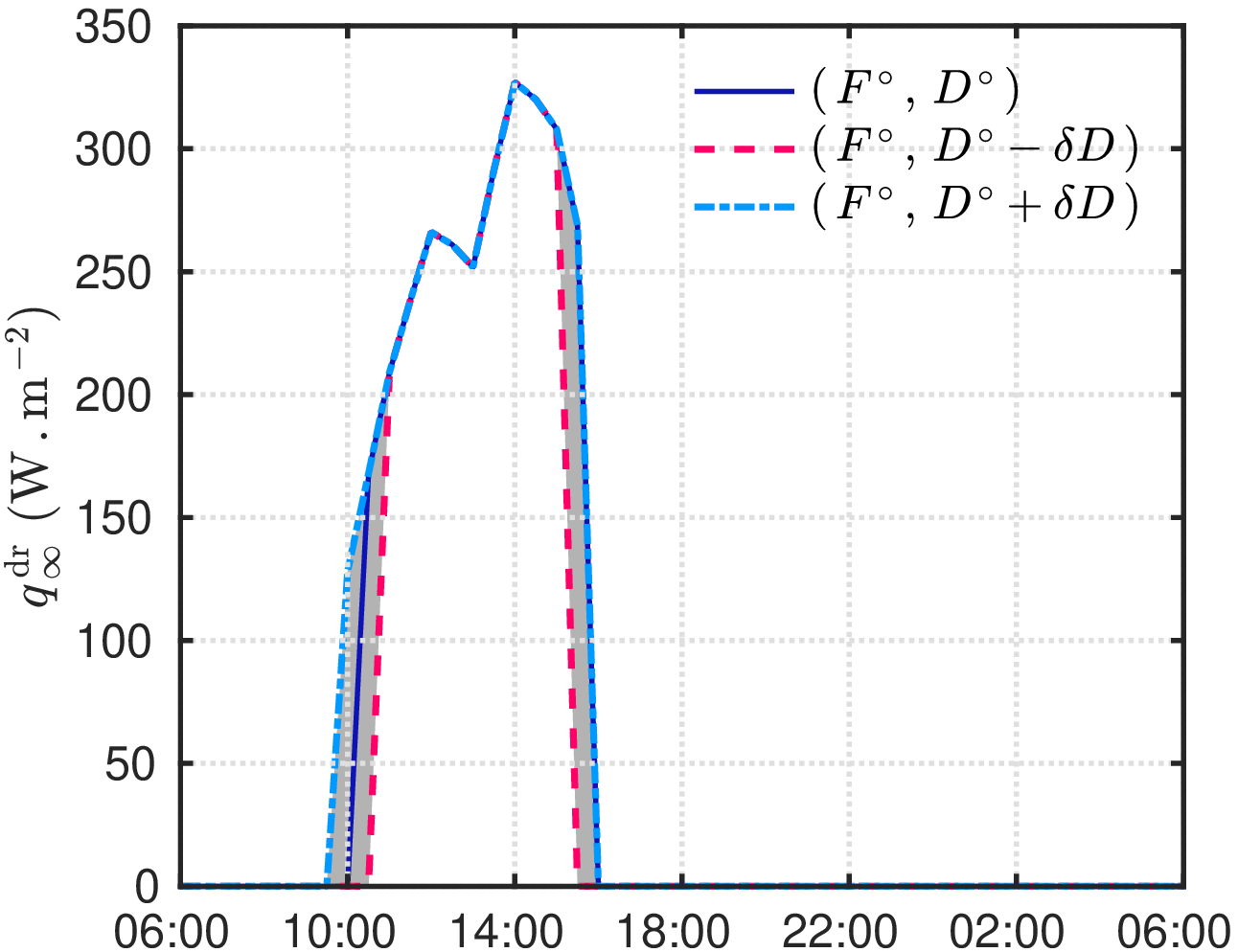}} 
\caption{Variation of the shadow height according to a slight variation of the front building facade height \emph{(a)} and distance \emph{(b)}. Taylor extension of the direct flux incident on the facade at $y \egal 0.6 \ \mathsf{m}$ according to a slight variation of the front building facade height \emph{(c)} and distance \emph{(d)}.
All results are plotted for February $7^{\,th}\,$.}
\label{fig:q1inf}
\end{figure}

First, the uncertainties on the boundary condition is presented. The modification of the surface transfer coefficient of the facade is illustrated in Figure~\ref{fig:h1inf_ft} according to slight variations of both parameters $h_{\,11}$ and $\beta\,$. The parameters variation have a higher impact on the modeling of the heat transfer coefficient between $12:00$ and $16:00$ due to higher wind velocity, as noted in Figure~\ref{fig:vinffeb7d_ft}. Then, Figures~\ref{fig:hF_ft} and \ref{fig:hD_ft} show the variation of shadow height on the facade during one day in winter according to slight variations of the front building facade height and distance, respectively. An increase of the height or a decrease of the distance induces a larger shadow on the building facade. Around noon, the shadow increases scales with $15 \ \mathsf{cm}$ and $3.4 \ \mathsf{cm}$ due to facade height and distance variations, respectively. As a consequence, the incident direct radiation flux on the facade is altered. Figure~\ref{fig:q1infF_ft} gives the variation of the flux for a point located at $y \egal 0.6 \ \mathsf{m}$ according to a slight variation of the front building facade height. At this specific height, the influence only occurs in the morning and afternoon, when the sun is rising or decreasing. At noon, there is no variation since the point $y \egal 0.6 \ \mathsf{m}$ is in the sunlit with or without variations of the front building height. Similar results are obtained for the variation of the incident flux according to a variation of the front building distance. One can note that $F$ and $D$ have opposite effects. An increase of the height and a decrease of the distance leads to an increase of the shadow height, respectively.

\begin{figure}
\centering
\subfigure[\label{fig:Tfeb7d_ft}]{\includegraphics[width=.45\textwidth]{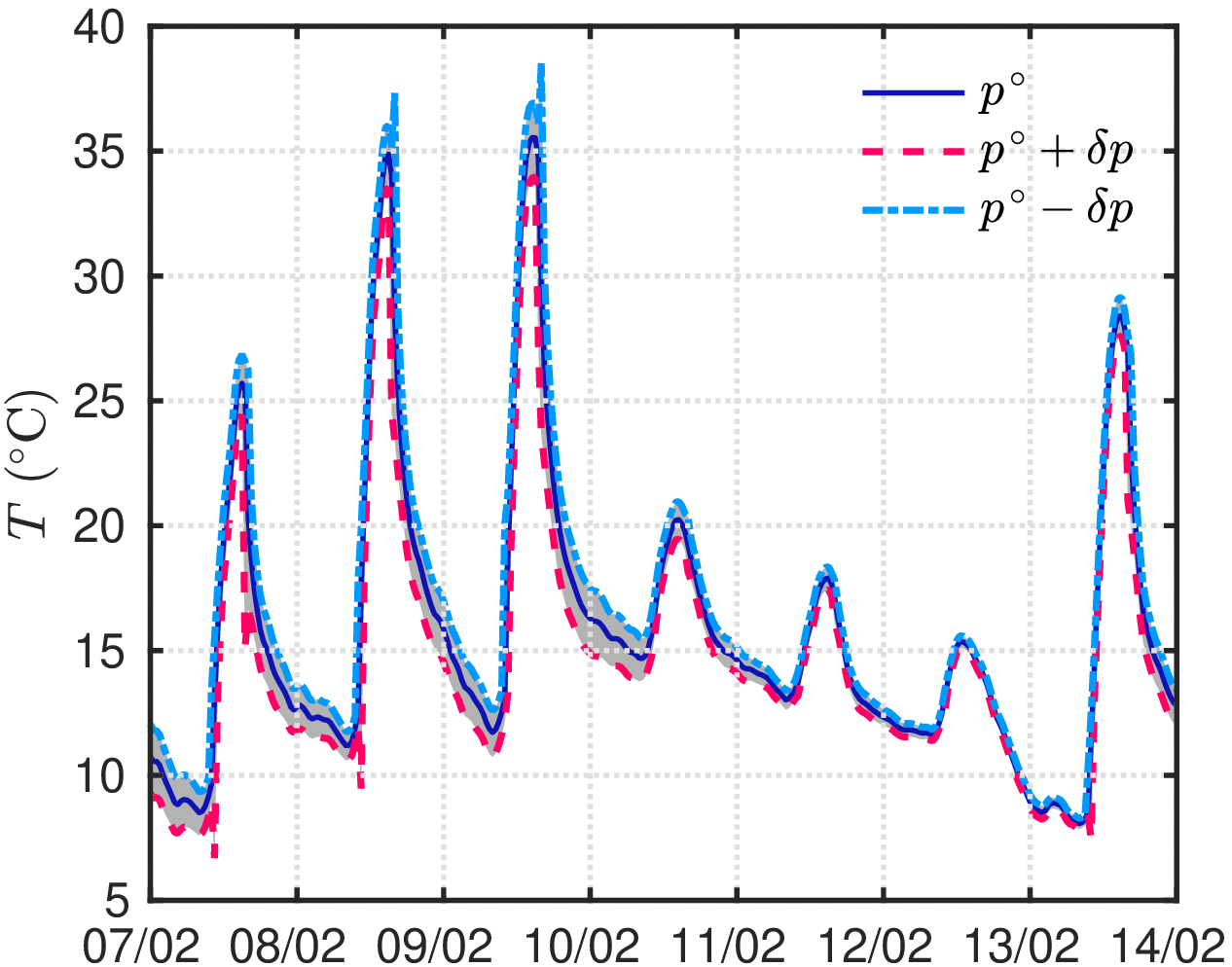}} \hspace{0.5cm}
\subfigure[\label{fig:Twinter01feb_ft}]{\includegraphics[width=.45\textwidth]{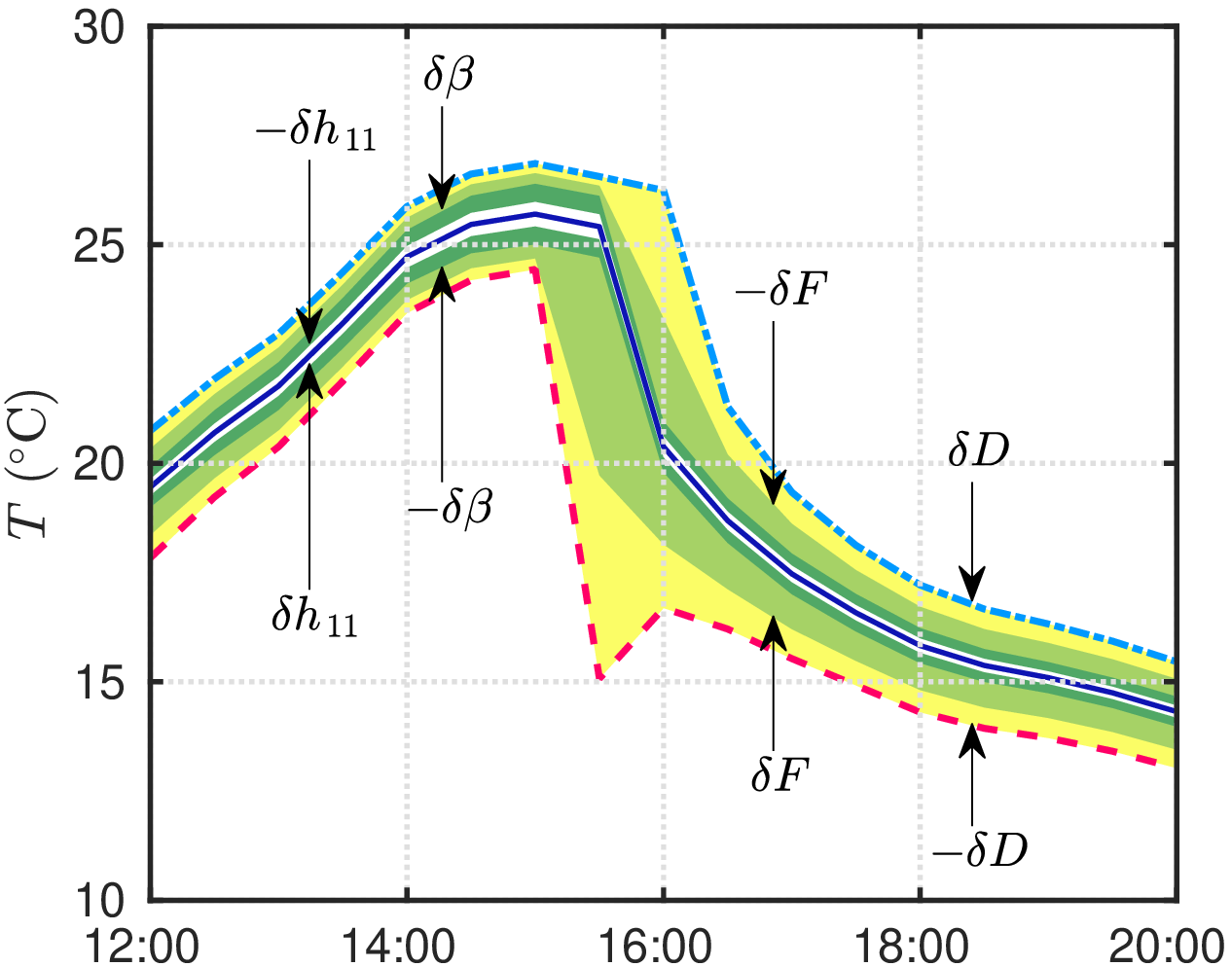}} \\
\subfigure[\label{fig:jRtotW_ft}]{\includegraphics[width=.45\textwidth]{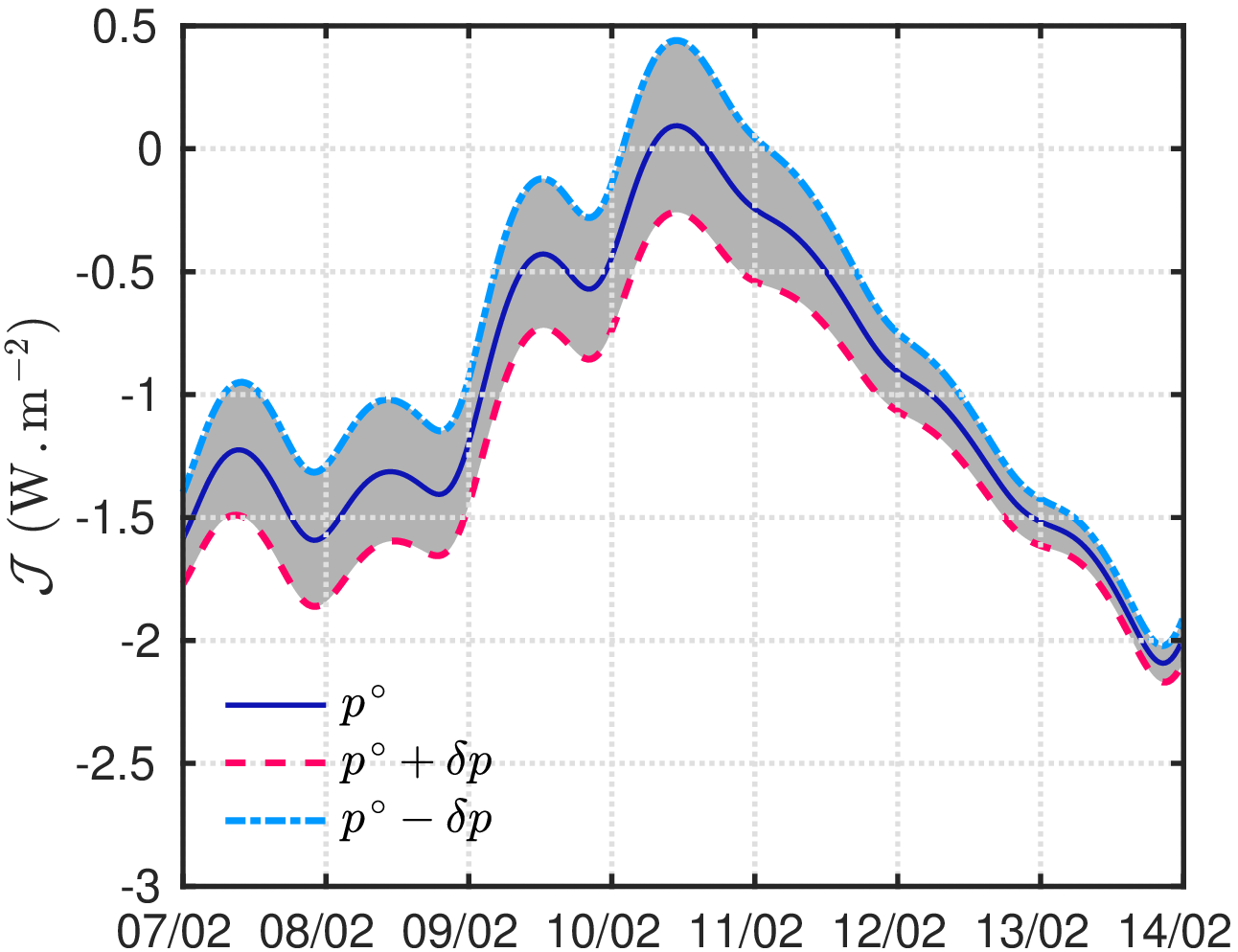}} \hspace{0.5cm}
\subfigure[\label{fig:jRtotS_ft}]{\includegraphics[width=.45\textwidth]{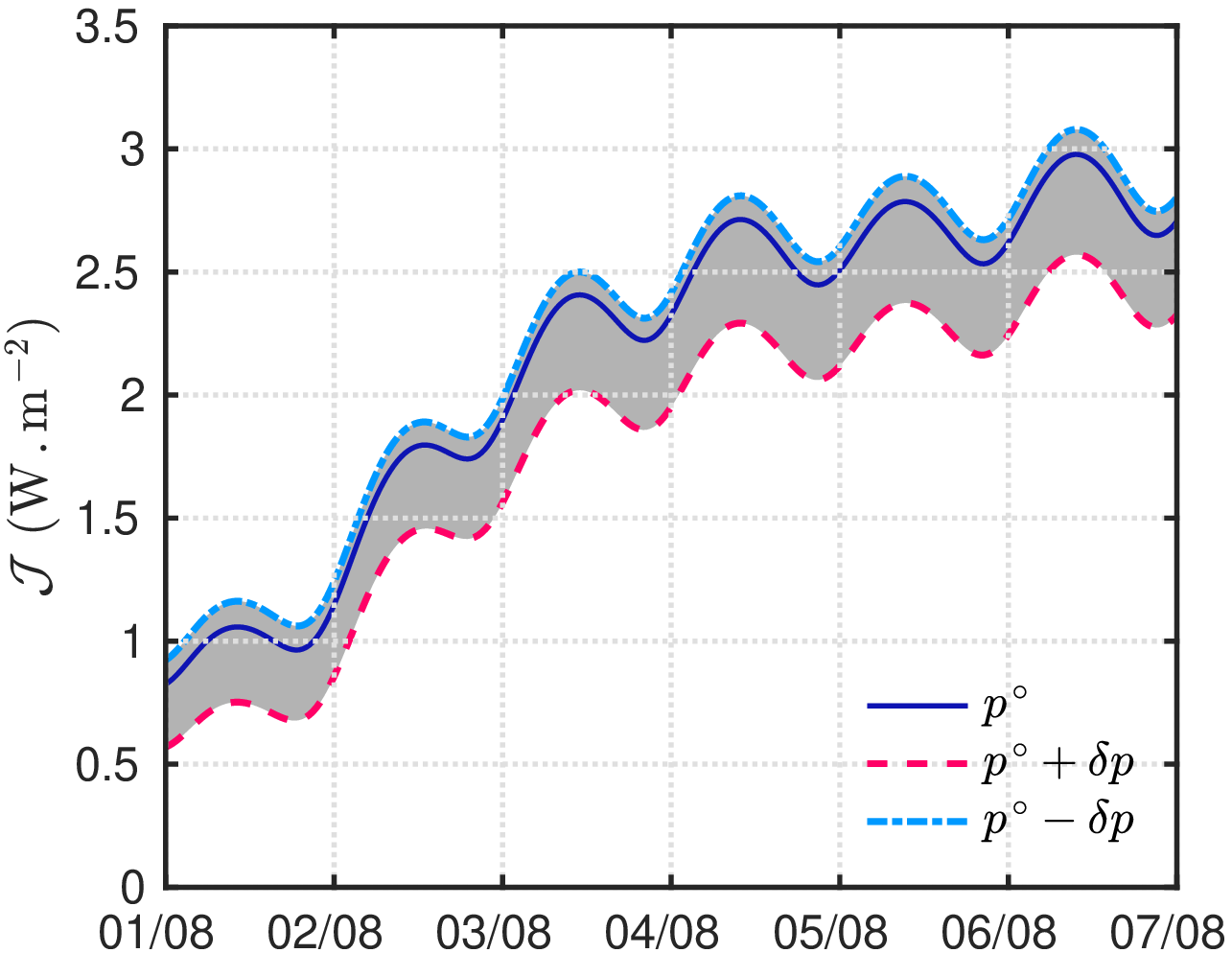}}
\caption{Taylor extension of the temperature on the facade at $(\,x\,,\,y\,) \egal (\,0\,,\,0.6\,) \ \mathsf{m}$ according to a slight variation of the four parameters of interests \emph{(a)}. Detailed contribution of the sensitivities to each of the four parameters on the temperature on the facade at $(\,x\,,\,y\,) \egal (\,0\,,\,0.6\,) \ \mathsf{m}$ for February $7^{\,th}$ \emph{(b)}. Taylor extension of the total heat flux on the inside boundary for a slight variation of the four parameters of interests in winter \emph{(c)} and summer \emph{(d)} periods.}
\end{figure}

\begin{figure}
\centering
\subfigure[\label{fig:E_ft}]{\includegraphics[width=.95\textwidth]{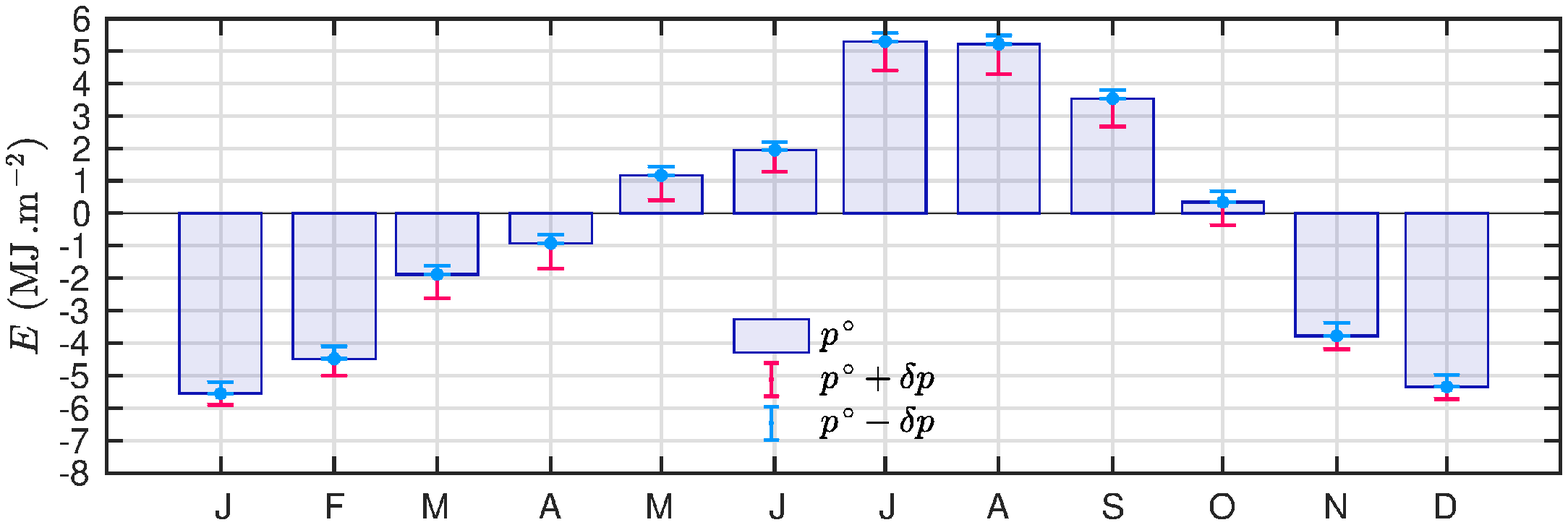}} \\
\subfigure[\label{fig:dEm_ft}]{\includegraphics[width=.45\textwidth]{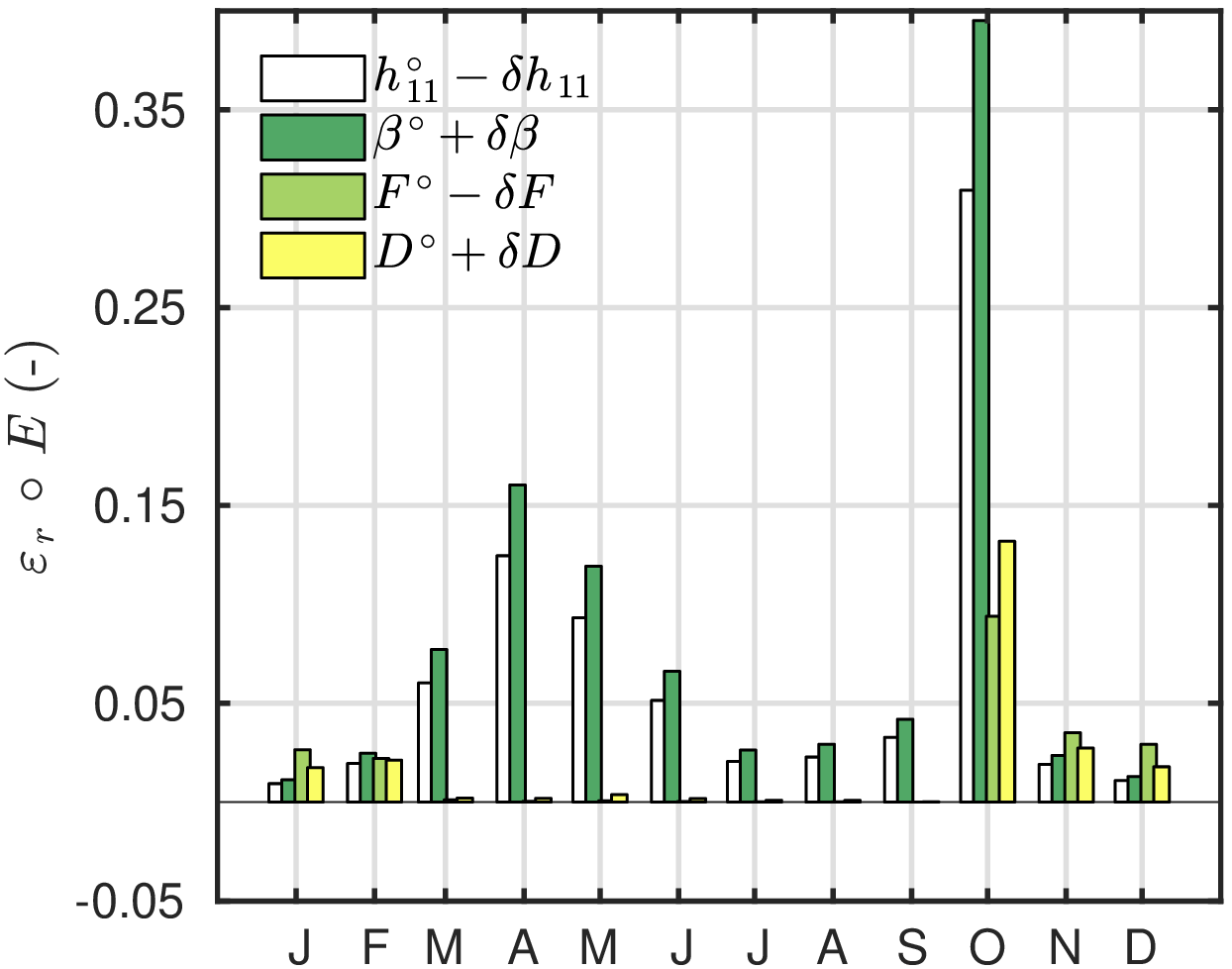}} \hspace{0.5cm}
\subfigure[\label{fig:dEp_ft}]{\includegraphics[width=.45\textwidth]{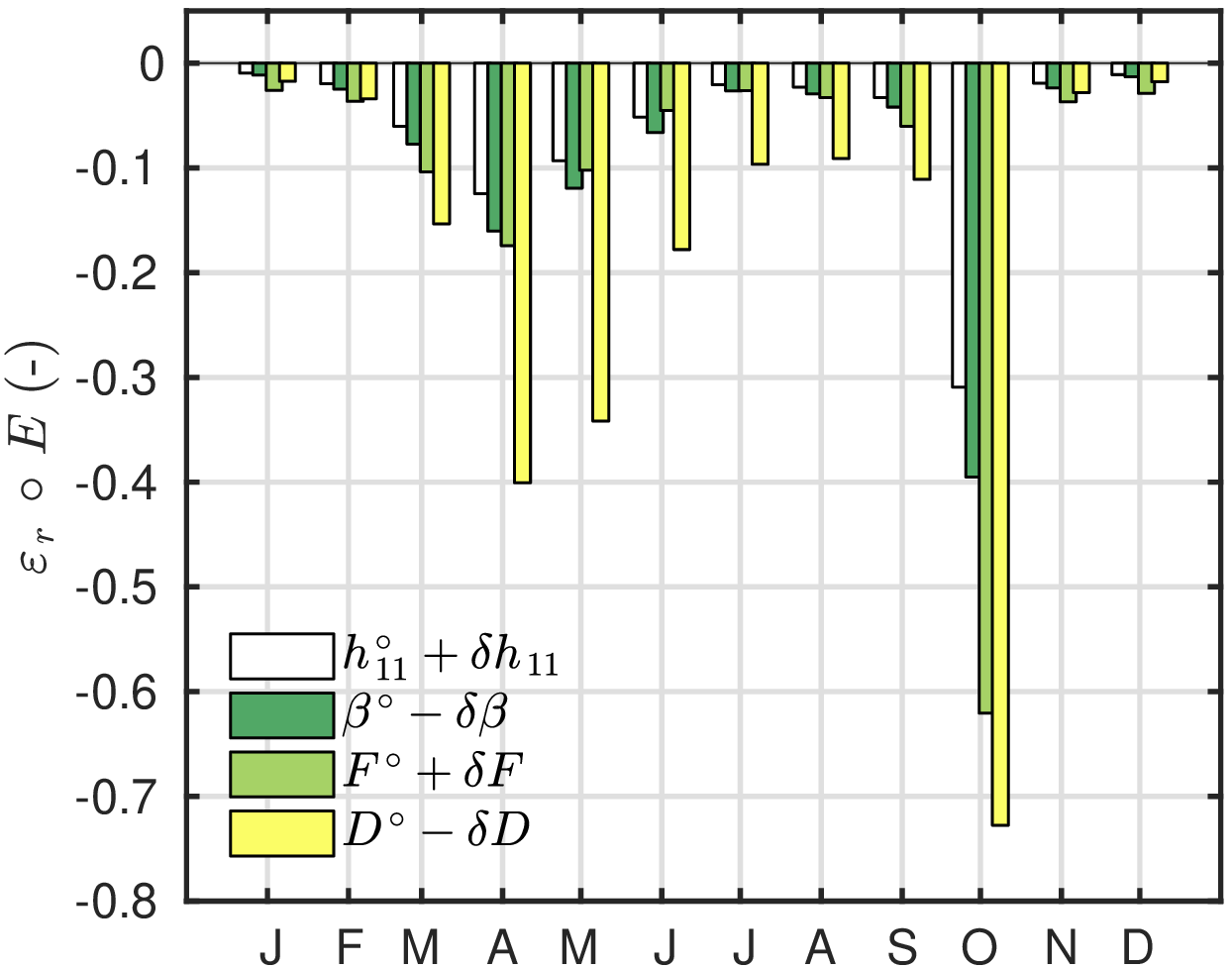}} 
\caption{Taylor extension of the thermal loads of the facade at according to a slight variation of the four parameters of interests \emph{(a)}. Detailed contribution of the relative variation of each parameters on the thermal loads of the facade\emph{(b,c)}.}
\end{figure}

Figures~\ref{fig:h1inf} and ~\ref{fig:q1inf} highlight the time and space changes of the boundary conditions modeling due to uncertainties in the input parameters. Those modifications impact the heat transfer process through the whole facade. Using the \textsc{Taylor} expansion and the sensitivity coefficients of each parameters, it is possible to evaluate the impact of those variations on the fields. Figure~\ref{fig:Tfeb7d_ft} shows the temperature variation during one week according to a modification on the all parameters. One can observe that using this approach a time varying temperature sensitivity is computed. A detailed analysis is carried in Figure~\ref{fig:Twinter01feb_ft} to evaluate the impact of the change of each parameter on the temperature. For this day and this point, the change of the facade height contributes to $5 \ \mathsf{^{\,\circ}C}$ of variation. This parameter has a strong impact compared to the three others. Similarly, the extension can be carried for the total flux $\mathcal{J}$ on the inside part of the facade. Figures~\ref{fig:jRtotW_ft} and \ref{fig:jRtotS_ft} give the variation according to change of all parameters for winter and summer weeks. The sensitivity of the flux can reach $0.5 \ \mathsf{W}$ for this period. It varies according to time since it depends on the magnitude of the radiation flux and wind velocity. 
Similarly, the influence of the boundary conditions uncertainties on the thermal loads of the facade can be investigated in Figure~\ref{fig:E_ft}. The variability of the facade energy efficiency depending on the modeling of the boundary conditions is presented. The maximum variation of $E$ occurs in September and reaches $0.91 \ \mathsf{MJ\,.\,m^{\,-2}}\,$. The variation $+\,\delta p$ has a higher impact on the thermal loads. It is probably due to the decrease of the incident radiation flux in such configuration, which strongly affects the heat transfer process through the facade. A detailed investigation is presented in Figures~\ref{fig:dEm_ft} and \ref{fig:dEp_ft}. The relative variation of the thermal loads is given according to positive or negative change of each parameter. The sensitivity varies according to the month. In Figure~\ref{fig:dEm_ft}, the parameters $h_{\,11}$ and $\beta$ have a higher impact than other parameters, particularly in April, May and October. It corresponds to months with important wind velocity values. Furthermore, the thermal loads are almost not sensible to a decrease and increase of the building front height and distance, respectively. Indeed, those changes only modify the incident radiation flux in the morning as noted in Figures~\ref{fig:q1infF_ft} and \ref{fig:q1infD_ft}. In Figure~\ref{fig:dEp_ft}, those two parameters have more impact when the height and distance increases and decreases, respectively. Particularly in October, the relative variation can reach $-70\,\%$ due to a reduction of the building front distance. Except for the months of high wind velocity, the parameters $h_{\,11}\,$, $\beta$ and $F$ have a comparable sensitivity on the thermal loads. 

\section{Conclusion}
\label{sec:conclusion}

The development of more accurate numerical models is essential to assess the energy efficiency of buildings considering the influence of urban environment. Due to computational issues, most of the today's building simulation programs proposes a one-dimensional approach to predict the phenomena of heat transfer within the building envelope. It is worth of investigation to propose innovative numerical methods to build a reliable model with high reliability. This paper presented an efficient numerical model for two--dimensional heat transfer in building facade, considering complex outside boundary conditions with shading effects and varying surface heat transfer coefficient. The main advantage is the fast computation of the solution, \emph{i.e.} the temperature field, and its sensitivity on the modeling of the boundary conditions. Using a \textsc{Taylor} expansion of the solution and the sensitivity functions, it is possible to evaluate the two-dimensional modeling of the boundary conditions on the energy efficiency.  

The numerical models are described in Section~\ref{sec:numerical_model}. It is based on the \DF ~scheme. It provides an explicit formulation, which enables a more direct treatment of the nonlinearities. An important advantage is the relaxed stability condition of the scheme compared to the traditional \Eu ~explicit approach. A first case is considered with an analytical solution in Section~\ref{sec:validation_case}. It validates the theoretical results and highlights the efficiency of the proposed numerical model. A perfect agreement is remarked with the analytical solution and the three other schemes: ADI, \Eu ~implicit and the \Eu ~explicit. The \DF ~model is the one proposing the best compromise between high accuracy of the solution and reduced computational efforts.

In Section~\ref{sec:real_case_study}, a more realistic case study was presented and investigated. The heat transfer occurs in a whole building facade. Shading effects are induced by the facing buildings of a urban environment. Thus, the incident radiation flux on the facade varies according to the height of the shadow. Both are computed using the pixel counting technique implemented in the \texttt{Domus} building simulation program. The outside heat transfer coefficient is varying according to the wind velocity and to the height of the facade using an empirical correlation obtained from the literature. In this way, the boundary conditions are modeled in two-dimensions, \emph{i.e.} depending on time and height. The model enables to compute accurately the two-dimensional fields with a reduced computational effort. Then, a comparison is carried for almost $120$ cities in France between the two-dimensional approach and the traditional one-dimensional one. The highest error on the prediction of the physical phenomena occurs in regions with high magnitude of wind and high short--wave radiation flux. Last, a sensitivity analysis is carried out using a derivative-based approach to highlight the most influencing parameters in the modeling of the two-dimensional boundary conditions. The influence of each parameter can be analyzed according to the time line. The model of the surface convective heat transfer coefficient has a significant effect on the solution for months with high wind velocity. A combined increase and decrease of the height and distance front building can induce a relative variation of $-70\ \%$ on the prediction of the thermal loads. 

Further work should be dedicated to implement such efficient numerical models in building simulation programs to simulate in a city scale. Particularly, the explicit formulation of the proposed model is a promising feature for future implementation and coupling with other numerical tools.

\section*{Acknowledgments}

The authors acknowledge the French and Brazilian agencies for their financial support through the project CAPES--COFECUB, as well as the CNPQ of the Brazilian  Ministry of Science, Technology and Innovation, for co-funding. 

\section*{Nomenclature and symbols}

\begin{tabular}{|cll|}
\hline
\multicolumn{3}{|c|}{\emph{Physical parameters}} \\ \hline
\multicolumn{3}{|c|}{Latin letters} \\ 
$c$ & volumetric heat capacity & $\unit{J\,.\,m^{\,-3}\,.\,K^{\,-1}}$ \\
$D$ & front building distance & $\unit{m}$ \\
$E$ & thermal loads & $\unit{J\,.\,m^{\,-2}}$ \\
$F$ & front building height & $\unit{m}$ \\
$\mathsf{h}$ & shadow height & $\unit{m}$ \\
$h\,,\, \overline{h}\,,\,h_{\,10} \,,\, h_{\,11}$ & surface heat transfer coefficient & $\unit{W\,.\,m^{\,-2}\,.\,K^{\,-1}}$ \\
$H$ & building facade height & $\unit{m}$ \\
$I$ & total direct solar radiation & $\unit{W\,.\,m^{\,-2}}$ \\
$j\,,\, \mathcal{J}$ & heat flux & $\unit{W\,.\,m^{\,-2}}$ \\
$k$ & thermal conductivity & $\unit{W\,.\,m^{\,-1} \,.\, K^{\,-1}}$ \\
$L$ & wall length & $\unit{m}$ \\
$q$ & radiation flux & $\unit{W\,.\,m^{\,-2}}$ \\
$S$ & sunlit area ratio & $\unit{-}$ \\
$t\,,\, \tf$ & time & $\unit{s}$ \\
$T$ & temperature & $\unit{K}$ \\
$x$ & horizontal space coordinate & $\unit{m}$ \\
$y$ & vertical space coordinate & $\unit{m}$ \\
$v$ & air velocity & $\unit{m\,.\,s^{\,-1}}$ \\
\hline
\multicolumn{3}{|c|}{Greek letters} \\
$\beta$ & surface transfer coefficient & $\unit{-}$ \\
$\chi$ & Indicator function of sunlit & $\unit{-}$ \\
$\Omega_{\,t}\,,\, \overline{\Omega}_{\,t}$ & time domain & $\unit{s}$ \\
$\Omega_{\,x}\,,\,\Omega_{\,y}$ & space domain & $\unit{m}$ \\
$\Gamma_{\,1}\,,\,\Gamma_{\,2}\,,\,\Gamma_{\,3}\,,\,\Gamma_{\,4}$ & Spatial boundary & $\unit{m}$ \\
$\Phi$ & interesting output & variable \\
$\theta$ & angle between wall normal and sun position & $\unit{-}$ \\
$\Theta$ & sensitivity function & variable \\
$\varepsilon_{\,2} $ & error & unit of $\Phi$ \\
$\varepsilon_{\,r}$ & relative error & $\unit{-}$\\
\hline
\end{tabular}

\hspace{2cm}

\begin{tabular}{|cll|}
\hline
\multicolumn{3}{|c|}{\emph{Mathematical notations}} \\
\hline
\multicolumn{3}{|c|}{Latin letters} \\
$\Bi$ & \textsc{Biot} number & \\
$\Fo$ & \textsc{Fourier} number & \\
$N_{\,x}\,,\,N_{\,y}\,,\,N_{\,t}$ & number of elements & \\
$R_{\,\mathrm{cpu}}$ & CPU time ratio & \\
$u$ & dimensionless temperature & \\
\hline
\multicolumn{3}{|c|}{Greek letters} \\
$\delta$ & slight variation & \\
$\Delta x$ & space mesh & \\
$\Delta t$ & time step & \\
$\lambda_{\,x}\,,\,\lambda_{\,y}$ & \DF ~scheme coefficients & \\
$\phi$ & piece wise function &  \\
$\Sigma_{\,x}\,,\,\Sigma_{\,y}\,,\,\Sigma_{\,xy}$ & \DF ~scheme coefficients & \\
$\tau$ & accuracy \DF ~scheme coefficients & \\
\hline
\multicolumn{3}{|c|}{Subscripts and superscripts} \\
$0$ & reference value or initial condition &  \\
$\mathrm{dr}$ & direct flux component &  \\
$\mathrm{df}$ & diffuse flux component &  \\
$\mathrm{rf}$ & reflective flux component &  \\
$\star$ & dimensionless value &  \\
$\infty$ & boundary &  \\
\hline
\end{tabular}

\bibliographystyle{unsrt}  
\bibliography{references}

\newpage

\appendix

\section{\ADI ~numerical scheme}
\label{sec:ADI}

The idea of the \ADI ~(ADI) numerical scheme is to split the time step into two intermediate stages. For the first stage $ t^{\,n+\half} \egal t^{\,n} \plus \frac{1}{2}\, \Delta t\,$, the scheme considers an implicit formulation in the $x$ direction and an explicit one in the $y$ direction. Using central finite differences for the space discretisation of Eq.~\eqref{eq:heat_2D}, it yields to:
\begin{align}
\label{eq:ADI_first_stage}
u_{\,j\,i}^{\,n+\half} \moins u_{\,j\,i}^{\,n} \egal \Lambda_{\,x} \, 
\biggl(\, u_{\,j+1\,i}^{\,n+\half} \moins 2 \, u_{\,j\,i}^{\,n+\half} \plus u_{\,j-1\,i}^{\,n+\half}\,\biggr)
\plus \Lambda_{\,y} \,
\biggl(\, u_{\,j\,i+1}^{\,n} \moins 2 \, u_{\,j\,i}^{\,n} \plus u_{\,j\,i-1}^{\,n}\,\biggr) \,,
\end{align}
with
\begin{align*}
\Lambda_{\,x} \, \eqdef \, \frac{\Delta t}{2 \, \Delta x^{\,2}} \ \frac{kx}{c} \,, \qquad
\Lambda_{\,y} \, \eqdef \, \frac{\Delta t}{2 \, \Delta y^{\,2}} \ \frac{ky}{c} \,.
\end{align*}
Thus, Eq.~\eqref{eq:ADI_first_stage} gives an implicit formulation to compute $u_{\,j\,i}^{\,n+\half}\,$:
\begin{align*}
\Bigl(\, 1 \plus 2 \, \Lambda_{\,x}\,\Bigr) \, u_{\,j\,i}^{\,n+\half}
\moins \Lambda_{\,x} \, u_{\,j+1\,i}^{\,n+\half}
\moins \Lambda_{\,x} \, u_{\,j-1\,i}^{\,n+\half}
\egal
u_{\,j\,i}^{\,n}
\plus \Lambda_{\,y} \,
\biggl(\, u_{\,j\,i+1}^{\,n} \moins 2 \, u_{\,j\,i}^{\,n} \plus u_{\,j\,i-1}^{\,n}\,\biggr) \,.
\end{align*}
This system can be written in a matrix formulation:
\begin{align*}
\boldsymbol{A}_{\,x} \ \boldsymbol{u}_{\,i}^{\,n+\half} \egal \boldsymbol{b}_{\,i}
\end{align*}
with 
\begin{align*}
\boldsymbol{u}_{\,i}^{\,n+\half} \egal
\begin{bmatrix}
u_{\,1\,i}^{\,n+\half} \\
\vdots \\
u_{\,N_{\,x}\,i}^{\,n+\half}
\end{bmatrix} \,.
\end{align*}
The system is then solved for $i \, \in \, \bigl\{\, 1 \,,\, \ldots \,,\, N_{\,y}\,\bigr\}\,$. The second stage enables to compute $u_{\,j\,i}^{\,n+1}$ from $u_{\,j\,i}^{\,n+\half}$. For this, it assumes an implicit formulation on the $y$ direction and an explicit one on the $x$ direction:
\begin{align*}
u_{\,j\,i}^{\,n+1} \moins u_{\,j\,i}^{\,n+\half} \egal \Lambda_{\,x} \, 
\biggl(\, u_{\,j+1\,i}^{\,n+\half} \moins 2 \, u_{\,j\,i}^{\,n+\half} \plus u_{\,j-1\,i}^{\,n+\half}\,\biggr)
\plus \Lambda_{\,y} \,
\biggl(\, u_{\,j\,i+1}^{\,n+1} \moins 2 \, u_{\,j\,i}^{\,n+1} \plus u_{\,j\,i-1}^{\,n+1}\,\biggr) \,,
\end{align*}
which can be formulated into an implicit expression:
\begin{align*}
\Bigl(\, 1 \plus 2 \, \Lambda_{\,y}\,\Bigr) \, u_{\,j\,i}^{\,n+1}
\moins \Lambda_{\,y} \, u_{\,j+1\,i}^{\,n+1}
\moins \Lambda_{\,y} \, u_{\,j-1\,i}^{\,n+1}
\egal
u_{\,j\,i}^{\,n+\half}
\plus \Lambda_{\,x} \,
\biggl(\, u_{\,j\,i+1}^{\,n+\half} \moins 2 \, u_{\,j\,i}^{\,n+\half} \plus u_{\,j\,i-1}^{\,n+\half}\,\biggr) \,.
\end{align*}
Again the system is formulated as:
\begin{align*}
\boldsymbol{A}_{\,y} \ \boldsymbol{u}_{\,j}^{\,n+1} \egal \boldsymbol{b}_{\,j}
\end{align*}
with 
\begin{align*}
\boldsymbol{u}_{\,j}^{\,n+1} \egal
\begin{bmatrix}
u_{\,j\,1}^{\,n+1} \\
\vdots \\
u_{\,j\,N_{\,y}}^{\,n+1}
\end{bmatrix} \,,
\end{align*}
and solved for $j \, \in \, \bigl\{\, 1 \,,\, \ldots \,,\, N_{\,x}\,\bigr\}\,$. The accuracy of the scheme is second order $\mathcal{O}\bigl(\, \Delta x^{\,2} \,,\, \Delta y^{\,2} \,,\, \Delta t^{\,2} \,\bigr)\,$. The scheme is unconditionally stable.

\end{document}